%% file: main.tex
\pdfoutput=1

\RequirePackage{etex}

\documentclass[a4paper]{article}
\usepackage{anysize}
\marginsize{2cm}{2cm}{2cm}{2cm}
\usepackage{authblk}
\usepackage[utf8]{inputenc}

\usepackage{lipsum}

\usepackage{hyperref,physics}
\usepackage{ifthen}
\usepackage{amsmath,amsthm,amssymb}
\usepackage{verbatim}
\usepackage{array}
\usepackage{mathtools}

\usepackage{multicol}

\usepackage{multirow}

\usepackage{enumitem}

\usepackage[export]{adjustbox}

\usepackage{soul}

\usepackage{subcaption}

\usepackage{algorithm}
\usepackage{algpseudocode}

\usepackage{lipsum}
\usepackage[dvipsnames,table]{xcolor}

\usepackage{nicematrix}

\usepackage{capt-of,etoolbox}

\usepackage[nowidering]{yhmath}

\newcommand{\R}[0]{\mathbb{R}}

\newcommand{\Span}[1]{\operatorname{span}\!\left(#1\right)}

\newcommand\restr[2]{{%
    \left.\kern-\nulldelimiterspace %
    #1 %
    \vphantom{\big|} %
    \right|_{#2} %
}}

\newcommand{\noinitial}[1]{}

\definecolor{SO}{HTML}{e66101}
\definecolor{SOR}{HTML}{fdb863}
\definecolor{ST}{HTML}{5e3c99}
\definecolor{STR}{HTML}{b2abd2}

\usepackage{tikz,pgfplots}
\pgfplotsset{compat = newest}
\usetikzlibrary{pgfplots.groupplots}
\usetikzlibrary{graphs, graphs.standard}
\usetikzlibrary{quotes}
\usepackage{tikz-network}

\usepackage[compat=0.6]{yquant}
\usetikzlibrary{fit}
\usetikzlibrary{tikzmark}

\usetikzlibrary{fadings,shapes.arrows,shadows,intersections}

\tikzfading[name=arrowfading, top color=transparent!0, bottom color=transparent!95]
\tikzset{arrowfill/.style={top color=black!2, bottom color=black!80, general shadow={fill=black, shadow yshift=-0.8ex, path fading=arrowfading}}}
\tikzset{arrowstyle/.style={draw=black,arrowfill, double arrow,minimum height=#1, double arrow,
single arrow head extend=.4cm,}}

\usepackage{booktabs}

\usepackage{simplewick}     

\usepackage[
backend=biber,
style=numeric,
url=false,
doi=false
]{biblatex}

\usepackage[rightmargin=2cm]{thmbox}
\newtheorem[L]{theorem}{Theorem}
\newtheorem[L]{proposition}{Proposition}
\newtheorem[L]{lemma}{Lemma}
\newtheorem[L]{remark}{Remark}
\newtheorem[L]{corollary}{Corollary}
\newtheorem[L]{problem}{Problem}
\newtheorem[L]{condition}{Condition}
\newtheorem[L]{example}{Example}
\newtheorem[L]{definition}{Definition}

\title{Quantum reservoir computing using the stabilizer formalism for encoding classical data}
\date{\today}

\author[$1,2$]{Franz G. Fuchs}
\author[$1$]{Alexander J. Stasik}
\author[$1$]{Stanley Miao}
\author[$3$]{Ola Tangen Kulseng}
\author[$1$]{Ruben Pariente Bassa}
\affil[$1$]{SINTEF AS, Department of Mathematics and Cybernetics, Oslo, Norway}
\affil[$2$]{Department of Mathematics, University of Oslo, Norway
}
\affil[$3$]{NTNU, Department of Physics, Trondheim, Norway}

\usepackage{capt-of,etoolbox}

\begin{document}

\maketitle

\abstract{
Utilizing a quantum system for reservoir computing has recently received a lot of attention.
Key challenges are related to how on can optimally en- and decode classical information, as well as what constitutes a good reservoir.
Our main contribution is a generalization of the standard way to robustly en- and decode time series into subspaces defined by the cosets of a given stabilizer.
A key observation is the necessity to perform the decoding step, which in turn ensures a consistent way of encoding.
This provides a systematic way to encode classical information in a robust way.
We provide a numerical analysis on a discrete time series given by two standard maps, namely the logistic and the Hénon map.
Our numerical findings indicate that the system's performance is increasing with the length of the training data.
}

\vspace{1\baselineskip}

\begin{center}
  \captionsetup{hypcap=false}        
    \includegraphics[width=.6\linewidth]{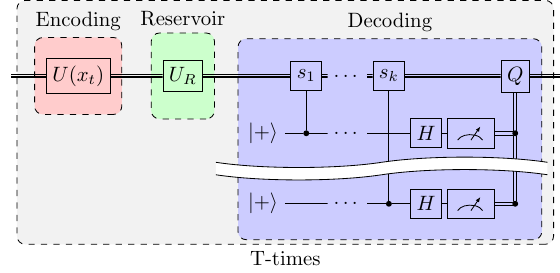}
    \captionof{figure}{
    The proposed encoding uses the cosets of a given stabilizer to encode information and ancillary qubits to perform syndrome measurements in order to project into a specific subspace.}
    \label{fig:reservoir-types}
    \captionsetup{hypcap=true}
\end{center}
\section{Introduction}

A machine learning technique called reservoir computing, also known as extreme learning machines, is an appealing approach for time series prediction\cite{nakajima2021reservoir, yan2024emerging}. The utilization of quantum systems, often in the form of quantum computers,  as a reservoir has inspired a significant amount of recent work~\cite{fujii2017harnessing,mujal2021opportunities}.
In this paper, we focus specifically on temporal tasks, where we are given a training sequence $\{x_t\}_{t=1}^{T}$ and $\{y_t\}_{t=1}^{T}$ consisting of multi-dimensional vectors. The aim of reservoir computing is to model a non-linear mapping $F$ such that
\begin{equation}
    \hat{y}_t = F(\{x_i\}_{i=1}^{t}).
\end{equation}
For each time step, quantum reservoir computing (QRC) can be described by a quantum channel of the form
\begin{equation}
    \rho_{t+1} = D(R(E(\rho_t))),
\end{equation}
where $D$, $R$, and $E$ are the decoding, reservoir, and encoding channels, respectively.
Finally, one performs \textit{linear regression} by minimizing the loss function
\begin{equation}
    L(v) = \sum_{t=1}^T \left( \hat{y}_t - y_t \right)^2
    = \sum_{t=1}^T \left( F_v(\{x_i\}_{i=1}^{t}) - y_t \right)^2
\end{equation}
where $F_v(\{x_i\}_{i=1}^{t}) = \sum_{j=1}^k v_j \langle O_j \rangle_{R(E(\rho_{t-1}))}$ and $O_j$ are given observables. This can thus be viewed as learning the observable $\widehat{O} = \sum_{j=1}^k v_j \langle O_j \rangle$ that best predicts the time series.

Choosing a ``good'' quantum reservoir has naturally received a lot of attention in the literature.
One way is to choose random Ising Hamiltonians, where it has been established
that optimal information processing capabilities for quantum reservoirs can be achieved in the ergodic phase as well as in its onset~\cite{PhysRevLett.127.100502}.

It is natural to analyze the Fourier spectrum of the encoding and reservoirs for extreme machine learning \cite{xiong2023fundamental}.
A common technique is to combine multiple small, parallel quantum systems driven by a common input sequence. 
This technique, commonly referred to as ``spatial multiplexing'', has been shown to boost the computational power of a single reservoir~\cite{nakajima2019boosting}. A similar technique is called ``temporal multiplexing''~\cite{nakajima2019boosting} where one measures the system multiple times during its evolution from one input to the next.

Regarding quantum noise, amplitude damping noise has been demonstrated to be potentially beneficial for a quantum reservoir, while depolarizing and phase damping noises have a negative effect on the performance~\cite{Domingo2023taking}.

In order to allow for the quantum reservoir to retain some of its memory, one can perform weak measurements~\cite{mujal2023time} in an online fashion, instead of the usual projective measurements.

In this paper, we propose a generalized way of en/decoding, that we show to be robust.
A recent preprint~\cite{yasuda2023quantum} introduces measuring $Z_1Z_2$ observables. However, as we show in this work, this does not scale beyond more than two such observables unless one correctly performs the decoding step, see Figure~\ref{fig:encodingVSnone}.
Furthermore, we suggest a generalization by encoding into cosets of a stabilizer subspace, allowing for more observables.
The general framework is presented in Section~\ref{sec:encoding}, followed by a description of the numerical benchmarks in Section~\ref{sec:numericalexamples}.
The results are presented in Section~\ref{sec:results}.

\section{Encoding of classical data into cosets}
\label{sec:encoding}

A standard way to extract an output signal from a quantum reservoir is to measure the first $k$ of $n$ qubits to estimate the observables $Z_1, \cdots, Z_k$.
Since by measuring we collapse only part of the state, this allows to retain some memory of the previous input. However, this has to be balanced with the number of fresh ancilla qubits provided, in order to control how mixed the state is during the evolution.
Measuring $Z_1, \cdots, Z_k$, either directly or indirectly with ancilla qubits, effectively divides the Hilbert space of the reservoir into $2^k$ cosets given by $V_{S\langle X_1, \cdots, X_k\rangle}$.
As an example measuring the two first qubits of a 3 qubit reservoir leads to the cosets $V_{S} = \Span{\ket{000}, \ket{001}}$,
$V_{SX_1} = \Span{\ket{100}, \ket{101}}$, 
$V_{SX_2} = \Span{\ket{010}, \ket{011}}$, 
$V_{SX_1X_2} = \Span{\ket{110}, \ket{111}}$.
An encoding of a time series into these cosets is given e.g., by rotations $RX_1$, $RX_2$.
This can be generalized in the following way:

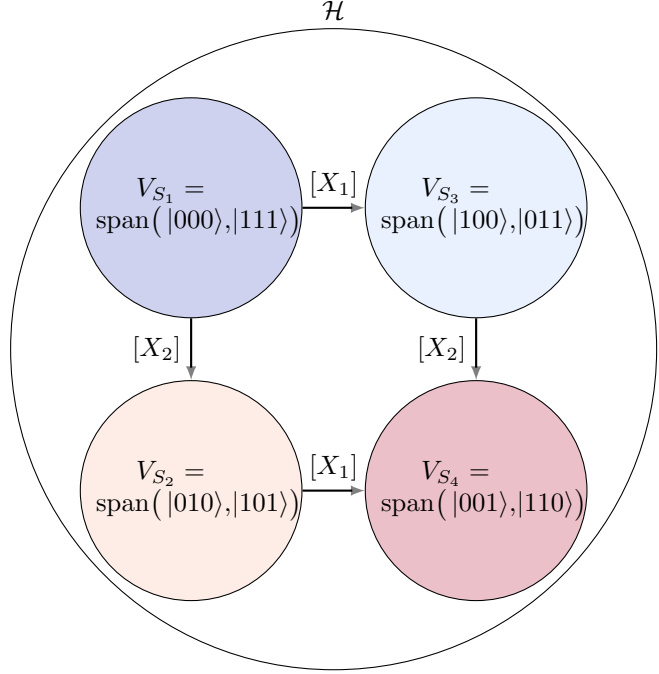
\begin{figure}
    \centering
    \input{figures/exampleStabilizerEncoding}
    \caption{A simple example to illustrate the encoding is given by $S=\langle Z_1Z_2, Z_2Z_3 \rangle$.
    The mapping between $V_{S_i}$ and $V_{S_j}$ is modulo the logical operator(s) of $S$. In this case the cosets are defined as $[P] = [P]_A = \{P\cdot A | A \in \{I,X_1X_2X_3\}$, so for instance $[X_2] = [X_1X_3]$ modulo $X_1X_2X_3$.
    In error-correction lingo those mappings would be called correctable errors.
    }
    \label{fig:exampleStab}
\end{figure}

The stabilizer formalism can be used to encode the time series into the cosets of a given stabilizer $S$.
A stabilizer group $S=\langle s_1, \cdots, s_k \rangle$ is given by $k$ independent generators from the Pauli group $G_n$ on $n$ qubits such that $-I\notin S$.
This defines a subspace, or ``code space'' $V_S=\Span{\ket{z} \ : \ s \ket{z} = \ket{z} \forall s \in S }$ with dimension $\operatorname{dim}(V_S)=2^{n-k}$.
This subspace can be used to encode $n-k$ logical qubits and there are $2^k-1$ orthogonal subspaces to $V_S$,
which we can enumerate as $V_{S_1}, \cdots, V_{S_{2^k}}$.
There are $k$ unique mappings $\{F_1, \cdots, F_k\}$ between these subspaces modulo logical X-operators, see Figure~\ref{fig:exampleStab} for an example.
We also know that $[F_i, F_j] = 0$.
These mappings are illustrated in Figure~\ref{fig:reservoir-types} and are used to encode information in the following way.

\paragraph{Encoding}
An input vector $x_t\in \R^\text{enc}$ of the trainining time series at time $t$ is encoded into the orthogonal subspaces defined by a stabilizer group $S$ in the following way.
Assuming we are in subspace $\ket{\phi} \in V_S = V_{S_1}$ we apply the following circuit:
\begin{equation}
\label{eq:encoding}
    E(\rho, t) = U(t) \rho U^\dagger(t), \quad U(t) = \prod_{j=1}^k e^{-i\beta_j x_t F_j}, \quad \beta_j \in \R .
\end{equation}
Note that there is no "Trotter-error" since all $F_j$ mutually commute~\cite{childs2021theory}.
The encoding is formulated for a one dimensional time series $x_t$, but is easy to extend to the multidimensional case.
If $x_t$ is a binary variable and $\beta=\pi$, this corresponds to a mapping from $V_S$ to $V_{S_{x_t}}$.
In the general case, the overlap with the subspaces is
$ \bra{\psi} U(t) \ket{\psi_1} = \text{Should the equation sign be removed or is something missing here?}$

\paragraph{Reservoir}
The time evolution, which is necessary to obtain a nonlinear behavior with respect to the input valuable $x_t$, is taken to be a unitary evolution operator $U_R$.
\begin{equation}
    R(\rho) = U_R \rho U_R^\dagger.
\end{equation}
The unitary evolution $U_R$ is typically obtained from the dynamics generated by Hamiltonian $H$. Often one uses Ising models for $H$ due to their easy structure and rich dynamics in terms of spectrum properties and dynamical response.

\paragraph{Decoding}
After the time evolution of the reservoir, one performs $k$ syndrome measurements observables $\{S_1, \cdots, S_k\}$.
We define the projectors onto the $\pm1$ eigenstates of $s_j$ as $\pi_j^{\pm1} = (I\pm S_k)/2$.
Then the resulting quantum channel is given by
\begin{equation}
    D(\rho,a) = Q(a)\left( \prod_{j=1}^k \pi_j^{a_j} \rho \pi_j^{a_j}\right) Q^\dagger(a),
\end{equation}
where $a_j = +1$ if auxiliary during $s_j$ syndrome-measurement is $0$, and $a_j = -1$ if syndrome measurement is $1$.
The unitary $Q(a)$ is a correction operator (modulo logical errors) such that $D(\rho) \in V_S$. For each readout vector given by $a$, we will have a density matrix $\rho_{t+1}(a)$ dependent on the syndrome outcomes. The dynamics of the model over all the possible syndrome outcomes can be expressed in terms of an ensemble density matrix as follows:
\begin{equation}
\rho_{t+1} = \sum_{a_t} p(a_t) D(U_{R} U(t) \rho_{t} U(t)^{\dag} U_{R}, a_t),
\end{equation}
where $p(a_t)$ is the probability of measuring the readout vector $a_t$.

In conclusion, the expectation values of the observables at time $t$ can be expressed as the mean of the observable over the readout vector distribution $p(a_t)$:
\begin{equation}
\left\langle O \right\rangle = \mathrm{Tr}(O \rho_{t+1}) = \sum_{a_t} p(a_t) \left\langle O_{a_t} \right\rangle = \sum_{a_t} p(a_t) \mathrm{Tr}(O D(U_{R} U(t) \rho_{t} U(t)^{\dag} U_{R}^\dagger, a_t)).
\end{equation}

\section{Numerical examples}
\label{sec:numericalexamples}
\begin{figure}
    \centering
   \begin{subfigure}[b]{.59\textwidth}
       \centering
            \input{figures/effect_of_correction2_ep3_3_num_qubits4_num_meas3_degree3_num_reservoirs20_timeplex10_methodquantum_stab_noiseTrue.pickle}
       \caption{This Figure shows the training sequence of the logistical map with gray background, as well as the predicted time series.}
    \end{subfigure}
   \hfill
    \begin{subfigure}[b]{.40\textwidth}
        \centering
            \input{figures/effect_of_correction_ep3_3_num_qubits4_num_meas3_degree3_num_reservoirs20_timeplex10_methodquantum_stab_noiseTrue.pickle}
         \caption{The Poincaré plot of the time series of the prediction step.}
   \end{subfigure}
    \caption{This figure shows the necessity to encode data in the correct coset. Both methods achieve a training score close to one. Despite this, only the method with the "correction" step is able to give good predictions.}
    \label{fig:encodingVSnone}
\end{figure}
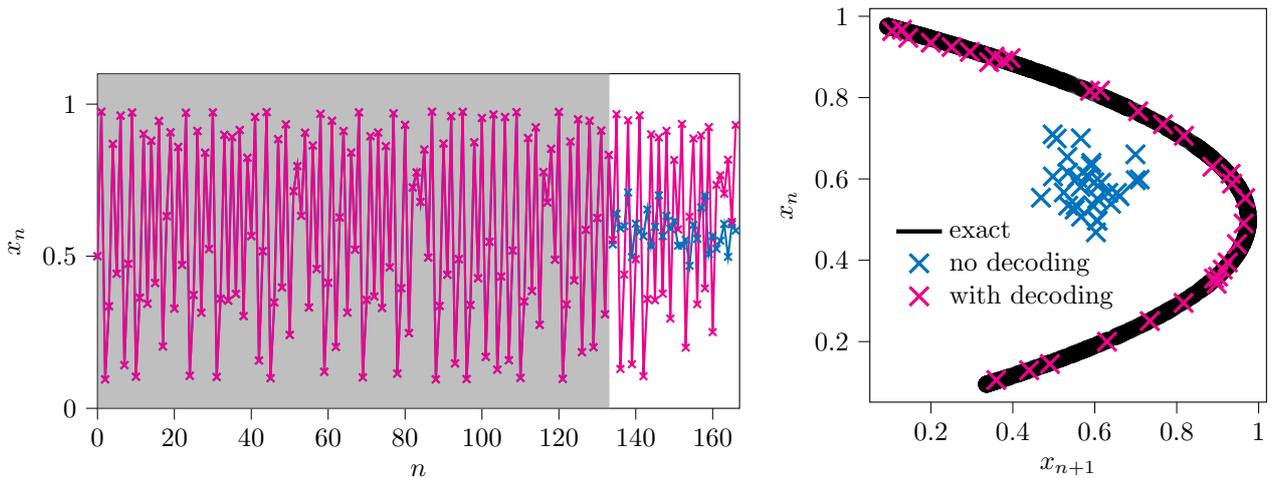

In this paper we benchmark the performance of a given quantum reservoir system on discrete time series given by two mappings. Discrete time systems are easier to study since we do not need to optimize the time discretization as we would need in a continuous case and can test if the model actually learns the time difference equation. 
\paragraph{Logistic map}
The first benchmark system is the logistic map~\cite{may1976simple}, dynamics given by
\begin{equation}
x_{n+1} = F(x_n), \quad F(x) = r x(1-x),
\label{eq:logisticmap}
\end{equation}
with parameter $r=3.9$ and initial condition $x_0=0.5$. This is in the chaotic domain of the logistic map, therefore presenting a challenging prediction problem.
The points $x_{n+1}, x_n$ must lie on a parabola, see for instance in the Poincaré plot\cite{poincare1890probleme} in Figure~\ref{fig:encodingVSnone}.
\paragraph{Hénon map}
The second benchmark system is the Hénon map~\cite{henon1976two}, which can be written as  a one-dimensional map given by
\begin{equation}
x_{n+1} = F(x_n,x_{n-1}), \quad F(x,y) = 1 - ax^2 + b y,
\label{eq:Henonmap}
\end{equation}
with parameters $a=1.4, b=0.3$ and initial condition $x_0=x_1=0$.
The Hénon attractor is a fractal, smooth in one direction and a Cantor set in another, see also Figure~\ref{fig:Henonattractor}.
\paragraph{Performance measure}
We measure how well a reservoir computing model has ``learned'' a given map $F$, by defining the error of a predicted time series $\{\widehat x_i\}_i$ as
\begin{equation}
\label{eq:error}
    E_F(\{\widehat x_i\}_i) = \sqrt{\sum_i \left(\widehat x_{i+1} - F(\widehat x_i, \widehat x_{i-1},\cdots)\right)^2}.
\end{equation}
This error does not measure if a given time series matches another.
Instead, it is very related to the Poincaré plot: it gives a measure how good a given time series relates the last $k$ points to the next one according to $F$.

In the following we briefly describe how we set up the quantum reservoirs model for the numerical experiments.
\paragraph{The reservoir}
All reservoirs use randomly generated Ising-Hamiltonians given by
\begin{equation}
    H_\text{res} = 
    \sum_i (h^x_i X_i + h^y_i Y_i + h^z_i Z_i)
    +\sum_{\{i,j\} \in E} (J^z_{i,j} Z_i Z_j + J^x_{i,j} X_i X_j),
\end{equation}
where the first sum is over all qubits, and the second, i.e., $E$ reflects the connectivity/topology of a given quantum chip.
In our simulations we assume all to all connectivity.
The remaining parameters are drawn from the continuous uniform distributions according to
$h^x_i, h^x_i, h^x_i \sim U(-\frac{1}{2},\frac{1}{2}) $
and 
$J^x_{i,j}, J^Z_{i,j} \sim U(-1,1)$.
These choices are made in order to be in the ergodic region~\cite{PhysRevLett.127.100502}, which ensures that the reservoir given by a Trotterized evolution of $e^{-i t H_\text{res} }$ has the right properties. We choose one Trotter step with $\Delta t = 1.645$, but other choices work equally well.

\paragraph{Multiplexing}
In all examples we use 20 parallel reservoirs in order to perform \textit{spatial multiplexing}~\cite{nakajima2019boosting}. This boosts the performance of the results considerably.
To be able to compare between setups, we use the same (randomly generated) reservoirs irrespective of the stabilizers which are measured, i.e. the same for $Z_i$ and $Z_iZ_j$.
As a comparison we also run classical reservoirs with an RNN structure, using the \href{https://github.com/reservoirpy/reservoirpy}{reservoirpy} implementation available on github with equally many neurons as observables in the quantum reservoir.
Additionally, we perform \textit{temporal multiplexing} in its most simple form. We perform linear regression on the observables of the last $l$ time steps, which comes basically for free. In the numerical examples we observed that using $l>1$ boost the performance, but increasing $l$ beyond a certain point does not lead to any further  improvement. Therefore, all examples in this article use $l=10$.

\paragraph{Exponential encoding}
As described in \cite{shin2023exponential}, the choice of the encoding parameters $\beta_j$ in Equation~\eqref{eq:encoding} influences the expressivity of the quantum reservoir. An analysis can be performed in terms of Fourier frequencies. Notice that the logical X-operators $F_i$ in Equation~\eqref{eq:encoding} all have eigenvalues $\pm 1$ and commute mutually. This means that for $\beta=\frac{1}{2}, \forall i$, we obtain the following spectrum:
\begin{equation}
\Omega = \{ -k, -k+1, \ldots, +k-1, k \},
\end{equation}
where $k$ is the number of cosets.

A better choice in terms of expressivity, and the one employed in this work, is the exponential encoding strategy given by $\beta_k = 3^{k-1}$. This results in a spectrum with exponentially scaled spacing:
\begin{equation}
\Omega = \{ -\frac{3^k - 1}{2}, -\frac{3^k - 1}{2} + 1, \ldots, \frac{3^k - 1}{2} - 1, \frac{3^k - 1}{2} \}.
\end{equation}

\paragraph{Higher order observables}
In general, the more output nodes/observables a reservoir has, the better its performance will be.
In order to increase the number of observables without extra cost we can do the following procedure.
Given a minimal generating set $\{s_1, \cdots s_k\}$ of a stabilizer $S$, observe that $[s_i,s_j]=0$ and $\expval{s_i}\expval{s_j} \neq \expval{s_is_j}$, and similarly for observables including 3, 4, up to $k$ terms.
Since $\expval{s_{i_1}s_{i_2}}$, $\expval{s_{i_1}s_{i_2}s_{i_3}}$, $\cdots$, $\expval{s_{1}\cdots s_{k}}$ can be easily estimated from the samples for $\{\expval{s_i}\}_{j=1}^k$, this means simple classical post processing leads to $2^k-1$ distinct observables purely with classical postprocessing.

\begin{table}
    \centering
    \begin{tabular}{r|rrr|rrr}
    \multicolumn{4}{c}{logistic map} & 
    \multicolumn{3}{c}{Hénon map} \\
    \toprule
training length & classical & $\langle Z_i\rangle$ &  $\langle Z_iZ_j\rangle$ & classical & $\langle Z_i\rangle$ &  $\langle Z_iZ_j\rangle$ \\
\midrule \multicolumn{7}{c}{3 qubits, measure 2 stabilizers } \\ \midrule
35 & 1.12e+00 & \cellcolor{lightgray}7.27e-01 & \cellcolor{lightgray}5.52e-01 & 5.71e+00 & 1.35e+00 & 1.50e+00 \\
68 & \cellcolor{lightgray}4.78e-01 & \cellcolor{lightgray}9.75e-01 & \cellcolor{lightgray}4.95e-01 & 1.04e+02 & \cellcolor{lightgray}9.89e-01 & 1.27e+00 \\
101 & 7.34e+02 & \cellcolor{lightgray}3.09e-01 & \cellcolor{lightgray}2.91e-01 & 1.77e+03 & \cellcolor{lightgray}3.70e-01 & \cellcolor{lightgray}6.00e-01 \\
134 & 6.73e+02 & \cellcolor{lightgray}1.95e-01 & \cellcolor{lightgray}2.27e-01 & 1.02e+03 & \cellcolor{lightgray}6.64e-01 & \cellcolor{lightgray}2.57e-01 \\
167 & \cellcolor{lightgray}4.73e-02 & \cellcolor{lightgray}2.05e-01 & \cellcolor{lightgray}2.78e-01 & 9.97e+03 & \cellcolor{lightgray}3.34e-01 & \cellcolor{lightgray}5.05e-01 \\
\midrule \multicolumn{7}{c}{4 qubits, measure 3 stabilizers } \\ \midrule
35 & \cellcolor{lightgray}6.94e-01 & 1.40e+00 & \cellcolor{lightgray}2.34e-01 & 7.82e+00 & 3.69e+00 & 3.68e+00 \\
68 & \cellcolor{lightgray}6.01e-02 & \cellcolor{lightgray}3.54e-01 & \cellcolor{lightgray}2.50e-01 & 6.53e+02 & 1.58e+00 & 3.15e+00 \\
101 & 4.18e+02 & \cellcolor{lightgray}1.86e-01 & \cellcolor{lightgray}1.43e-01 & \cellcolor{lightgray}2.87e-01 & 1.69e+00 & 2.71e+00 \\
134 & \cellcolor{lightgray}2.66e-02 & \cellcolor{lightgray}1.12e-01 & \cellcolor{lightgray}9.50e-02 & 6.65e+02 & 1.05e+00 & 1.81e+00 \\
167 & 2.81e+03 & \cellcolor{lightgray}7.85e-02 & \cellcolor{lightgray}1.09e-01 & 7.39e+02 & 1.33e+00 & 1.84e+00 \\
\midrule \multicolumn{7}{c}{4 qubits, measure 2 stabilizers } \\ \midrule
35 & \cellcolor{lightgray}1.10e-01 & \cellcolor{lightgray}8.10e-01 & \cellcolor{lightgray}7.23e-01 & 4.45e+01 & 2.88e+00 & 2.54e+00 \\
68 & \cellcolor{lightgray}8.29e-01 & \cellcolor{lightgray}4.67e-01 & \cellcolor{lightgray}4.71e-01 & 1.17e+00 & 1.17e+00 & 1.28e+00 \\
101 & 1.46e+03 & \cellcolor{lightgray}2.65e-01 & \cellcolor{lightgray}6.52e-01 & 8.59e+02 & \cellcolor{lightgray}8.78e-01 & \cellcolor{lightgray}8.47e-01 \\
134 & 4.84e+03 & \cellcolor{lightgray}2.48e-01 & \cellcolor{lightgray}2.43e-01 & 2.83e+01 & \cellcolor{lightgray}6.60e-01 & \cellcolor{lightgray}4.30e-01 \\
167 & 9.09e+01 & \cellcolor{lightgray}1.99e-01 & \cellcolor{lightgray}2.14e-01 & 4.92e+03 & \cellcolor{lightgray}3.62e-01 & \cellcolor{lightgray}4.81e-01 \\
\midrule \multicolumn{7}{c}{5 qubits, measure 4 stabilizers } \\ \midrule
35 & \cellcolor{lightgray}7.88e-01 & 1.76e+00 & 2.04e+00 & 3.40e+00 & 4.06e+00 & 4.09e+00 \\
68 & \cellcolor{lightgray}2.97e-01 & 1.05e+00 & 1.11e+00 & 2.01e+03 & 3.28e+00 & 3.52e+00 \\
101 & \cellcolor{lightgray}1.54e-02 & \cellcolor{lightgray}7.19e-01 & 1.03e+00 & 7.15e+02 & 2.08e+00 & 3.74e+00 \\
134 & \cellcolor{lightgray}4.13e-03 & \cellcolor{lightgray}7.91e-01 & \cellcolor{lightgray}7.88e-01 & \cellcolor{lightgray}3.10e-02 & 1.60e+00 & 2.62e+00 \\
167 & \cellcolor{lightgray}3.49e-01 & \cellcolor{lightgray}5.68e-01 & \cellcolor{lightgray}1.81e-01 & 2.31e+03 & 1.66e+00 & 2.56e+00 \\
\midrule \multicolumn{7}{c}{5 qubits, measure 3 stabilizers } \\ \midrule
35 & \cellcolor{lightgray}4.73e-01 & 1.59e+00 & 1.58e+00 & 2.27e+01 & 4.03e+00 & 3.99e+00 \\
68 & \cellcolor{lightgray}2.06e-01 & \cellcolor{lightgray}3.79e-01 & \cellcolor{lightgray}3.81e-01 & \cellcolor{lightgray}4.39e-01 & 3.00e+00 & 2.63e+00 \\
101 & \cellcolor{lightgray}7.92e-02 & \cellcolor{lightgray}2.21e-01 & \cellcolor{lightgray}3.37e-01 & 7.97e+03 & 2.58e+00 & 2.19e+00 \\
134 & 3.47e+03 & \cellcolor{lightgray}1.92e-01 & \cellcolor{lightgray}2.56e-01 & 2.04e+03 & 1.53e+00 & 1.68e+00 \\
167 & 2.30e+03 & \cellcolor{lightgray}1.24e-01 & \cellcolor{lightgray}1.82e-01 & 1.91e+01 & \cellcolor{lightgray}9.26e-01 & 1.50e+00 \\
\midrule \multicolumn{7}{c}{5 qubits, measure 2 stabilizers } \\ \midrule
35 & 2.27e+03 & \cellcolor{lightgray}7.95e-01 & \cellcolor{lightgray}5.31e-01 & 1.78e+00 & 2.85e+00 & 3.31e+00 \\
68 & \cellcolor{lightgray}1.86e-01 & 2.23e+00 & \cellcolor{lightgray}4.59e-01 & 1.64e+00 & 1.93e+00 & 1.35e+00 \\
101 & 8.45e+01 & \cellcolor{lightgray}2.22e-01 & \cellcolor{lightgray}3.05e-01 & 3.18e+02 & 1.02e+00 & 1.50e+00 \\
134 & \cellcolor{lightgray}6.38e-02 & \cellcolor{lightgray}3.25e-01 & \cellcolor{lightgray}3.77e-01 & 1.93e+00 & \cellcolor{lightgray}7.42e-01 & \cellcolor{lightgray}8.52e-01 \\
167 & \cellcolor{lightgray}9.03e-01 & \cellcolor{lightgray}2.99e-01 & \cellcolor{lightgray}3.60e-01 & 9.98e+03 & \cellcolor{lightgray}3.57e-01 & \cellcolor{lightgray}8.79e-01 \\
\bottomrule
    \end{tabular}
    \caption{
    This plot shows how well the reservoirs can learn a discrete-time dynamical system $F$ for the logistic map and the  Hénon map as defined by equation~\eqref{eq:error}.
    Values below $1$ have a gray background for ease of inspection.
    Measuring $k$ stabilizers gives $2^k -1 $ observables. Since we apply spatial multiplexing of 20 reservoirs, this means $2, 3, 4$ stabilizers result in $60, 140, 300$ observables, respectively. For the classical reservoir this equals the number of (external) neurons to make the comparison fair.
    In the table $\langle Z_i\rangle$ refers to direct measurement of the qubits, whereas  $\langle Z_iZ_j\rangle$ refers to measurement of stabilizers through ancilla qubits.
    }
    \label{tab:funapprox}
\end{table}

\section{Results}
\label{sec:results}
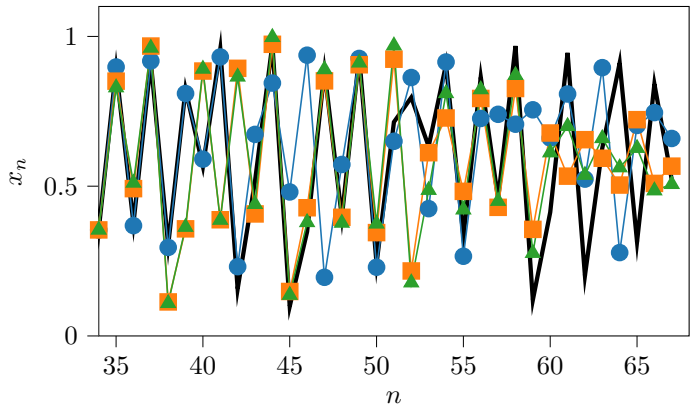
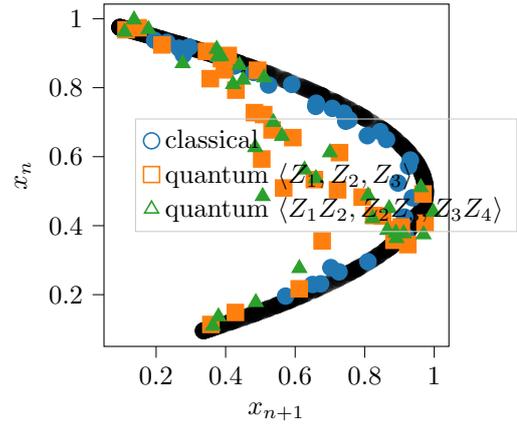
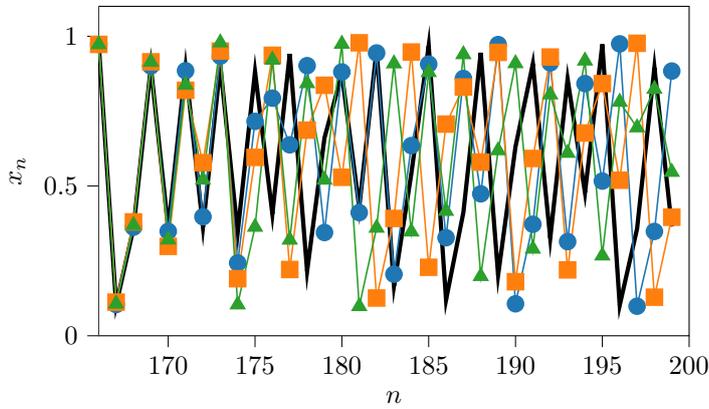
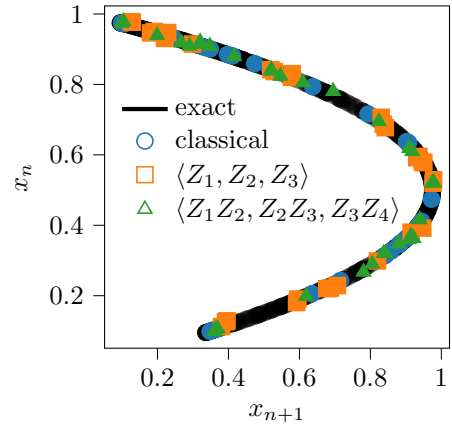
\begin{figure}
    \centering
    \begin{subfigure}[b]{.50\textwidth}
        \centering
            \input{figures/convergence_ep0_0_num_qubits4_num_meas3_degree3_num_reservoirs20_timeplex10_methodquantum_stab_noiseTrue.pickle}
         \caption{Predictions for the logistic map with 35 training points.}
   \end{subfigure}
   \hfill
    \begin{subfigure}[b]{.450\textwidth}
        \centering
            \input{figures/convergence2_ep0_0_num_qubits4_num_meas3_degree3_num_reservoirs20_timeplex10_methodquantum_stab_noiseTrue.pickle}
         \caption{Poincaré plot for predictions with 35 training points.}
   \end{subfigure}
   \vfill
   \begin{subfigure}[b]{.50\textwidth}
       \centering
            \input{figures/convergence_ep4_4_num_qubits4_num_meas3_degree3_num_reservoirs20_timeplex10_methodquantum_stab_noiseTrue.pickle}
         \caption{Predictions for the logistic map with 167 training points.}
    \end{subfigure}
   \hfill
   \begin{subfigure}[b]{.450\textwidth}
       \centering
            \input{figures/convergence2_ep4_4_num_qubits4_num_meas3_degree3_num_reservoirs20_timeplex10_methodquantum_stab_noiseTrue.pickle}
         \caption{Poincaré plot for predictions with 167 training points.}
    \end{subfigure}
    \caption{Comparison for logistic map using 4 qubits.
    Longer training time shows an increase in accuracy.
    }
    \label{fig:comparisonLongShort}
\end{figure}

\begin{figure}
   \begin{subfigure}[b]{.50\textwidth}
       \centering
       \input{figures/convergence_ep4_4_casenamehenon_num_qubits4_num_meas2_degree2_num_reservoirs20_timeplex10_methodquantum_stab_noiseNone.pickle}
         \caption{Predictions with 167 training points.}
    \end{subfigure}
   \hfill
   \begin{subfigure}[b]{.450\textwidth}
       \centering
        \centering
        \includegraphics[width=.8\linewidth]{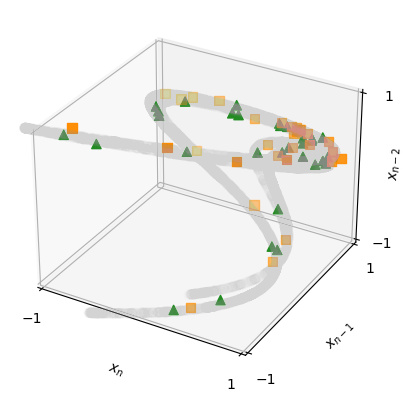}
        \caption{Poincaré plot for predictions with 167 training points.}
    \end{subfigure}
    \caption{Results for the Hénon map using 4 qubits.}
    \label{fig:Henonattractor}
\end{figure}

We perform various experiments on both the logistic map and the Hénon map with classical and quantum reservoirs of different size and training duration. The results in terms of error is given in Table~\ref{tab:funapprox}. For classical reservoirs, we observe there is a threshold value of reservoir size in terms of nodes, therefore too small classical reservoirs are simply not capable of modeling the respective systems. As expected, we see a general trend with increasing reservoir model performance with training duration if the reservoir has the capacity for modeling. 
This is also shown in Figure~\ref{fig:comparisonLongShort}, where a the predictions and the Poincaré plots are shown.
Figure~\ref{fig:Henonattractor} shows one example for the Hénon map.
We observe that classical reservoirs typically require more node than quantum reservoirs to achieve equal performance, pointing to the exponentially large Hilbert space of quantum systems in general.

Finally, we also observe that the correction step, i.e., that one projects consistently into the same coset before the encoding step is performed, is essential for being able to successfully make predictions. This is exemplified in Figure~\ref{fig:encodingVSnone} uses a reservoir comprised of 4 qubits with $S = \langle Z_1Z_2, \cdots, Z_3Z_4\rangle$.

\section{Conclusion and future work}
We presented a new systematic way to encode data into a reservoir based on choosing a stabilizer and using the cosets as encoding. 
A suit of numerical examples are performed to show that the methods works well.
Future work includes running larger instances, high dimensional examples time series, as well as run benchmarks on real hardware.

\section{Availability of Data and Code}
All data and the python/jupyter notebook source code for reproducing the results obtained in this article are available at \url{https://github.com/OpenQuantumComputing/quantumreservoirpy}.

\section{Author contributions}
Franz G. Fuchs formulated the concept of using stabilizers, developed the methodology, performed the experiments, made the formal analysis and investigation, wrote the article, made the visualizations.
Alexander J. Stasik contributed with the concept, developed the methodology, did parts of the formal analysis, and wrote parts of the article.
Stanley Miao wrote parts of the code and conducted experiments.
Ola Tangen Kulseng wrote the original code and conducted experiments.
Ruben Pariente Bassa contributed with the concept and methodology, and wrote parts of the article.

\section{Acknowledgment}
We would like to thank for funding of the work by the Research Council of Norway through project number 332023 (80\%) and 324944 (20\%).

\printbibliography

\end{document}

%% file: figures/exampleStabilizerEncoding.tex
\begin{tikzpicture}[scale=.5]

\begin{scope}[fill opacity=0.5,text opacity=1]

    \definecolor{IXII}{RGB}{58.6004535,76.17308133,192.189204015};
    \definecolor{IIXX}{RGB}{170.14949570500002,198.68999653499998,253.204599315};
    \definecolor{XIXI}{RGB}{246.891866745,183.8152455,156.13471279};
    \definecolor{XIII}{RGB}{179.94665529,3.9668208,38.30936706};
    
    \def\sx{-16}
    
    \node[] at (\sx+8,0) {\phantom{$\cdots \longrightarrow$}};
    
    \def\sx{0}
    
    \draw[draw = black] (\sx,0) circle (8.5);
    
    \node[] at (\sx,9.0) {$\mathcal{H}$};
    
    \node [
    circle,
    text width=2.5cm,
    draw,
    fill=IXII, fill opacity=0.25] (V1) at (\sx-3.75,+3.75) {$\phantom{spa}V_{S_1}=$\\$\operatorname{span}\!\big( \ket{000}\!, \!\ket{111}\!\big)$};

    \node [
    circle,
    text width=2.5cm,
    draw,
    fill=XIXI, fill opacity=0.25] (V2) at (\sx-3.75,-3.75) {$\phantom{spa}V_{S_2}=$\\$\operatorname{span}\!\big( \ket{010}\!, \!\ket{101}\!\big)$};
    
    \node [
    circle,
    text width=2.5cm,
    draw,
    fill=IIXX, fill opacity=0.25] (V3) at (\sx+3.75,+3.75) {$\phantom{spa}V_{S_3}=$\\$\operatorname{span}\!\big( \ket{100}\!, \!\ket{011}\!\big)$};
    
    \node [
    circle,
    text width=2.5cm,
    draw,
    fill=XIII, fill opacity=0.25] (V4) at (\sx+3.75,-3.75) {$\phantom{spa}V_{S_4}=$\\$\operatorname{span}\!\big( \ket{001}\!, \!\ket{110}\!\big)$};
    
    \draw [thick, -latex] (V1) to [] node [above] {$[X_1]$} (V3);
    \draw [thick, -latex] (V1) to [] node [left] {$[X_2]$} (V2);
    \draw [thick, -latex] (V3) to [] node [left] {$[X_2]$} (V4);
    \draw [thick, -latex] (V2) to [] node [above] {$[X_1]$} (V4);

\end{scope}
\end{tikzpicture}

%% file: figures/effect_of_correction2_ep3_3_num_qubits4_num_meas3_degree3_num_reservoirs20_timeplex10_methodquantum_stab_noiseTrue.pickle.tex
\begin{tikzpicture}

\definecolor{darkblue}{RGB}{0,0,139}
\definecolor{darkgray176}{RGB}{176,176,176}
\definecolor{gray}{RGB}{128,128,128}
\definecolor{lightblue}{RGB}{173,216,230}
\definecolor{lightgray204}{RGB}{204,204,204}

\begin{axis}[
width=1\textwidth,
height=.6\textwidth,
tick align=outside,
tick pos=left,
x grid style={darkgray176},
xmin=0, xmax=167,
xtick style={color=black},
y grid style={darkgray176},
ymin=0, ymax=1.1,
ytick style={color=black},
xlabel=$n$,
ylabel=$x_n$
]
\path [draw=gray, fill=gray, opacity=0.5]
(axis cs:0,2)
--(axis cs:0,-2)
--(axis cs:1,-2)
--(axis cs:2,-2)
--(axis cs:3,-2)
--(axis cs:4,-2)
--(axis cs:5,-2)
--(axis cs:6,-2)
--(axis cs:7,-2)
--(axis cs:8,-2)
--(axis cs:9,-2)
--(axis cs:10,-2)
--(axis cs:11,-2)
--(axis cs:12,-2)
--(axis cs:13,-2)
--(axis cs:14,-2)
--(axis cs:15,-2)
--(axis cs:16,-2)
--(axis cs:17,-2)
--(axis cs:18,-2)
--(axis cs:19,-2)
--(axis cs:20,-2)
--(axis cs:21,-2)
--(axis cs:22,-2)
--(axis cs:23,-2)
--(axis cs:24,-2)
--(axis cs:25,-2)
--(axis cs:26,-2)
--(axis cs:27,-2)
--(axis cs:28,-2)
--(axis cs:29,-2)
--(axis cs:30,-2)
--(axis cs:31,-2)
--(axis cs:32,-2)
--(axis cs:33,-2)
--(axis cs:34,-2)
--(axis cs:35,-2)
--(axis cs:36,-2)
--(axis cs:37,-2)
--(axis cs:38,-2)
--(axis cs:39,-2)
--(axis cs:40,-2)
--(axis cs:41,-2)
--(axis cs:42,-2)
--(axis cs:43,-2)
--(axis cs:44,-2)
--(axis cs:45,-2)
--(axis cs:46,-2)
--(axis cs:47,-2)
--(axis cs:48,-2)
--(axis cs:49,-2)
--(axis cs:50,-2)
--(axis cs:51,-2)
--(axis cs:52,-2)
--(axis cs:53,-2)
--(axis cs:54,-2)
--(axis cs:55,-2)
--(axis cs:56,-2)
--(axis cs:57,-2)
--(axis cs:58,-2)
--(axis cs:59,-2)
--(axis cs:60,-2)
--(axis cs:61,-2)
--(axis cs:62,-2)
--(axis cs:63,-2)
--(axis cs:64,-2)
--(axis cs:65,-2)
--(axis cs:66,-2)
--(axis cs:67,-2)
--(axis cs:68,-2)
--(axis cs:69,-2)
--(axis cs:70,-2)
--(axis cs:71,-2)
--(axis cs:72,-2)
--(axis cs:73,-2)
--(axis cs:74,-2)
--(axis cs:75,-2)
--(axis cs:76,-2)
--(axis cs:77,-2)
--(axis cs:78,-2)
--(axis cs:79,-2)
--(axis cs:80,-2)
--(axis cs:81,-2)
--(axis cs:82,-2)
--(axis cs:83,-2)
--(axis cs:84,-2)
--(axis cs:85,-2)
--(axis cs:86,-2)
--(axis cs:87,-2)
--(axis cs:88,-2)
--(axis cs:89,-2)
--(axis cs:90,-2)
--(axis cs:91,-2)
--(axis cs:92,-2)
--(axis cs:93,-2)
--(axis cs:94,-2)
--(axis cs:95,-2)
--(axis cs:96,-2)
--(axis cs:97,-2)
--(axis cs:98,-2)
--(axis cs:99,-2)
--(axis cs:100,-2)
--(axis cs:101,-2)
--(axis cs:102,-2)
--(axis cs:103,-2)
--(axis cs:104,-2)
--(axis cs:105,-2)
--(axis cs:106,-2)
--(axis cs:107,-2)
--(axis cs:108,-2)
--(axis cs:109,-2)
--(axis cs:110,-2)
--(axis cs:111,-2)
--(axis cs:112,-2)
--(axis cs:113,-2)
--(axis cs:114,-2)
--(axis cs:115,-2)
--(axis cs:116,-2)
--(axis cs:117,-2)
--(axis cs:118,-2)
--(axis cs:119,-2)
--(axis cs:120,-2)
--(axis cs:121,-2)
--(axis cs:122,-2)
--(axis cs:123,-2)
--(axis cs:124,-2)
--(axis cs:125,-2)
--(axis cs:126,-2)
--(axis cs:127,-2)
--(axis cs:128,-2)
--(axis cs:129,-2)
--(axis cs:130,-2)
--(axis cs:131,-2)
--(axis cs:132,-2)
--(axis cs:133,-2)
--(axis cs:133,2)
--(axis cs:133,2)
--(axis cs:132,2)
--(axis cs:131,2)
--(axis cs:130,2)
--(axis cs:129,2)
--(axis cs:128,2)
--(axis cs:127,2)
--(axis cs:126,2)
--(axis cs:125,2)
--(axis cs:124,2)
--(axis cs:123,2)
--(axis cs:122,2)
--(axis cs:121,2)
--(axis cs:120,2)
--(axis cs:119,2)
--(axis cs:118,2)
--(axis cs:117,2)
--(axis cs:116,2)
--(axis cs:115,2)
--(axis cs:114,2)
--(axis cs:113,2)
--(axis cs:112,2)
--(axis cs:111,2)
--(axis cs:110,2)
--(axis cs:109,2)
--(axis cs:108,2)
--(axis cs:107,2)
--(axis cs:106,2)
--(axis cs:105,2)
--(axis cs:104,2)
--(axis cs:103,2)
--(axis cs:102,2)
--(axis cs:101,2)
--(axis cs:100,2)
--(axis cs:99,2)
--(axis cs:98,2)
--(axis cs:97,2)
--(axis cs:96,2)
--(axis cs:95,2)
--(axis cs:94,2)
--(axis cs:93,2)
--(axis cs:92,2)
--(axis cs:91,2)
--(axis cs:90,2)
--(axis cs:89,2)
--(axis cs:88,2)
--(axis cs:87,2)
--(axis cs:86,2)
--(axis cs:85,2)
--(axis cs:84,2)
--(axis cs:83,2)
--(axis cs:82,2)
--(axis cs:81,2)
--(axis cs:80,2)
--(axis cs:79,2)
--(axis cs:78,2)
--(axis cs:77,2)
--(axis cs:76,2)
--(axis cs:75,2)
--(axis cs:74,2)
--(axis cs:73,2)
--(axis cs:72,2)
--(axis cs:71,2)
--(axis cs:70,2)
--(axis cs:69,2)
--(axis cs:68,2)
--(axis cs:67,2)
--(axis cs:66,2)
--(axis cs:65,2)
--(axis cs:64,2)
--(axis cs:63,2)
--(axis cs:62,2)
--(axis cs:61,2)
--(axis cs:60,2)
--(axis cs:59,2)
--(axis cs:58,2)
--(axis cs:57,2)
--(axis cs:56,2)
--(axis cs:55,2)
--(axis cs:54,2)
--(axis cs:53,2)
--(axis cs:52,2)
--(axis cs:51,2)
--(axis cs:50,2)
--(axis cs:49,2)
--(axis cs:48,2)
--(axis cs:47,2)
--(axis cs:46,2)
--(axis cs:45,2)
--(axis cs:44,2)
--(axis cs:43,2)
--(axis cs:42,2)
--(axis cs:41,2)
--(axis cs:40,2)
--(axis cs:39,2)
--(axis cs:38,2)
--(axis cs:37,2)
--(axis cs:36,2)
--(axis cs:35,2)
--(axis cs:34,2)
--(axis cs:33,2)
--(axis cs:32,2)
--(axis cs:31,2)
--(axis cs:30,2)
--(axis cs:29,2)
--(axis cs:28,2)
--(axis cs:27,2)
--(axis cs:26,2)
--(axis cs:25,2)
--(axis cs:24,2)
--(axis cs:23,2)
--(axis cs:22,2)
--(axis cs:21,2)
--(axis cs:20,2)
--(axis cs:19,2)
--(axis cs:18,2)
--(axis cs:17,2)
--(axis cs:16,2)
--(axis cs:15,2)
--(axis cs:14,2)
--(axis cs:13,2)
--(axis cs:12,2)
--(axis cs:11,2)
--(axis cs:10,2)
--(axis cs:9,2)
--(axis cs:8,2)
--(axis cs:7,2)
--(axis cs:6,2)
--(axis cs:5,2)
--(axis cs:4,2)
--(axis cs:3,2)
--(axis cs:2,2)
--(axis cs:1,2)
--(axis cs:0,2)
--cycle;

\addplot [thick, RoyalBlue, mark=x, mark size=2, mark options={solid}]
table {%
0 0.5
1 0.975
2 0.0950625000000001
3 0.335499922265625
4 0.869464925259
5 0.442633109113109
6 0.962165255336889
7 0.141972779361614
8 0.475084386199614
9 0.972578927536905
10 0.104009713267468
11 0.363447601972601
12 0.90227842611257
13 0.343871064749135
14 0.879932646751981
15 0.412039617334933
16 0.944825587217519
17 0.2033077681307
18 0.631697506238813
19 0.907357490716863
20 0.327833511551757
21 0.859398930996064
22 0.471246392755656
23 0.971775597274708
24 0.10696836468276
25 0.372551921195438
26 0.911652250115202
27 0.314115457402856
28 0.840243053611457
29 0.52351519142969
30 0.972843439510898
31 0.103034418674866
32 0.360431476248473
33 0.899030445993496
34 0.354021342363904
35 0.891891902907578
36 0.376040872098363
37 0.915073124978476
38 0.30308577359035
39 0.823776671006208
40 0.566157802517338
41 0.9579302661477
42 0.157169498248997
43 0.516622263569707
44 0.973922431379895
45 0.0990503632363789
46 0.348033616238569
47 0.884934251005247
48 0.397119927371816
49 0.933721193558477
50 0.24135511240702
51 0.714101006275858
52 0.796226960535493
53 0.632773392622424
54 0.906247782224973
55 0.331354683805434
56 0.864079153569976
57 0.458040842749502
58 0.968133773579029
59 0.120318003135167
60 0.412782166901259
61 0.945332893399284
62 0.201546594820824
63 0.627609703254123
64 0.911491478178039
65 0.314631577208725
66 0.840990336544313
67 0.521529802495244
68 0.973192223657611
69 0.101747565932864
70 0.356440495162445
71 0.894623607426105
72 0.367661613001829
73 0.906697550174217
74 0.329928700460932
75 0.862195436985061
76 0.463376415166083
77 0.96976898083226
78 0.114336708126496
79 0.394928918675041
80 0.931944264689895
81 0.247354193585875
82 0.72606337635529
83 0.775691864496302
84 0.678576583817155
85 0.85063057447756
86 0.495526980941992
87 0.974921969191976
88 0.0953515803973814
89 0.336412660401102
90 0.870632811059524
91 0.439262145527849
92 0.960612560833067
93 0.147560668330936
94 0.490567418221388
95 0.97465300296386
96 0.0963476544318197
97 0.339552657277495
98 0.874600935831819
99 0.427729141608309
100 0.95462999980658
101 0.16891509677589
102 0.547492868742619
103 0.966203266932526
104 0.127352604215616
105 0.433422281818955
106 0.957712889023041
107 0.157945753766508
108 0.518695681271186
109 0.973636838857044
110 0.100105765022275
111 0.351329943243161
112 0.888799135473281
113 0.385457405795393
114 0.923831977040227
115 0.274429175428874
116 0.776559432098978
117 0.676708034016653
118 0.853219655784481
119 0.488419911593223
120 0.974477016054716
121 0.0969990888196709
122 0.341602035792516
123 0.877149331246205
124 0.420257689568106
125 0.950200539315235
126 0.184545950161058
127 0.586906095516819
128 0.945544589191884
129 0.200811074263158
130 0.625895348194095
131 0.913186409082042
132 0.309180266264562
133 0.832992533946832
134 0.538795799812285
135 0.639687490097443
136 0.591813036321478
137 0.598746655100378
138 0.709671004663524
139 0.497238654471472
140 0.607092357746446
141 0.580236499372394
142 0.565535062634486
143 0.654043711872644
144 0.533078464917981
145 0.596959818189684
146 0.700181891322699
147 0.565990550222539
148 0.6328680390405
149 0.590896361691596
150 0.615702286681458
151 0.534401257631748
152 0.535105440965114
153 0.553927421994248
154 0.468683977855696
155 0.602836661222961
156 0.557307375397452
157 0.660452667277966
158 0.699255131559479
159 0.507005018225519
160 0.566661713641444
161 0.524396003680959
162 0.551061050615747
163 0.60606317504587
164 0.497908603262945
165 0.605072860196422
166 0.584340000340011
};
\addplot [thick, magenta, mark=x, mark size=2, mark options={solid}]
table {%
0 0.5
1 0.975
2 0.0950625000000001
3 0.335499922265625
4 0.869464925259
5 0.442633109113109
6 0.962165255336889
7 0.141972779361614
8 0.475084386199614
9 0.972578927536905
10 0.104009713267468
11 0.363447601972601
12 0.90227842611257
13 0.343871064749135
14 0.879932646751981
15 0.412039617334933
16 0.944825587217519
17 0.2033077681307
18 0.631697506238813
19 0.907357490716863
20 0.327833511551757
21 0.859398930996064
22 0.471246392755656
23 0.971775597274708
24 0.10696836468276
25 0.372551921195438
26 0.911652250115202
27 0.314115457402856
28 0.840243053611457
29 0.52351519142969
30 0.972843439510898
31 0.103034418674866
32 0.360431476248473
33 0.899030445993496
34 0.354021342363904
35 0.891891902907578
36 0.376040872098363
37 0.915073124978476
38 0.30308577359035
39 0.823776671006208
40 0.566157802517338
41 0.9579302661477
42 0.157169498248997
43 0.516622263569707
44 0.973922431379895
45 0.0990503632363789
46 0.348033616238569
47 0.884934251005247
48 0.397119927371816
49 0.933721193558477
50 0.24135511240702
51 0.714101006275858
52 0.796226960535493
53 0.632773392622424
54 0.906247782224973
55 0.331354683805434
56 0.864079153569976
57 0.458040842749502
58 0.968133773579029
59 0.120318003135167
60 0.412782166901259
61 0.945332893399284
62 0.201546594820824
63 0.627609703254123
64 0.911491478178039
65 0.314631577208725
66 0.840990336544313
67 0.521529802495244
68 0.973192223657611
69 0.101747565932864
70 0.356440495162445
71 0.894623607426105
72 0.367661613001829
73 0.906697550174217
74 0.329928700460932
75 0.862195436985061
76 0.463376415166083
77 0.96976898083226
78 0.114336708126496
79 0.394928918675041
80 0.931944264689895
81 0.247354193585875
82 0.72606337635529
83 0.775691864496302
84 0.678576583817155
85 0.85063057447756
86 0.495526980941992
87 0.974921969191976
88 0.0953515803973814
89 0.336412660401102
90 0.870632811059524
91 0.439262145527849
92 0.960612560833067
93 0.147560668330936
94 0.490567418221388
95 0.97465300296386
96 0.0963476544318197
97 0.339552657277495
98 0.874600935831819
99 0.427729141608309
100 0.95462999980658
101 0.16891509677589
102 0.547492868742619
103 0.966203266932526
104 0.127352604215616
105 0.433422281818955
106 0.957712889023041
107 0.157945753766508
108 0.518695681271186
109 0.973636838857044
110 0.100105765022275
111 0.351329943243161
112 0.888799135473281
113 0.385457405795393
114 0.923831977040227
115 0.274429175428874
116 0.776559432098978
117 0.676708034016653
118 0.853219655784481
119 0.488419911593223
120 0.974477016054716
121 0.0969990888196709
122 0.341602035792516
123 0.877149331246205
124 0.420257689568106
125 0.950200539315235
126 0.184545950161058
127 0.586906095516819
128 0.945544589191884
129 0.200811074263158
130 0.625895348194095
131 0.913186409082042
132 0.309180266264562
133 0.832992533946832
134 0.553156765174898
135 0.967138710581483
136 0.129467728973731
137 0.439799951564868
138 0.946999863979551
139 0.144946772996457
140 0.490461401950633
141 0.963455818426014
142 0.105455021907201
143 0.359576501152593
144 0.900815383661233
145 0.355275060596562
146 0.890819977630871
147 0.377048060696877
148 0.912286744653274
149 0.29511227342811
150 0.816657157993081
151 0.588372346190304
152 0.93497409826702
153 0.199753505083902
154 0.629883132658195
155 0.887062882767759
156 0.342183751568319
157 0.897361055252411
158 0.394491220852889
159 0.924428038957591
160 0.250449339472705
161 0.734360126339841
162 0.766283393209023
163 0.706015745983646
164 0.817498761498924
165 0.612863133636368
166 0.931496652193939
};
\end{axis}

\end{tikzpicture}

%% file: figures/effect_of_correction_ep3_3_num_qubits4_num_meas3_degree3_num_reservoirs20_timeplex10_methodquantum_stab_noiseTrue.pickle.tex
\begin{tikzpicture}

\definecolor{darkblue}{RGB}{0,0,139}
\definecolor{darkgray176}{RGB}{176,176,176}
\definecolor{gray}{RGB}{128,128,128}
\definecolor{lightblue}{RGB}{173,216,230}
\definecolor{lightgray204}{RGB}{204,204,204}

\begin{axis}[
width=1\textwidth,
height=1\textwidth,
legend cell align={left},
legend style={
  fill opacity=0,
  draw opacity=1,
  text opacity=1,
  at={(0.04,0.35)},
  anchor=west,
  draw=none
},
tick align=outside,
tick pos=left,
x grid style={darkgray176},
xmin=0.0510656250000001, xmax=1.018996875,
xtick style={color=black},
y grid style={darkgray176},
ymin=0.0510656250000001, ymax=1.018996875,
ytick style={color=black},
xlabel=$x_{n+1}$,
ylabel=$x_n$
]
\addplot [ultra thick, black]
table {%
-10 10
};
\addlegendentry{exact}

\addplot [only marks, black, opacity=0.5, mark=*, mark size=3, mark options={solid}, forget plot]
table {%
0.975 0.5
0.0950625000000001 0.975
0.335499922265625 0.0950625000000001
0.869464925259 0.335499922265625
0.442633109113109 0.869464925259
0.962165255336889 0.442633109113109
0.141972779361614 0.962165255336889
0.475084386199614 0.141972779361614
0.972578927536905 0.475084386199614
0.104009713267468 0.972578927536905
0.363447601972601 0.104009713267468
0.90227842611257 0.363447601972601
0.343871064749135 0.90227842611257
0.879932646751981 0.343871064749135
0.412039617334933 0.879932646751981
0.944825587217519 0.412039617334933
0.2033077681307 0.944825587217519
0.631697506238813 0.2033077681307
0.907357490716863 0.631697506238813
0.327833511551757 0.907357490716863
0.859398930996064 0.327833511551757
0.471246392755656 0.859398930996064
0.971775597274708 0.471246392755656
0.10696836468276 0.971775597274708
0.372551921195438 0.10696836468276
0.911652250115202 0.372551921195438
0.314115457402856 0.911652250115202
0.840243053611457 0.314115457402856
0.52351519142969 0.840243053611457
0.972843439510898 0.52351519142969
0.103034418674866 0.972843439510898
0.360431476248473 0.103034418674866
0.899030445993496 0.360431476248473
0.354021342363904 0.899030445993496
0.891891902907578 0.354021342363904
0.376040872098363 0.891891902907578
0.915073124978476 0.376040872098363
0.30308577359035 0.915073124978476
0.823776671006208 0.30308577359035
0.566157802517338 0.823776671006208
0.9579302661477 0.566157802517338
0.157169498248997 0.9579302661477
0.516622263569707 0.157169498248997
0.973922431379895 0.516622263569707
0.0990503632363789 0.973922431379895
0.348033616238569 0.0990503632363789
0.884934251005247 0.348033616238569
0.397119927371816 0.884934251005247
0.933721193558477 0.397119927371816
0.24135511240702 0.933721193558477
0.714101006275858 0.24135511240702
0.796226960535493 0.714101006275858
0.632773392622424 0.796226960535493
0.906247782224973 0.632773392622424
0.331354683805434 0.906247782224973
0.864079153569976 0.331354683805434
0.458040842749502 0.864079153569976
0.968133773579029 0.458040842749502
0.120318003135167 0.968133773579029
0.412782166901259 0.120318003135167
0.945332893399284 0.412782166901259
0.201546594820824 0.945332893399284
0.627609703254123 0.201546594820824
0.911491478178039 0.627609703254123
0.314631577208725 0.911491478178039
0.840990336544313 0.314631577208725
0.521529802495244 0.840990336544313
0.973192223657611 0.521529802495244
0.101747565932864 0.973192223657611
0.356440495162445 0.101747565932864
0.894623607426105 0.356440495162445
0.367661613001829 0.894623607426105
0.906697550174217 0.367661613001829
0.329928700460932 0.906697550174217
0.862195436985061 0.329928700460932
0.463376415166083 0.862195436985061
0.96976898083226 0.463376415166083
0.114336708126496 0.96976898083226
0.394928918675041 0.114336708126496
0.931944264689895 0.394928918675041
0.247354193585875 0.931944264689895
0.72606337635529 0.247354193585875
0.775691864496302 0.72606337635529
0.678576583817155 0.775691864496302
0.85063057447756 0.678576583817155
0.495526980941992 0.85063057447756
0.974921969191976 0.495526980941992
0.0953515803973814 0.974921969191976
0.336412660401102 0.0953515803973814
0.870632811059524 0.336412660401102
0.439262145527849 0.870632811059524
0.960612560833067 0.439262145527849
0.147560668330936 0.960612560833067
0.490567418221388 0.147560668330936
0.97465300296386 0.490567418221388
0.0963476544318197 0.97465300296386
0.339552657277495 0.0963476544318197
0.874600935831819 0.339552657277495
0.427729141608309 0.874600935831819
0.95462999980658 0.427729141608309
0.16891509677589 0.95462999980658
0.547492868742619 0.16891509677589
0.966203266932526 0.547492868742619
0.127352604215616 0.966203266932526
0.433422281818955 0.127352604215616
0.957712889023041 0.433422281818955
0.157945753766508 0.957712889023041
0.518695681271186 0.157945753766508
0.973636838857044 0.518695681271186
0.100105765022275 0.973636838857044
0.351329943243161 0.100105765022275
0.888799135473281 0.351329943243161
0.385457405795393 0.888799135473281
0.923831977040227 0.385457405795393
0.274429175428874 0.923831977040227
0.776559432098978 0.274429175428874
0.676708034016653 0.776559432098978
0.853219655784481 0.676708034016653
0.488419911593223 0.853219655784481
0.974477016054716 0.488419911593223
0.0969990888196709 0.974477016054716
0.341602035792516 0.0969990888196709
0.877149331246205 0.341602035792516
0.420257689568106 0.877149331246205
0.950200539315235 0.420257689568106
0.184545950161058 0.950200539315235
0.586906095516819 0.184545950161058
0.945544589191884 0.586906095516819
0.200811074263158 0.945544589191884
0.625895348194095 0.200811074263158
0.913186409082042 0.625895348194095
0.309180266264562 0.913186409082042
0.832992533946832 0.309180266264562
0.542552292109105 0.832992533946832
0.967938279501419 0.542552292109105
0.121031689651281 0.967938279501419
0.414893777030605 0.121031689651281
0.946752030166346 0.414893777030605
0.196609231814768 0.946752030166346
0.616020762941902 0.196609231814768
0.922502812008878 0.616020762941902
0.278816357993905 0.922502812008878
0.784203406384786 0.278816357993905
0.659990852817209 0.784203406384786
0.875171415357807 0.659990852817209
0.426060995483839 0.875171415357807
0.953678792083521 0.426060995483839
0.172284659093187 0.953678792083521
0.556150355803892 0.172284659093187
0.962703836418076 0.556150355803892
0.14003012307958 0.962703836418076
0.4696445820686 0.14003012307958
0.971406339548619 0.4696445820686
0.108326645830151 0.971406339548619
0.376708736170027 0.108326645830151
0.915717130626511 0.376708736170027
0.300999142484275 0.915717130626511
0.820554768961224 0.300999142484275
0.574254096375244 0.820554768961224
0.953496683768834 0.574254096375244
0.172928955461613 0.953496683768834
0.557795674115817 0.172928955461613
0.961972674208644 0.557795674115817
0.142666868309607 0.961972674208644
0.47702082868571 0.142666868309607
0.972940634974263 0.47702082868571
0.102675907581549 0.972940634974263
0.359320905777032 0.102675907581549
0.897816630549561 0.359320905777032
0.357793520986957 0.897816630549561
0.896131537574181 0.357793520986957
0.363011239262551 0.896131537574181
0.90181290978331 0.363011239262551
0.345330903572732 0.90181290978331
0.881702135380453 0.345330903572732
0.406783571399409 0.881702135380453
0.941111720011909 0.406783571399409
0.216139756825728 0.941111720011909
0.660751113145668 0.216139756825728
0.874220410527471 0.660751113145668
0.428840428944138 0.874220410527471
0.955251630243868 0.428840428944138
0.16670921732517 0.955251630243868
0.541777291317597 0.16670921732517
0.968193165927642 0.541777291317597
0.120101121576741 0.968193165927642
0.412139684473726 0.120101121576741
0.944894203326932 0.412139684473726
0.203069676599769 0.944894203326932
0.631146293877194 0.203069676599769
0.907922533448879 0.631146293877194
0.326036906148129 0.907922533448879
0.856973683712192 0.326036906148129
0.478022177634096 0.856973683712192
0.973116203763806 0.478022177634096
0.102028125170887 0.973116203763806
0.357311708695503 0.102028125170887
0.895596200945952 0.357311708695503
0.364664218608807 0.895596200945952
0.903568482473417 0.364664218608807
0.339816671821003 0.903568482473417
0.874931075356649 0.339816671821003
0.426764086054436 0.874931075356649
0.954082353543376 0.426764086054436
0.170855943181984 0.954082353543376
0.552490340459377 0.170855943181984
0.964254580217989 0.552490340459377
0.134423970511815 0.964254580217989
0.453781249988249 0.134423970511815
0.96666892587467 0.453781249988249
0.12565844312963 0.96666892587467
0.428486755319482 0.12565844312963
0.955054837757531 0.428486755319482
0.167407869070521 0.955054837757531
0.543591650330775 0.167407869070521
0.967589095283614 0.543591650330775
0.122305708090221 0.967589095283614
0.418653385249205 0.122305708090221
0.949192640247485 0.418653385249205
0.188081290595226 0.949192640247485
0.595556203020715 0.188081290595226
0.939389147050629 0.595556203020715
0.222054992071069 0.939389147050629
0.673711633012808 0.222054992071069
0.857314647368492 0.673711633012808
0.477072346826128 0.857314647368492
0.972949858607761 0.477072346826128
0.102641881847307 0.972949858607761
0.359215451158787 0.102641881847307
0.897700872149547 0.359215451158787
0.358152663536811 0.897700872149547
0.896529399239365 0.358152663536811
0.361781298601585 0.896529399239365
0.90049280327657 0.361781298601585
0.349461506642331 0.90049280327657
0.886618831868652 0.349461506642331
0.392050927493625 0.886618831868652
0.929553291205552 0.392050927493625
0.255387483056464 0.929553291205552
0.741642394562739 0.255387483056464
0.747274917284944 0.741642394562739
0.736534949598723 0.747274917284944
0.756799748711485 0.736534949598723
0.717810167340701 0.756799748711485
0.789979050911761 0.717810167340701
0.647057385126026 0.789979050911761
0.890659089371594 0.647057385126026
0.379803355976293 0.890659089371594
0.91865579038521 0.379803355976293
0.291436583790049 0.91865579038521
0.805355075533455 0.291436583790049
0.611357283599236 0.805355075533455
0.926638266018657 0.611357283599236
0.265121180877515 0.926638266018657
0.759844567277746 0.265121180877515
0.71167512333934 0.759844567277746
0.800255204421174 0.71167512333934
0.623402567650197 0.800255204421174
0.915610044559621 0.623402567650197
0.301346334358485 0.915610044559621
0.821093212396038 0.301346334358485
0.572906680917453 0.821093212396038
0.954270001922643 0.572906680917453
0.170191184877488 0.954270001922643
0.550781967323191 0.170191184877488
0.964942647999667 0.550781967323191
0.131930502877126 0.964942647999667
0.446646896622088 0.131930502877126
0.963898440803789 0.446646896622088
0.135713122817273 0.963898440803789
0.457450777338581 0.135713122817273
0.967939298238545 0.457450777338581
0.121027971339671 0.967939298238545
0.414882785822991 0.121027971339671
0.946744733417906 0.414882785822991
0.196634658337119 0.946744733417906
0.616080930963277 0.196634658337119
0.922448348120126 0.616080930963277
0.278995833365251 0.922448348120126
0.784512917487312 0.278995833365251
0.659304359153146 0.784512917487312
0.876026272503741 0.659304359153146
0.423556545309076 0.876026272503741
0.952209953116175 0.423556545309076
0.177474017380401 0.952209953116175
0.569310263087524 0.177474017380401
0.956264740979879 0.569310263087524
0.163107695940399 0.956264740979879
0.532363944315112 0.163107695940399
0.970915042922637 0.532363944315112
0.110132187161768 0.970915042922637
0.38221204519966 0.110132187161768
0.920891391045417 0.38221204519966
0.284116704081031 0.920891391045417
0.793238169918335 0.284116704081031
0.639644365241489 0.793238169918335
0.898947859899576 0.639644365241489
0.354278359818038 0.898947859899576
0.892184293972447 0.354278359818038
0.3751467702892 0.892184293972447
0.91420551702005 0.3751467702892
0.305891779713596 0.91420551702005
0.828055795387256 0.305891779713596
0.555279640940057 0.828055795387256
0.9630822290604 0.555279640940057
0.13866391160096 0.9630822290604
0.465801301759877 0.13866391160096
0.970438751250856 0.465801301759877
0.111880787153987 0.970438751250856
0.387517558817187 0.111880787153987
0.925656031659665 0.387517558817187
0.268386076575811 0.925656031659665
0.765784462856611 0.268386076575811
0.699498615285689 0.765784462856611
0.819781179746462 0.699498615285689
0.576185988611848 0.819781179746462
0.952363211043018 0.576185988611848
0.176933348649914 0.952363211043018
0.567948961263225 0.176933348649914
0.95699346078667 0.567948961263225
0.160512209513066 0.95699346078667
0.525517356430168 0.160512209513066
0.972460571631181 0.525517356430168
0.104445932190355 0.972460571631181
0.364794219813047 0.104445932190355
0.903705648315746 0.364794219813047
0.339384823120057 0.903705648315746
0.874390783327716 0.339384823120057
0.428343011301112 0.874390783327716
0.954974576285369 0.428343011301112
0.167692726242401 0.954974576285369
0.544330315650389 0.167692726242401
0.967335810145914 0.544330315650389
0.123229238165524 0.967335810145914
0.421370792804007 0.123229238165524
0.950888046325346 0.421370792804007
0.18212988175556 0.950888046325346
0.580938492916337 0.18212988175556
0.949450945421285 0.580938492916337
0.187176005873661 0.949450945421285
0.593350479925492 0.187176005873661
0.941014182800953 0.593350479925492
0.216475313216789 0.941014182800953
0.661493632739481 0.216475313216789
0.873287245679962 0.661493632739481
0.431560865629405 0.873287245679962
0.956732731057754 0.431560865629405
0.161441328284048 0.956732731057754
0.527974300643128 0.161441328284048
0.971948010163759 0.527974300643128
0.106333795239628 0.971948010163759
0.370604984995304 0.106333795239628
0.909702027358544 0.370604984995304
0.320362570235365 0.909702027358544
0.849148535927468 0.320362570235365
0.499571669452852 0.849148535927468
0.974999284478475 0.499571669452852
0.0950651510052528 0.974999284478475
0.335508295471445 0.0950651510052528
0.869475668651033 0.335508295471445
0.442602148072 0.869475668651033
0.962151397716801 0.442602148072
0.142022733794789 0.962151397716801
0.47522387983295 0.142022733794789
0.972605961090925 0.47522387983295
0.103910061611161 0.972605961090925
0.36313976675779 0.103910061611161
0.901950178571862 0.36313976675779
0.344900610389587 0.901950178571862
0.881182299435661 0.344900610389587
0.408330212928074 0.881182299435661
0.942226935538932 0.408330212928074
0.212297816186997 0.942226935538932
0.652187068373993 0.212297816186997
0.884672475256945 0.652187068373993
0.397905638440811 0.884672475256945
0.934349291217504 0.397905638440811
0.239228703553523 0.934349291217504
0.709793490703532 0.239228703553523
0.803348095907865 0.709793490703532
0.616121737565385 0.803348095907865
0.922411394052704 0.616121737565385
0.279117595280359 0.922411394052704
0.784722756812549 0.279117595280359
0.658838511836943 0.784722756812549
0.876604275913959 0.658838511836943
0.42185995551696 0.876604275913959
0.951187120447935 0.42185995551696
0.181077711133412 0.951187120447935
0.578325437289978 0.181077711133412
0.951073990906002 0.578325437289978
0.181475793439716 0.951073990906002
0.579315086357052 0.181475793439716
0.950465556597076 0.579315086357052
0.183615051046777 0.950465556597076
0.584612199895879 0.183615051046777
0.947079024952241 0.584612199895879
0.195469347246239 0.947079024952241
0.613318217980137 0.195469347246239
0.924920027747844 0.613318217980137
0.270827583073207 0.924920027747844
0.770172012947736 0.270827583073207
0.690327625337097 0.770172012947736
0.833724040630813 0.690327625337097
0.54065023234967 0.833724040630813
0.96855547857868 0.54065023234967
0.118777477626228 0.96855547857868
0.408210614896416 0.118777477626228
0.942141364250993 0.408210614896416
0.212592954671255 0.942141364250993
0.652849042152062 0.212592954671255
0.883884964221469 0.652849042152062
0.400266103554257 0.883884964221469
0.936207284609024 0.400266103554257
0.232920498930684 0.936207284609024
0.696807306423405 0.232920498930684
0.823940848139619 0.696807306423405
0.565743074935678 0.823940848139619
0.958143607582207 0.565743074935678
0.156407295843087 0.958143607582207
0.514581809235545 0.156407295843087
0.974170746273731 0.514581809235545
0.0981322031750315 0.974170746273731
0.345158868112679 0.0981322031750315
0.881494473115828 0.345158868112679
0.407401671230101 0.881494473115828
0.941559643085184 0.407401671230101
0.21459781823409 0.941559643085184
0.657327819108707 0.21459781823409
0.878467033604542 0.657327819108707
0.416374547450856 0.878467033604542
0.947726456375208 0.416374547450856
0.193209979020625 0.947726456375208
0.607931543807152 0.193209979020625
0.929568049220479 0.607931543807152
0.255338035246759 0.929568049220479
0.741548039712027 0.255338035246759
0.747452723593981 0.741548039712027
0.736191883385092 0.747452723593981
0.757432237469712 0.736191883385092
0.716541708134217 0.757432237469712
0.79212778568943 0.716541708134217
0.642179291629942 0.79212778568943
0.896161691223271 0.642179291629942
0.36291806618776 0.896161691223271
0.901713319346957 0.36291806618776
0.345642995331075 0.901713319346957
0.882078268927586 0.345642995331075
0.405663166011867 0.882078268927586
0.940292190837072 0.405663166011867
0.218956868082775 0.940292190837072
0.666957556208413 0.218956868082775
0.866288180257167 0.666957556208413
0.451748579115184 0.866288180257167
0.965920021492126 0.451748579115184
0.12838228093383 0.965920021492126
0.436411056416623 0.12838228093383
0.959230140390406 0.436411056416623
0.152519944812332 0.959230140390406
0.504104683862424 0.152519944812332
0.974934291124519 0.504104683862424
0.0953059345447966 0.974934291124519
0.336268582202824 0.0953059345447966
0.870448889021894 0.336268582202824
0.439793720427534 0.870448889021894
0.960863295210164 0.439793720427534
0.146659590199314 0.960863295210164
0.488087163727345 0.146659590199314
0.97444652889457 0.488087163727345
0.0971119157575874 0.97444652889457
0.341956647144367 0.0971119157575874
0.877586964610784 0.341956647144367
0.418969528208456 0.877586964610784
0.949392844300835 0.418969528208456
0.1873796788157 0.949392844300835
0.593847285652239 0.1873796788157
0.940651479205258 0.593847285652239
0.217722468109452 0.940651479205258
0.664245640460144 0.217722468109452
0.869791141400364 0.664245640460144
0.441692595793081 0.869791141400364
0.961740961797139 0.441692595793081
0.143501608374748 0.961740961797139
0.479344697397572 0.143501608374748
0.973336098050168 0.479344697397572
0.101216459302302 0.973336098050168
0.354789581907569 0.101216459302302
0.892764344461944 0.354789581907569
0.37337106190558 0.892764344461944
0.91246393694461 0.37337106190558
0.311506652808595 0.91246393694461
0.836434006451864 0.311506652808595
0.533567421280713 0.836434006451864
0.970605590091396 0.533567421280713
0.111268476441442 0.970605590091396
0.385662430108186 0.111268476441442
0.924014988433815 0.385662430108186
0.27382402937554 0.924014988433815
0.775493278217125 0.27382402937554
0.679003469263008 0.775493278217125
0.850035256168049 0.679003469263008
0.497153745813537 0.850035256168049
0.974968405464714 0.497153745813537
0.0951795538601992 0.974968405464714
0.335869584909973 0.0951795538601992
0.869938706685265 0.335869584909973
0.441266877854532 0.869938706685265
0.961546639415877 0.441266877854532
0.144201328611249 0.961546639415877
0.481288491208199 0.144201328611249
0.973634529811064 0.481288491208199
0.100114295465545 0.973634529811064
0.351356550905031 0.100114295465545
0.888829987660491 0.351356550905031
0.385363838714176 0.888829987660491
0.923748347050037 0.385363838714176
0.27470561965221 0.923748347050037
0.77704552451645 0.27470561965221
0.67565853164708 0.77704552451645
0.854661913012409 0.67565853164708
0.484438217087677 0.854661913012409
0.9740555405591 0.484438217087677
0.0985582434143563 0.9740555405591
0.346493612670788 0.0985582434143563
0.883099577291622 0.346493612670788
0.402615384128024 0.883099577291622
0.938013322706724 0.402615384128024
0.226762883612521 0.938013322706724
0.683831765090195 0.226762883612521
0.843202940360911 0.683831765090195
0.515625792737737 0.843202940360911
0.974047754945138 0.515625792737737
0.0985870315227519 0.974047754945138
0.346583752079307 0.0985870315227519
0.883207474008352 0.346583752079307
0.402292924270138 0.883207474008352
0.937767976674044 0.402292924270138
0.227600874534982 0.937767976674044
0.685614994138984 0.227600874534982
0.840633588408062 0.685614994138984
0.522478157948164 0.840633588408062
0.973029456419504 0.522478157948164
0.102348220101929 0.973029456419504
0.358304941581194 0.102348220101929
0.896697790636795 0.358304941581194
0.361260365325153 0.896697790636795
0.899930123704131 0.361260365325153
0.351217995000593 0.899930123704131
0.88866926845459 0.351217995000593
0.385851179059998 0.88866926845459
0.924183182044228 0.385851179059998
0.273267649476249 0.924183182044228
0.774510520781302 0.273267649476249
0.681111498523476 0.774510520781302
0.847074637900066 0.681111498523476
0.505202863333499 0.847074637900066
0.974894427831218 0.505202863333499
0.0954536014179529 0.974894427831218
0.336734624437752 0.0954536014179529
0.871043226855821 0.336734624437752
0.438075002837237 0.871043226855821
0.960044649432928 0.438075002837237
0.149599790059724 0.960044649432928
0.496156802207861 0.149599790059724
0.974942396339849 0.496156802207861
0.0952759086199513 0.974942396339849
0.336173798440716 0.0952759086199513
0.870327805162362 0.336173798440716
0.440143535222147 0.870327805162362
0.961027094134761 0.440143535222147
0.146070672047268 0.961027094134761
0.486462720178218 0.146070672047268
0.974285294014605 0.486462720178218
0.0977084935377726 0.974285294014605
0.343830020930572 0.0977084935377726
0.879882656786072 0.343830020930572
0.412187751585307 0.879882656786072
0.94492713521059 0.412187751585307
0.202955392977861 0.94492713521059
0.630881555612345 0.202955392977861
0.908193071761921 0.630881555612345
0.325175823045712 0.908193071761921
0.855802377893798 0.325175823045712
0.481278204751964 0.855802377893798
0.973633028092493 0.481278204751964
0.100119843329748 0.973633028092493
0.351373855175362 0.100119843329748
0.888850049390806 0.351373855175362
0.385302992446193 0.888850049390806
0.923693926186987 0.385302992446193
0.274885481957797 0.923693926186987
0.777361469689845 0.274885481957797
0.674975399012809 0.777361469689845
0.855596077987202 0.674975399012809
0.481850574348466 0.855596077987202
0.973715333559226 0.481850574348466
0.099815752728399 0.973715333559226
0.350425016119082 0.099815752728399
0.887746564368391 0.350425016119082
0.388645147099987 0.887746564368391
0.926640377268904 0.388645147099987
0.265114155087003 0.926640377268904
0.75983169545208 0.265114155087003
0.71170121115014 0.75983169545208
0.800212129070499 0.71170121115014
0.623503442479937 0.800212129070499
0.91551290881286 0.623503442479937
0.301661188178518 0.91551290881286
0.821580691328446 0.301661188178518
0.571684849962403 0.821580691328446
0.954959000914884 0.571684849962403
0.167747999197468 0.954959000914884
0.544473571054586 0.167747999197468
0.967286195762845 0.544473571054586
0.123410083873002 0.967286195762845
0.421902136778697 0.123410083873002
0.95121282266504 0.421902136778697
0.180987255784324 0.95121282266504
0.578100389209137 0.180987255784324
0.951211283900987 0.578100389209137
0.180992671393648 0.951211283900987
0.578113864752213 0.180992671393648
0.951203074120545 0.578113864752213
0.181021565026263 0.951203074120545
0.578185756284652 0.181021565026263
0.951159251305368 0.578185756284652
0.181175786850161 0.951159251305368
0.578569372326607 0.181175786850161
0.950924729555592 0.578569372326607
0.182000864273343 0.950924729555592
0.580618543740687 0.182000864273343
0.949652536580011 0.580618543740687
0.186469125753933 0.949652536580011
0.591623724488537 0.186469125753933
0.94225986313231 0.591623724488537
0.212184232502544 0.94225986313231
0.651932127521417 0.212184232502544
0.884974851644582 0.651932127521417
0.396998018044804 0.884974851644582
0.933623307681875 0.396998018044804
0.241686225436615 0.933623307681875
0.714768576096186 0.241686225436615
0.795110389014305 0.714768576096186
0.63534844735372 0.795110389014305
0.903555111415856 0.63534844735372
0.339858760995531 0.903555111415856
0.874983655923443 0.339858760995531
0.426610305382128 0.874983655923443
0.953994415623192 0.426610305382128
0.171167375273529 0.953994415623192
0.55328950917045 0.171167375273529
0.963924890028253 0.55328950917045
0.135617416007867 0.963924890028253
0.457178796684543 0.135617416007867
0.967848743731803 0.457178796684543
0.121358456655436 0.967848743731803
0.415859268449222 0.121358456655436
0.94738931544699 0.415859268449222
0.194386921653109 0.94738931544699
0.610742520739014 0.194386921653109
0.927170766991439 0.610742520739014
0.263348029728988 0.927170766991439
0.756583795370707 0.263348029728988
0.718242548217336 0.756583795370707
0.789243741575654 0.718242548217336
0.648718426041333 0.789243741575654
0.888743036047575 0.648718426041333
0.385627522505604 0.888743036047575
0.923983851927996 0.385627522505604
0.273927003886766 0.923983851927996
0.775674901670704 0.273927003886766
0.678613059496506 0.775674901670704
0.850579762411461 0.678613059496506
0.495665937731343 0.850579762411461
0.974926742026581 0.495665937731343
0.0953338998612693 0.974926742026581
0.336356854854192 0.0953338998612693
0.870561592082473 0.336356854854192
0.43946801524588 0.870561592082473
0.960709927404735 0.43946801524588
0.147210814883822 0.960709927404735
0.489605184373747 0.147210814883822
0.974578596451575 0.489605184373747
0.0966231075812061 0.974578596451575
0.340419622383971 0.0966231075812061
0.87568300201171 0.340419622383971
0.42456289979793 0.87568300201171
0.952806051261101 0.42456289979793
0.175370051771183 0.952806051261101
0.564000047180526 0.175370051771183
0.959025576447473 0.564000047180526
0.153252528651555 0.959025576447473
0.506088145342492 0.153252528651555
0.974855444496526 0.506088145342492
0.0955979966448264 0.974855444496526
0.337190176761056 0.0955979966448264
0.871622549681925 0.337190176761056
0.43639705421483 0.871622549681925
0.95922319462105 0.43639705421483
0.152544824335946 0.95922319462105
0.50417211352662 0.152544824335946
0.974932114528012 0.50417211352662
0.0953139977008105 0.974932114528012
0.336294034218092 0.0953139977008105
0.870481391392909 0.336294034218092
0.439699800663139 0.870481391392909
0.960819155243746 0.439699800663139
0.146818254025719 0.960819155243746
0.488524351811177 0.146818254025719
0.974486407044721 0.488524351811177
0.096964333166187 0.974486407044721
0.341492779913312 0.096964333166187
0.877014298603522 0.341492779913312
0.420654852729127 0.877014298603522
0.950446955657797 0.420654852729127
0.183680406540627 0.950446955657797
0.584773467695419 0.183680406540627
0.946972490782086 0.584773467695419
0.195840810687827 0.946972490782086
0.61420003147215 0.195840810687827
0.924137575965864 0.61420003147215
0.273418534949824 0.924137575965864
0.774777274813292 0.273418534949824
0.680540052060104 0.774777274813292
0.847880629448326 0.680540052060104
0.503018363853083 0.847880629448326
0.974964468970637 0.503018363853083
0.0951941375402206 0.974964468970637
0.335915633500957 0.0951941375402206
0.86999765061537 0.335915633500957
0.441096780302514 0.86999765061537
0.961468601766152 0.441096780302514
0.144482245377572 0.961468601766152
0.482067791978085 0.144482245377572
0.973745900070289 0.482067791978085
0.0997028064497072 0.973745900070289
0.350072411659462 0.0997028064497072
0.887334701192121 0.350072411659462
0.389890134084399 0.887334701192121
0.927715687969389 0.389890134084399
0.261531232033003 0.927715687969389
0.75321732214678 0.261531232033003
0.724935852282774 0.75321732214678
0.777675063195506 0.724935852282774
0.674296581189549 0.777675063195506
0.856520736963976 0.674296581189549
0.479282560047188 0.856520736963976
0.973326071959026 0.479282560047188
0.10125347545497 0.973326071959026
0.354904715736714 0.10125347545497
0.892894698089769 0.354904715736714
0.372971649230501 0.892894698089769
0.912068812593046 0.372971649230501
0.312777245383812 0.912068812593046
0.838295796600313 0.312777245383812
0.528668220609983 0.838295796600313
0.971794719195524 0.528668220609983
0.106897997462946 0.971794719195524
0.372336180845296 0.106897997462946
0.911437602187457 0.372336180845296
0.314804488074321 0.911437602187457
0.841240227214084 0.314804488074321
0.520864918590432 0.841240227214084
0.973302155171637 0.520864918590432
0.101341772648546 0.973302155171637
0.355179309283487 0.101341772648546
0.893205173407529 0.355179309283487
0.372019797261665 0.893205173407529
0.911122164057513 0.372019797261665
0.315816408260602 0.911122164057513
0.842697977682502 0.315816408260602
0.516976574760061 0.842697977682502
0.973876004046723 0.516976574760061
0.0992219778759708 0.973876004046723
0.348570210231169 0.0992219778759708
0.885569173205214 0.348570210231169
0.395212009428005 0.885569173205214
0.932175960424346 0.395212009428005
0.246573363002048 0.932175960424346
0.724522264673642 0.246573363002048
0.778400035396694 0.724522264673642
0.672724339135367 0.778400035396694
0.858648580413978 0.672724339135367
0.47334766349145 0.858648580413978
0.972229646538677 0.47334766349145
0.105296727626831 0.972229646538677
0.367416374433855 0.105296727626831
0.906444170703772 0.367416374433855
0.3307322307936 0.906444170703772
0.863258847000767 0.3307322307936
0.460367739295125 0.863258847000767
0.968874207254541 0.460367739295125
0.11761221330856 0.968874207254541
0.404740364297964 0.11761221330856
0.93960984704307 0.404740364297964
0.221298411292798 0.93960984704307
0.672069155363119 0.221298411292798
0.85952960251323 0.672069155363119
0.470880013175049 0.85952960251323
0.971692902832527 0.470880013175049
0.107272641127954 0.971692902832527
0.37348436421421 0.107272641127954
0.912575796216697 0.37348436421421
0.311146728267017 0.912575796216697
0.835904322847417 0.311146728267017
0.53495631498043 0.835904322847417
0.970234418567657 0.53495631498043
0.112630407217917 0.970234418567657
0.389784714492588 0.112630407217917
0.927625104278026 0.389784714492588
0.261833403745708 0.927625104278026
0.753779022471715 0.261833403745708
0.723825210237873 0.753779022471715
0.779618873521691 0.723825210237873
0.670071813724797 0.779618873521691
0.862194754887796 0.670071813724797
0.4633783421739 0.862194754887796
0.969769531294465 0.4633783421739
0.114334691122776 0.969769531294465
0.394922851161677 0.114334691122776
0.931939291888835 0.394922851161677
0.24737094767803 0.931939291888835
0.726096391499323 0.24737094767803
0.77563364482884 0.726096391499323
0.678701765969638 0.77563364482884
0.850456147473397 0.678701765969638
0.496003905922607 0.850456147473397
0.974937721805286 0.496003905922607
0.0952932255849784 0.974937721805286
0.336228464296096 0.0952932255849784
0.870397647963422 0.336228464296096
0.439941771294344 0.870397647963422
0.960932735742483 0.439941771294344
0.146409951171707 0.960932735742483
0.487398901741462 0.146409951171707
0.974380728058447 0.487398901741462
0.0973554069022781 0.974380728058447
0.342721593431763 0.0973554069022781
0.878527661026689 0.342721593431763
0.416195558366888 0.878527661026689
0.947609580693993 0.416195558366888
0.193618086756695 0.947609580693993
0.608907480625751 0.193618086756695
0.928742726588631 0.608907480625751
0.25810073014965 0.928742726588631
0.746790498658884 0.25810073014965
0.737468354109628 0.746790498658884
0.755074245106211 0.737468354109628
0.721254804985637 0.755074245106211
0.784080613955994 0.721254804985637
0.660262998620103 0.784080613955994
0.874831507965843 0.660262998620103
0.427055228480605 0.874831507965843
0.954248335201134 0.427055228480605
0.170267954871292 0.954248335201134
0.550979435819434 0.170267954871292
0.964864278781776 0.550979435819434
0.132214689019528 0.964864278781776
0.447462463605292 0.132214689019528
0.964235248351341 0.447462463605292
0.134493973333857 0.964235248351341
0.453980843435842 0.134493973333857
0.966740725193582 0.453980843435842
0.125397072238501 0.966740725193582
0.427723321398802 0.125397072238501
0.954626718748479 0.427723321398802
0.168926731739967 0.954626718748479
0.54752291507012 0.168926731739967
0.966192132918629 0.54752291507012
0.12739309129864 0.966192132918629
0.433539957193263 0.12739309129864
0.957773944569494 0.433539957193263
0.157727761125815 0.957773944569494
0.518113886534605 0.157727761125815
0.973720359746985 0.518113886534605
0.0997971809686305 0.973720359746985
0.350367044193445 0.0997971809686305
0.887678916292715 0.350367044193445
0.388849725662213 0.887678916292715
0.926817904407075 0.388849725662213
0.264523258262457 0.926817904407075
0.758747745992614 0.264523258262457
0.713893455380592 0.758747745992614
0.796573400006867 0.713893455380592
0.631972451792631 0.796573400006867
0.907074760674583 0.631972451792631
0.328731542964752 0.907074760674583
0.860601750936585 0.328731542964752
0.467868871163731 0.860601750936585
0.970973603182858 0.467868871163731
0.109917073909329 0.970973603182858
0.381557712012908 0.109917073909329
0.920288555223892 0.381557712012908
0.286094368356471 0.920288555223892
0.796553084929614 0.286094368356471
0.632019444493044 0.796553084929614
0.907026378475417 0.632019444493044
0.328885156178227 0.907026378475417
0.860806869873016 0.328885156178227
0.467291770344503 0.860806869873016
0.970827669679933 0.467291770344503
0.11045309130868 0.970827669679933
0.383187503123243 0.11045309130868
0.921783878236328 0.383187503123243
0.281183604233696 0.921783878236328
0.788265601280993 0.281183604233696
0.650921478160619 0.788265601280993
0.886168558976273 0.650921478160619
0.393407991826935 0.886168558976273
0.930688760795169 0.393407991826935
0.251578046166413 0.930688760795169
0.734317478128663 0.251578046166413
0.760871745829352 0.734317478128663
0.709589135688989 0.760871745829352
0.803682337384455 0.709589135688989
0.615330448046784 0.803682337384455
0.923125662237979 0.615330448046784
0.276762228427118 0.923125662237979
0.780643099638378 0.276762228427118
0.667833857438917 0.780643099638378
0.865144005558975 0.667833857438917
0.455012435296954 0.865144005558975
0.967106864186148 0.455012435296954
0.124063591977716 0.967106864186148
0.42382008678088 0.124063591977716
0.952366821205517 0.42382008678088
0.176920610382423 0.952366821205517
0.567916861216903 0.176920610382423
0.957010469853531 0.567916861216903
0.160451578732592 0.957010469853531
0.525356791497763 0.160451578732592
0.972492429187262 0.525356791497763
0.104328527006808 0.972492429187262
0.364431933291653 0.104328527006808
0.903323067226948 0.364431933291653
0.340588963426322 0.903323067226948
0.875893673532171 0.340588963426322
0.423945390174109 0.875893673532171
0.952441215664503 0.423945390174109
0.176658090835304 0.952441215664503
0.567255038133146 0.176658090835304
0.957359363398265 0.567255038133146
0.159207409576541 0.957359363398265
0.522055600218626 0.159207409576541
0.973102846946085 0.522055600218626
0.102077415224886 0.973102846946085
0.357464704450981 0.102077415224886
0.895766389138739 0.357464704450981
0.364138964389528 0.895766389138739
0.903012938111115 0.364138964389528
0.341564229688683 0.903012938111115
0.877102616074852 0.341564229688683
0.42039510610306 0.877102616074852
0.950285937383862 0.42039510610306
0.184246040917914 0.950285937383862
0.586167805563554 0.184246040917914
0.94604292620901 0.586167805563554
0.199078261117725 0.94604292620901
0.621839817565464 0.199078261117725
0.917104729536896 0.621839817565464
0.296492213932015 0.917104729536896
0.813479865937859 0.296492213932015
0.59174845724117 0.813479865937859
0.942170660316074 0.59174845724117
0.212491917907023 0.942170660316074
0.652622500651751 0.212491917907023
0.884154851949745 0.652622500651751
0.399457693921532 0.884154851949745
0.935575854284852 0.399457693921532
0.235067333139683 0.935575854284852
0.701261659918092 0.235067333139683
0.817025602563355 0.701261659918092
0.583029592545432 0.817025602563355
0.948113738370784 0.583029592545432
0.191856902185106 0.948113738370784
0.604686541949258 0.191856902185106
0.932258838945355 0.604686541949258
0.246293954999094 0.932258838945355
0.723969646646994 0.246293954999094
0.779366629785201 0.723969646646994
0.670621716033588 0.779366629785201
0.861464097069239 0.670621716033588
0.465440455466688 0.861464097069239
0.970341987738435 0.465440455466688
0.112235816824005 0.970341987738435
0.38859185915883 0.112235816824005
0.926594082001824 0.38859185915883
0.265268207883981 0.926594082001824
0.760113844503008 0.265268207883981
0.711129072801674 0.760113844503008
0.801155607009831 0.711129072801674
0.621290671429506 0.801155607009831
0.9176254347943 0.621290671429506
0.2947970852302 0.9176254347943
0.810777878702915 0.2947970852302
0.598326729424774 0.810777878702915
0.937294231694446 0.598326729424774
0.229217644214379 0.937294231694446
0.689039971601238 0.229217644214379
};
\addplot [very thick, RoyalBlue, mark=x, mark size=5, mark options={solid}, only marks]
table {%
0.639687490097443 0.538795799812285
0.591813036321478 0.639687490097443
0.598746655100378 0.591813036321478
0.709671004663524 0.598746655100378
0.497238654471472 0.709671004663524
0.607092357746446 0.497238654471472
0.580236499372394 0.607092357746446
0.565535062634486 0.580236499372394
0.654043711872644 0.565535062634486
0.533078464917981 0.654043711872644
0.596959818189684 0.533078464917981
0.700181891322699 0.596959818189684
0.565990550222539 0.700181891322699
0.6328680390405 0.565990550222539
0.590896361691596 0.6328680390405
0.615702286681458 0.590896361691596
0.534401257631748 0.615702286681458
0.535105440965114 0.534401257631748
0.553927421994248 0.535105440965114
0.468683977855696 0.553927421994248
0.602836661222961 0.468683977855696
0.557307375397452 0.602836661222961
0.660452667277966 0.557307375397452
0.699255131559479 0.660452667277966
0.507005018225519 0.699255131559479
0.566661713641444 0.507005018225519
0.524396003680959 0.566661713641444
0.551061050615747 0.524396003680959
0.60606317504587 0.551061050615747
0.497908603262945 0.60606317504587
0.605072860196422 0.497908603262945
0.584340000340011 0.605072860196422
};
\addlegendentry{no decoding}
\addplot [very thick, magenta, mark=x, mark size=5, mark options={solid}, only marks]
table {%
0.967138710581483 0.553156765174898
0.129467728973731 0.967138710581483
0.439799951564868 0.129467728973731
0.946999863979551 0.439799951564868
0.144946772996457 0.946999863979551
0.490461401950633 0.144946772996457
0.963455818426014 0.490461401950633
0.105455021907201 0.963455818426014
0.359576501152593 0.105455021907201
0.900815383661233 0.359576501152593
0.355275060596562 0.900815383661233
0.890819977630871 0.355275060596562
0.377048060696877 0.890819977630871
0.912286744653274 0.377048060696877
0.29511227342811 0.912286744653274
0.816657157993081 0.29511227342811
0.588372346190304 0.816657157993081
0.93497409826702 0.588372346190304
0.199753505083902 0.93497409826702
0.629883132658195 0.199753505083902
0.887062882767759 0.629883132658195
0.342183751568319 0.887062882767759
0.897361055252411 0.342183751568319
0.394491220852889 0.897361055252411
0.924428038957591 0.394491220852889
0.250449339472705 0.924428038957591
0.734360126339841 0.250449339472705
0.766283393209023 0.734360126339841
0.706015745983646 0.766283393209023
0.817498761498924 0.706015745983646
0.612863133636368 0.817498761498924
0.931496652193939 0.612863133636368
};
\addlegendentry{with decoding}
\end{axis}

\end{tikzpicture}

%% file: figures/convergence_ep0_0_num_qubits4_num_meas3_degree3_num_reservoirs20_timeplex10_methodquantum_stab_noiseTrue.pickle.tex
\begin{tikzpicture}

\definecolor{crimson2143940}{RGB}{214,39,40}
\definecolor{darkgray176}{RGB}{176,176,176}
\definecolor{darkorange25512714}{RGB}{255,127,14}
\definecolor{forestgreen4416044}{RGB}{44,160,44}
\definecolor{gray}{RGB}{128,128,128}
\definecolor{lightgray204}{RGB}{204,204,204}
\definecolor{steelblue31119180}{RGB}{31,119,180}

\begin{axis}[
width=1.1\textwidth,
height=.7\textwidth,
legend cell align={left},
legend style={
  fill opacity=0,
  draw opacity=0,
  text opacity=0,
  at={(0.03,0.03)},
  anchor=south west,
  draw=lightgray204
},
tick align=outside,
tick pos=left,
x grid style={darkgray176},
xmin=34, xmax=68,
xtick style={color=black},
y grid style={darkgray176},
ymin=0, ymax=1.1,
ytick style={color=black},
xlabel=$n$,
ylabel=$x_n$
]
\path [draw=gray, fill=gray, opacity=0.5]
(axis cs:0,2)
--(axis cs:0,-2)
--(axis cs:1,-2)
--(axis cs:2,-2)
--(axis cs:3,-2)
--(axis cs:4,-2)
--(axis cs:5,-2)
--(axis cs:6,-2)
--(axis cs:7,-2)
--(axis cs:8,-2)
--(axis cs:9,-2)
--(axis cs:10,-2)
--(axis cs:11,-2)
--(axis cs:12,-2)
--(axis cs:13,-2)
--(axis cs:14,-2)
--(axis cs:15,-2)
--(axis cs:16,-2)
--(axis cs:17,-2)
--(axis cs:18,-2)
--(axis cs:19,-2)
--(axis cs:20,-2)
--(axis cs:21,-2)
--(axis cs:22,-2)
--(axis cs:23,-2)
--(axis cs:24,-2)
--(axis cs:25,-2)
--(axis cs:26,-2)
--(axis cs:27,-2)
--(axis cs:28,-2)
--(axis cs:29,-2)
--(axis cs:30,-2)
--(axis cs:31,-2)
--(axis cs:32,-2)
--(axis cs:33,-2)
--(axis cs:34,-2)
--(axis cs:34,2)
--(axis cs:34,2)
--(axis cs:33,2)
--(axis cs:32,2)
--(axis cs:31,2)
--(axis cs:30,2)
--(axis cs:29,2)
--(axis cs:28,2)
--(axis cs:27,2)
--(axis cs:26,2)
--(axis cs:25,2)
--(axis cs:24,2)
--(axis cs:23,2)
--(axis cs:22,2)
--(axis cs:21,2)
--(axis cs:20,2)
--(axis cs:19,2)
--(axis cs:18,2)
--(axis cs:17,2)
--(axis cs:16,2)
--(axis cs:15,2)
--(axis cs:14,2)
--(axis cs:13,2)
--(axis cs:12,2)
--(axis cs:11,2)
--(axis cs:10,2)
--(axis cs:9,2)
--(axis cs:8,2)
--(axis cs:7,2)
--(axis cs:6,2)
--(axis cs:5,2)
--(axis cs:4,2)
--(axis cs:3,2)
--(axis cs:2,2)
--(axis cs:1,2)
--(axis cs:0,2)
--cycle;

\addplot [ultra thick, black] 
table {%
0 0.5
1 0.975
2 0.0950625000000001
3 0.335499922265625
4 0.869464925259
5 0.442633109113109
6 0.962165255336889
7 0.141972779361614
8 0.475084386199614
9 0.972578927536905
10 0.104009713267468
11 0.363447601972601
12 0.90227842611257
13 0.343871064749135
14 0.879932646751981
15 0.412039617334933
16 0.944825587217519
17 0.2033077681307
18 0.631697506238813
19 0.907357490716863
20 0.327833511551757
21 0.859398930996064
22 0.471246392755656
23 0.971775597274708
24 0.10696836468276
25 0.372551921195438
26 0.911652250115202
27 0.314115457402856
28 0.840243053611457
29 0.52351519142969
30 0.972843439510898
31 0.103034418674866
32 0.360431476248473
33 0.899030445993496
34 0.354021342363904
35 0.891891902907578
36 0.376040872098363
37 0.915073124978476
38 0.30308577359035
39 0.823776671006208
40 0.566157802517338
41 0.9579302661477
42 0.157169498248997
43 0.516622263569707
44 0.973922431379895
45 0.0990503632363789
46 0.348033616238569
47 0.884934251005247
48 0.397119927371816
49 0.933721193558477
50 0.24135511240702
51 0.714101006275858
52 0.796226960535493
53 0.632773392622424
54 0.906247782224973
55 0.331354683805434
56 0.864079153569976
57 0.458040842749502
58 0.968133773579029
59 0.120318003135167
60 0.412782166901259
61 0.945332893399284
62 0.201546594820824
63 0.627609703254123
64 0.911491478178039
65 0.314631577208725
66 0.840990336544313
67 0.521529802495244
};

\addplot [semithick, steelblue31119180, mark=*, mark size=3, mark options={solid}]
table {%
0 0.5
1 0.975
2 0.0950625000000001
3 0.335499922265625
4 0.869464925259
5 0.442633109113109
6 0.962165255336889
7 0.141972779361614
8 0.475084386199614
9 0.972578927536905
10 0.104009713267468
11 0.363447601972601
12 0.90227842611257
13 0.343871064749135
14 0.879932646751981
15 0.412039617334933
16 0.944825587217519
17 0.2033077681307
18 0.631697506238813
19 0.907357490716863
20 0.327833511551757
21 0.859398930996064
22 0.471246392755656
23 0.971775597274708
24 0.10696836468276
25 0.372551921195438
26 0.911652250115202
27 0.314115457402856
28 0.840243053611457
29 0.52351519142969
30 0.972843439510898
31 0.103034418674866
32 0.360431476248473
33 0.899030445993496
34 0.354021342363904
35 0.897966504908837
36 0.368330771049688
37 0.918133537704314
38 0.295558417320631
39 0.809465280754736
40 0.59023008351814
41 0.931459699162453
42 0.231345126069112
43 0.672944916027848
44 0.844248764746705
45 0.480769223014381
46 0.937353482509415
47 0.195798547476823
48 0.57250615734559
49 0.926520907909534
50 0.229458379402281
51 0.64990392056038
52 0.86281083697337
53 0.424999926431383
54 0.914582062930172
55 0.266108118015229
56 0.726131326192984
57 0.739983641175623
58 0.706930227524104
59 0.755018997426433
60 0.66084159117678
61 0.80776177657274
62 0.523595577744072
63 0.895787247231855
64 0.278330035890999
65 0.70281829591552
66 0.745752673732866
67 0.659295169889397
};
\addlegendentry{classical}
\addplot [semithick, darkorange25512714, mark=square*, mark size=3, mark options={solid}]
table {%
0 0.5
1 0.975
2 0.0950625000000001
3 0.335499922265625
4 0.869464925259
5 0.442633109113109
6 0.962165255336889
7 0.141972779361614
8 0.475084386199614
9 0.972578927536905
10 0.104009713267468
11 0.363447601972601
12 0.90227842611257
13 0.343871064749135
14 0.879932646751981
15 0.412039617334933
16 0.944825587217519
17 0.2033077681307
18 0.631697506238813
19 0.907357490716863
20 0.327833511551757
21 0.859398930996064
22 0.471246392755656
23 0.971775597274708
24 0.10696836468276
25 0.372551921195438
26 0.911652250115202
27 0.314115457402856
28 0.840243053611457
29 0.52351519142969
30 0.972843439510898
31 0.103034418674866
32 0.360431476248473
33 0.899030445993496
34 0.354021342363904
35 0.85070360927882
36 0.491849795938633
37 0.967859190865356
38 0.113581380978474
39 0.357563583227189
40 0.884406841068515
41 0.387934250897835
42 0.893505596217641
43 0.407356878231103
44 0.97372037923968
45 0.148448703282967
46 0.42801585048125
47 0.851660063078141
48 0.396094135115538
49 0.905405285431639
50 0.345024295839106
51 0.924027509489122
52 0.216987350658235
53 0.611743173725331
54 0.727729657421753
55 0.482711696711145
56 0.79291333171595
57 0.429404082078727
58 0.826136736140733
59 0.35584976249855
60 0.677572670556818
61 0.5335182162737
62 0.655284104541455
63 0.593265757648044
64 0.503783242906937
65 0.722096453266802
66 0.509275739574639
67 0.566919279753287
};
\addlegendentry{quantum $\langle Z_1, Z_2, Z_3\rangle$}
\addplot [semithick, forestgreen4416044, mark=triangle*, mark size=3, mark options={solid}]
table {%
0 0.5
1 0.975
2 0.0950625000000001
3 0.335499922265625
4 0.869464925259
5 0.442633109113109
6 0.962165255336889
7 0.141972779361614
8 0.475084386199614
9 0.972578927536905
10 0.104009713267468
11 0.363447601972601
12 0.90227842611257
13 0.343871064749135
14 0.879932646751981
15 0.412039617334933
16 0.944825587217519
17 0.2033077681307
18 0.631697506238813
19 0.907357490716863
20 0.327833511551757
21 0.859398930996064
22 0.471246392755656
23 0.971775597274708
24 0.10696836468276
25 0.372551921195438
26 0.911652250115202
27 0.314115457402856
28 0.840243053611457
29 0.52351519142969
30 0.972843439510898
31 0.103034418674866
32 0.360431476248473
33 0.899030445993496
34 0.354021342363904
35 0.829258099060368
36 0.510900530212432
37 0.961528769765003
38 0.109186231062077
39 0.363434279689993
40 0.891293709062423
41 0.386944774330152
42 0.865115992785717
43 0.43977225704009
44 0.997209000396607
45 0.136506904232594
46 0.379168272930296
47 0.889786273211007
48 0.37715592163589
49 0.912020787025167
50 0.37430164020688
51 0.9692997964073
52 0.177888732309289
53 0.486792928648562
54 0.809929846316222
55 0.421386668911068
56 0.822802776726386
57 0.450184290459215
58 0.870530118641021
59 0.276433015689181
60 0.612068637559899
61 0.700037123474407
62 0.537091575878353
63 0.659659055059914
64 0.562145096746528
65 0.626955482439796
66 0.485148377646148
67 0.50642065413148
};
\addlegendentry{quantum $\langle Z_1Z_2, Z_2Z_3, Z_3Z_4\rangle$}
\end{axis}

\end{tikzpicture}

%% file: figures/convergence2_ep0_0_num_qubits4_num_meas3_degree3_num_reservoirs20_timeplex10_methodquantum_stab_noiseTrue.pickle.tex
\begin{tikzpicture}

\definecolor{darkgray176}{RGB}{176,176,176}
\definecolor{darkorange25512714}{RGB}{255,127,14}
\definecolor{forestgreen4416044}{RGB}{44,160,44}
\definecolor{gray}{RGB}{128,128,128}
\definecolor{lightgray204}{RGB}{204,204,204}
\definecolor{steelblue31119180}{RGB}{31,119,180}

\begin{axis}[
width=.8\textwidth,
height=.8\textwidth,
legend cell align={left},
legend style={
  fill opacity=0,
  draw opacity=9,
  text opacity=9,
  at={(0.09,0.5)},
  anchor=west,
  draw=lightgray204
},
tick align=outside,
tick pos=left,
x grid style={darkgray176},
xmin=0.0499551749801697, xmax=1.04231632541644,
xtick style={color=black},
y grid style={darkgray176},
ymin=0.0499551749801697, ymax=1.04231632541644,
ytick style={color=black},
xlabel=$x_{n+1}$,
ylabel=$x_n$
]
\addplot [only marks, black, opacity=0.25, mark=*, mark size=3, mark options={solid}, forget plot]
table {%
0.975 0.5
0.0950625000000001 0.975
0.335499922265625 0.0950625000000001
0.869464925259 0.335499922265625
0.442633109113109 0.869464925259
0.962165255336889 0.442633109113109
0.141972779361614 0.962165255336889
0.475084386199614 0.141972779361614
0.972578927536905 0.475084386199614
0.104009713267468 0.972578927536905
0.363447601972601 0.104009713267468
0.90227842611257 0.363447601972601
0.343871064749135 0.90227842611257
0.879932646751981 0.343871064749135
0.412039617334933 0.879932646751981
0.944825587217519 0.412039617334933
0.2033077681307 0.944825587217519
0.631697506238813 0.2033077681307
0.907357490716863 0.631697506238813
0.327833511551757 0.907357490716863
0.859398930996064 0.327833511551757
0.471246392755656 0.859398930996064
0.971775597274708 0.471246392755656
0.10696836468276 0.971775597274708
0.372551921195438 0.10696836468276
0.911652250115202 0.372551921195438
0.314115457402856 0.911652250115202
0.840243053611457 0.314115457402856
0.52351519142969 0.840243053611457
0.972843439510898 0.52351519142969
0.103034418674866 0.972843439510898
0.360431476248473 0.103034418674866
0.899030445993496 0.360431476248473
0.354021342363904 0.899030445993496
0.891891902907578 0.354021342363904
0.376040872098363 0.891891902907578
0.915073124978476 0.376040872098363
0.30308577359035 0.915073124978476
0.823776671006208 0.30308577359035
0.566157802517338 0.823776671006208
0.9579302661477 0.566157802517338
0.157169498248997 0.9579302661477
0.516622263569707 0.157169498248997
0.973922431379895 0.516622263569707
0.0990503632363789 0.973922431379895
0.348033616238569 0.0990503632363789
0.884934251005247 0.348033616238569
0.397119927371816 0.884934251005247
0.933721193558477 0.397119927371816
0.24135511240702 0.933721193558477
0.714101006275858 0.24135511240702
0.796226960535493 0.714101006275858
0.632773392622424 0.796226960535493
0.906247782224973 0.632773392622424
0.331354683805434 0.906247782224973
0.864079153569976 0.331354683805434
0.458040842749502 0.864079153569976
0.968133773579029 0.458040842749502
0.120318003135167 0.968133773579029
0.412782166901259 0.120318003135167
0.945332893399284 0.412782166901259
0.201546594820824 0.945332893399284
0.627609703254123 0.201546594820824
0.911491478178039 0.627609703254123
0.314631577208725 0.911491478178039
0.840990336544313 0.314631577208725
0.521529802495244 0.840990336544313
0.973192223657611 0.521529802495244
0.101747565932864 0.973192223657611
0.356440495162445 0.101747565932864
0.894623607426105 0.356440495162445
0.367661613001829 0.894623607426105
0.906697550174217 0.367661613001829
0.329928700460932 0.906697550174217
0.862195436985061 0.329928700460932
0.463376415166083 0.862195436985061
0.96976898083226 0.463376415166083
0.114336708126496 0.96976898083226
0.394928918675041 0.114336708126496
0.931944264689895 0.394928918675041
0.247354193585875 0.931944264689895
0.72606337635529 0.247354193585875
0.775691864496302 0.72606337635529
0.678576583817155 0.775691864496302
0.85063057447756 0.678576583817155
0.495526980941992 0.85063057447756
0.974921969191976 0.495526980941992
0.0953515803973814 0.974921969191976
0.336412660401102 0.0953515803973814
0.870632811059524 0.336412660401102
0.439262145527849 0.870632811059524
0.960612560833067 0.439262145527849
0.147560668330936 0.960612560833067
0.490567418221388 0.147560668330936
0.97465300296386 0.490567418221388
0.0963476544318197 0.97465300296386
0.339552657277495 0.0963476544318197
0.874600935831819 0.339552657277495
0.427729141608309 0.874600935831819
0.95462999980658 0.427729141608309
0.16891509677589 0.95462999980658
0.547492868742619 0.16891509677589
0.966203266932526 0.547492868742619
0.127352604215616 0.966203266932526
0.433422281818955 0.127352604215616
0.957712889023041 0.433422281818955
0.157945753766508 0.957712889023041
0.518695681271186 0.157945753766508
0.973636838857044 0.518695681271186
0.100105765022275 0.973636838857044
0.351329943243161 0.100105765022275
0.888799135473281 0.351329943243161
0.385457405795393 0.888799135473281
0.923831977040227 0.385457405795393
0.274429175428874 0.923831977040227
0.776559432098978 0.274429175428874
0.676708034016653 0.776559432098978
0.853219655784481 0.676708034016653
0.488419911593223 0.853219655784481
0.974477016054716 0.488419911593223
0.0969990888196709 0.974477016054716
0.341602035792516 0.0969990888196709
0.877149331246205 0.341602035792516
0.420257689568106 0.877149331246205
0.950200539315235 0.420257689568106
0.184545950161058 0.950200539315235
0.586906095516819 0.184545950161058
0.945544589191884 0.586906095516819
0.200811074263158 0.945544589191884
0.625895348194095 0.200811074263158
0.913186409082042 0.625895348194095
0.309180266264562 0.913186409082042
0.832992533946832 0.309180266264562
0.542552292109105 0.832992533946832
0.967938279501419 0.542552292109105
0.121031689651281 0.967938279501419
0.414893777030605 0.121031689651281
0.946752030166346 0.414893777030605
0.196609231814768 0.946752030166346
0.616020762941902 0.196609231814768
0.922502812008878 0.616020762941902
0.278816357993905 0.922502812008878
0.784203406384786 0.278816357993905
0.659990852817209 0.784203406384786
0.875171415357807 0.659990852817209
0.426060995483839 0.875171415357807
0.953678792083521 0.426060995483839
0.172284659093187 0.953678792083521
0.556150355803892 0.172284659093187
0.962703836418076 0.556150355803892
0.14003012307958 0.962703836418076
0.4696445820686 0.14003012307958
0.971406339548619 0.4696445820686
0.108326645830151 0.971406339548619
0.376708736170027 0.108326645830151
0.915717130626511 0.376708736170027
0.300999142484275 0.915717130626511
0.820554768961224 0.300999142484275
0.574254096375244 0.820554768961224
0.953496683768834 0.574254096375244
0.172928955461613 0.953496683768834
0.557795674115817 0.172928955461613
0.961972674208644 0.557795674115817
0.142666868309607 0.961972674208644
0.47702082868571 0.142666868309607
0.972940634974263 0.47702082868571
0.102675907581549 0.972940634974263
0.359320905777032 0.102675907581549
0.897816630549561 0.359320905777032
0.357793520986957 0.897816630549561
0.896131537574181 0.357793520986957
0.363011239262551 0.896131537574181
0.90181290978331 0.363011239262551
0.345330903572732 0.90181290978331
0.881702135380453 0.345330903572732
0.406783571399409 0.881702135380453
0.941111720011909 0.406783571399409
0.216139756825728 0.941111720011909
0.660751113145668 0.216139756825728
0.874220410527471 0.660751113145668
0.428840428944138 0.874220410527471
0.955251630243868 0.428840428944138
0.16670921732517 0.955251630243868
0.541777291317597 0.16670921732517
0.968193165927642 0.541777291317597
0.120101121576741 0.968193165927642
0.412139684473726 0.120101121576741
0.944894203326932 0.412139684473726
0.203069676599769 0.944894203326932
0.631146293877194 0.203069676599769
0.907922533448879 0.631146293877194
0.326036906148129 0.907922533448879
0.856973683712192 0.326036906148129
0.478022177634096 0.856973683712192
0.973116203763806 0.478022177634096
0.102028125170887 0.973116203763806
0.357311708695503 0.102028125170887
0.895596200945952 0.357311708695503
0.364664218608807 0.895596200945952
0.903568482473417 0.364664218608807
0.339816671821003 0.903568482473417
0.874931075356649 0.339816671821003
0.426764086054436 0.874931075356649
0.954082353543376 0.426764086054436
0.170855943181984 0.954082353543376
0.552490340459377 0.170855943181984
0.964254580217989 0.552490340459377
0.134423970511815 0.964254580217989
0.453781249988249 0.134423970511815
0.96666892587467 0.453781249988249
0.12565844312963 0.96666892587467
0.428486755319482 0.12565844312963
0.955054837757531 0.428486755319482
0.167407869070521 0.955054837757531
0.543591650330775 0.167407869070521
0.967589095283614 0.543591650330775
0.122305708090221 0.967589095283614
0.418653385249205 0.122305708090221
0.949192640247485 0.418653385249205
0.188081290595226 0.949192640247485
0.595556203020715 0.188081290595226
0.939389147050629 0.595556203020715
0.222054992071069 0.939389147050629
0.673711633012808 0.222054992071069
0.857314647368492 0.673711633012808
0.477072346826128 0.857314647368492
0.972949858607761 0.477072346826128
0.102641881847307 0.972949858607761
0.359215451158787 0.102641881847307
0.897700872149547 0.359215451158787
0.358152663536811 0.897700872149547
0.896529399239365 0.358152663536811
0.361781298601585 0.896529399239365
0.90049280327657 0.361781298601585
0.349461506642331 0.90049280327657
0.886618831868652 0.349461506642331
0.392050927493625 0.886618831868652
0.929553291205552 0.392050927493625
0.255387483056464 0.929553291205552
0.741642394562739 0.255387483056464
0.747274917284944 0.741642394562739
0.736534949598723 0.747274917284944
0.756799748711485 0.736534949598723
0.717810167340701 0.756799748711485
0.789979050911761 0.717810167340701
0.647057385126026 0.789979050911761
0.890659089371594 0.647057385126026
0.379803355976293 0.890659089371594
0.91865579038521 0.379803355976293
0.291436583790049 0.91865579038521
0.805355075533455 0.291436583790049
0.611357283599236 0.805355075533455
0.926638266018657 0.611357283599236
0.265121180877515 0.926638266018657
0.759844567277746 0.265121180877515
0.71167512333934 0.759844567277746
0.800255204421174 0.71167512333934
0.623402567650197 0.800255204421174
0.915610044559621 0.623402567650197
0.301346334358485 0.915610044559621
0.821093212396038 0.301346334358485
0.572906680917453 0.821093212396038
0.954270001922643 0.572906680917453
0.170191184877488 0.954270001922643
0.550781967323191 0.170191184877488
0.964942647999667 0.550781967323191
0.131930502877126 0.964942647999667
0.446646896622088 0.131930502877126
0.963898440803789 0.446646896622088
0.135713122817273 0.963898440803789
0.457450777338581 0.135713122817273
0.967939298238545 0.457450777338581
0.121027971339671 0.967939298238545
0.414882785822991 0.121027971339671
0.946744733417906 0.414882785822991
0.196634658337119 0.946744733417906
0.616080930963277 0.196634658337119
0.922448348120126 0.616080930963277
0.278995833365251 0.922448348120126
0.784512917487312 0.278995833365251
0.659304359153146 0.784512917487312
0.876026272503741 0.659304359153146
0.423556545309076 0.876026272503741
0.952209953116175 0.423556545309076
0.177474017380401 0.952209953116175
0.569310263087524 0.177474017380401
0.956264740979879 0.569310263087524
0.163107695940399 0.956264740979879
0.532363944315112 0.163107695940399
0.970915042922637 0.532363944315112
0.110132187161768 0.970915042922637
0.38221204519966 0.110132187161768
0.920891391045417 0.38221204519966
0.284116704081031 0.920891391045417
0.793238169918335 0.284116704081031
0.639644365241489 0.793238169918335
0.898947859899576 0.639644365241489
0.354278359818038 0.898947859899576
0.892184293972447 0.354278359818038
0.3751467702892 0.892184293972447
0.91420551702005 0.3751467702892
0.305891779713596 0.91420551702005
0.828055795387256 0.305891779713596
0.555279640940057 0.828055795387256
0.9630822290604 0.555279640940057
0.13866391160096 0.9630822290604
0.465801301759877 0.13866391160096
0.970438751250856 0.465801301759877
0.111880787153987 0.970438751250856
0.387517558817187 0.111880787153987
0.925656031659665 0.387517558817187
0.268386076575811 0.925656031659665
0.765784462856611 0.268386076575811
0.699498615285689 0.765784462856611
0.819781179746462 0.699498615285689
0.576185988611848 0.819781179746462
0.952363211043018 0.576185988611848
0.176933348649914 0.952363211043018
0.567948961263225 0.176933348649914
0.95699346078667 0.567948961263225
0.160512209513066 0.95699346078667
0.525517356430168 0.160512209513066
0.972460571631181 0.525517356430168
0.104445932190355 0.972460571631181
0.364794219813047 0.104445932190355
0.903705648315746 0.364794219813047
0.339384823120057 0.903705648315746
0.874390783327716 0.339384823120057
0.428343011301112 0.874390783327716
0.954974576285369 0.428343011301112
0.167692726242401 0.954974576285369
0.544330315650389 0.167692726242401
0.967335810145914 0.544330315650389
0.123229238165524 0.967335810145914
0.421370792804007 0.123229238165524
0.950888046325346 0.421370792804007
0.18212988175556 0.950888046325346
0.580938492916337 0.18212988175556
0.949450945421285 0.580938492916337
0.187176005873661 0.949450945421285
0.593350479925492 0.187176005873661
0.941014182800953 0.593350479925492
0.216475313216789 0.941014182800953
0.661493632739481 0.216475313216789
0.873287245679962 0.661493632739481
0.431560865629405 0.873287245679962
0.956732731057754 0.431560865629405
0.161441328284048 0.956732731057754
0.527974300643128 0.161441328284048
0.971948010163759 0.527974300643128
0.106333795239628 0.971948010163759
0.370604984995304 0.106333795239628
0.909702027358544 0.370604984995304
0.320362570235365 0.909702027358544
0.849148535927468 0.320362570235365
0.499571669452852 0.849148535927468
0.974999284478475 0.499571669452852
0.0950651510052528 0.974999284478475
0.335508295471445 0.0950651510052528
0.869475668651033 0.335508295471445
0.442602148072 0.869475668651033
0.962151397716801 0.442602148072
0.142022733794789 0.962151397716801
0.47522387983295 0.142022733794789
0.972605961090925 0.47522387983295
0.103910061611161 0.972605961090925
0.36313976675779 0.103910061611161
0.901950178571862 0.36313976675779
0.344900610389587 0.901950178571862
0.881182299435661 0.344900610389587
0.408330212928074 0.881182299435661
0.942226935538932 0.408330212928074
0.212297816186997 0.942226935538932
0.652187068373993 0.212297816186997
0.884672475256945 0.652187068373993
0.397905638440811 0.884672475256945
0.934349291217504 0.397905638440811
0.239228703553523 0.934349291217504
0.709793490703532 0.239228703553523
0.803348095907865 0.709793490703532
0.616121737565385 0.803348095907865
0.922411394052704 0.616121737565385
0.279117595280359 0.922411394052704
0.784722756812549 0.279117595280359
0.658838511836943 0.784722756812549
0.876604275913959 0.658838511836943
0.42185995551696 0.876604275913959
0.951187120447935 0.42185995551696
0.181077711133412 0.951187120447935
0.578325437289978 0.181077711133412
0.951073990906002 0.578325437289978
0.181475793439716 0.951073990906002
0.579315086357052 0.181475793439716
0.950465556597076 0.579315086357052
0.183615051046777 0.950465556597076
0.584612199895879 0.183615051046777
0.947079024952241 0.584612199895879
0.195469347246239 0.947079024952241
0.613318217980137 0.195469347246239
0.924920027747844 0.613318217980137
0.270827583073207 0.924920027747844
0.770172012947736 0.270827583073207
0.690327625337097 0.770172012947736
0.833724040630813 0.690327625337097
0.54065023234967 0.833724040630813
0.96855547857868 0.54065023234967
0.118777477626228 0.96855547857868
0.408210614896416 0.118777477626228
0.942141364250993 0.408210614896416
0.212592954671255 0.942141364250993
0.652849042152062 0.212592954671255
0.883884964221469 0.652849042152062
0.400266103554257 0.883884964221469
0.936207284609024 0.400266103554257
0.232920498930684 0.936207284609024
0.696807306423405 0.232920498930684
0.823940848139619 0.696807306423405
0.565743074935678 0.823940848139619
0.958143607582207 0.565743074935678
0.156407295843087 0.958143607582207
0.514581809235545 0.156407295843087
0.974170746273731 0.514581809235545
0.0981322031750315 0.974170746273731
0.345158868112679 0.0981322031750315
0.881494473115828 0.345158868112679
0.407401671230101 0.881494473115828
0.941559643085184 0.407401671230101
0.21459781823409 0.941559643085184
0.657327819108707 0.21459781823409
0.878467033604542 0.657327819108707
0.416374547450856 0.878467033604542
0.947726456375208 0.416374547450856
0.193209979020625 0.947726456375208
0.607931543807152 0.193209979020625
0.929568049220479 0.607931543807152
0.255338035246759 0.929568049220479
0.741548039712027 0.255338035246759
0.747452723593981 0.741548039712027
0.736191883385092 0.747452723593981
0.757432237469712 0.736191883385092
0.716541708134217 0.757432237469712
0.79212778568943 0.716541708134217
0.642179291629942 0.79212778568943
0.896161691223271 0.642179291629942
0.36291806618776 0.896161691223271
0.901713319346957 0.36291806618776
0.345642995331075 0.901713319346957
0.882078268927586 0.345642995331075
0.405663166011867 0.882078268927586
0.940292190837072 0.405663166011867
0.218956868082775 0.940292190837072
0.666957556208413 0.218956868082775
0.866288180257167 0.666957556208413
0.451748579115184 0.866288180257167
0.965920021492126 0.451748579115184
0.12838228093383 0.965920021492126
0.436411056416623 0.12838228093383
0.959230140390406 0.436411056416623
0.152519944812332 0.959230140390406
0.504104683862424 0.152519944812332
0.974934291124519 0.504104683862424
0.0953059345447966 0.974934291124519
0.336268582202824 0.0953059345447966
0.870448889021894 0.336268582202824
0.439793720427534 0.870448889021894
0.960863295210164 0.439793720427534
0.146659590199314 0.960863295210164
0.488087163727345 0.146659590199314
0.97444652889457 0.488087163727345
0.0971119157575874 0.97444652889457
0.341956647144367 0.0971119157575874
0.877586964610784 0.341956647144367
0.418969528208456 0.877586964610784
0.949392844300835 0.418969528208456
0.1873796788157 0.949392844300835
0.593847285652239 0.1873796788157
0.940651479205258 0.593847285652239
0.217722468109452 0.940651479205258
0.664245640460144 0.217722468109452
0.869791141400364 0.664245640460144
0.441692595793081 0.869791141400364
0.961740961797139 0.441692595793081
0.143501608374748 0.961740961797139
0.479344697397572 0.143501608374748
0.973336098050168 0.479344697397572
0.101216459302302 0.973336098050168
0.354789581907569 0.101216459302302
0.892764344461944 0.354789581907569
0.37337106190558 0.892764344461944
0.91246393694461 0.37337106190558
0.311506652808595 0.91246393694461
0.836434006451864 0.311506652808595
0.533567421280713 0.836434006451864
0.970605590091396 0.533567421280713
0.111268476441442 0.970605590091396
0.385662430108186 0.111268476441442
0.924014988433815 0.385662430108186
0.27382402937554 0.924014988433815
0.775493278217125 0.27382402937554
0.679003469263008 0.775493278217125
0.850035256168049 0.679003469263008
0.497153745813537 0.850035256168049
0.974968405464714 0.497153745813537
0.0951795538601992 0.974968405464714
0.335869584909973 0.0951795538601992
0.869938706685265 0.335869584909973
0.441266877854532 0.869938706685265
0.961546639415877 0.441266877854532
0.144201328611249 0.961546639415877
0.481288491208199 0.144201328611249
0.973634529811064 0.481288491208199
0.100114295465545 0.973634529811064
0.351356550905031 0.100114295465545
0.888829987660491 0.351356550905031
0.385363838714176 0.888829987660491
0.923748347050037 0.385363838714176
0.27470561965221 0.923748347050037
0.77704552451645 0.27470561965221
0.67565853164708 0.77704552451645
0.854661913012409 0.67565853164708
0.484438217087677 0.854661913012409
0.9740555405591 0.484438217087677
0.0985582434143563 0.9740555405591
0.346493612670788 0.0985582434143563
0.883099577291622 0.346493612670788
0.402615384128024 0.883099577291622
0.938013322706724 0.402615384128024
0.226762883612521 0.938013322706724
0.683831765090195 0.226762883612521
0.843202940360911 0.683831765090195
0.515625792737737 0.843202940360911
0.974047754945138 0.515625792737737
0.0985870315227519 0.974047754945138
0.346583752079307 0.0985870315227519
0.883207474008352 0.346583752079307
0.402292924270138 0.883207474008352
0.937767976674044 0.402292924270138
0.227600874534982 0.937767976674044
0.685614994138984 0.227600874534982
0.840633588408062 0.685614994138984
0.522478157948164 0.840633588408062
0.973029456419504 0.522478157948164
0.102348220101929 0.973029456419504
0.358304941581194 0.102348220101929
0.896697790636795 0.358304941581194
0.361260365325153 0.896697790636795
0.899930123704131 0.361260365325153
0.351217995000593 0.899930123704131
0.88866926845459 0.351217995000593
0.385851179059998 0.88866926845459
0.924183182044228 0.385851179059998
0.273267649476249 0.924183182044228
0.774510520781302 0.273267649476249
0.681111498523476 0.774510520781302
0.847074637900066 0.681111498523476
0.505202863333499 0.847074637900066
0.974894427831218 0.505202863333499
0.0954536014179529 0.974894427831218
0.336734624437752 0.0954536014179529
0.871043226855821 0.336734624437752
0.438075002837237 0.871043226855821
0.960044649432928 0.438075002837237
0.149599790059724 0.960044649432928
0.496156802207861 0.149599790059724
0.974942396339849 0.496156802207861
0.0952759086199513 0.974942396339849
0.336173798440716 0.0952759086199513
0.870327805162362 0.336173798440716
0.440143535222147 0.870327805162362
0.961027094134761 0.440143535222147
0.146070672047268 0.961027094134761
0.486462720178218 0.146070672047268
0.974285294014605 0.486462720178218
0.0977084935377726 0.974285294014605
0.343830020930572 0.0977084935377726
0.879882656786072 0.343830020930572
0.412187751585307 0.879882656786072
0.94492713521059 0.412187751585307
0.202955392977861 0.94492713521059
0.630881555612345 0.202955392977861
0.908193071761921 0.630881555612345
0.325175823045712 0.908193071761921
0.855802377893798 0.325175823045712
0.481278204751964 0.855802377893798
0.973633028092493 0.481278204751964
0.100119843329748 0.973633028092493
0.351373855175362 0.100119843329748
0.888850049390806 0.351373855175362
0.385302992446193 0.888850049390806
0.923693926186987 0.385302992446193
0.274885481957797 0.923693926186987
0.777361469689845 0.274885481957797
0.674975399012809 0.777361469689845
0.855596077987202 0.674975399012809
0.481850574348466 0.855596077987202
0.973715333559226 0.481850574348466
0.099815752728399 0.973715333559226
0.350425016119082 0.099815752728399
0.887746564368391 0.350425016119082
0.388645147099987 0.887746564368391
0.926640377268904 0.388645147099987
0.265114155087003 0.926640377268904
0.75983169545208 0.265114155087003
0.71170121115014 0.75983169545208
0.800212129070499 0.71170121115014
0.623503442479937 0.800212129070499
0.91551290881286 0.623503442479937
0.301661188178518 0.91551290881286
0.821580691328446 0.301661188178518
0.571684849962403 0.821580691328446
0.954959000914884 0.571684849962403
0.167747999197468 0.954959000914884
0.544473571054586 0.167747999197468
0.967286195762845 0.544473571054586
0.123410083873002 0.967286195762845
0.421902136778697 0.123410083873002
0.95121282266504 0.421902136778697
0.180987255784324 0.95121282266504
0.578100389209137 0.180987255784324
0.951211283900987 0.578100389209137
0.180992671393648 0.951211283900987
0.578113864752213 0.180992671393648
0.951203074120545 0.578113864752213
0.181021565026263 0.951203074120545
0.578185756284652 0.181021565026263
0.951159251305368 0.578185756284652
0.181175786850161 0.951159251305368
0.578569372326607 0.181175786850161
0.950924729555592 0.578569372326607
0.182000864273343 0.950924729555592
0.580618543740687 0.182000864273343
0.949652536580011 0.580618543740687
0.186469125753933 0.949652536580011
0.591623724488537 0.186469125753933
0.94225986313231 0.591623724488537
0.212184232502544 0.94225986313231
0.651932127521417 0.212184232502544
0.884974851644582 0.651932127521417
0.396998018044804 0.884974851644582
0.933623307681875 0.396998018044804
0.241686225436615 0.933623307681875
0.714768576096186 0.241686225436615
0.795110389014305 0.714768576096186
0.63534844735372 0.795110389014305
0.903555111415856 0.63534844735372
0.339858760995531 0.903555111415856
0.874983655923443 0.339858760995531
0.426610305382128 0.874983655923443
0.953994415623192 0.426610305382128
0.171167375273529 0.953994415623192
0.55328950917045 0.171167375273529
0.963924890028253 0.55328950917045
0.135617416007867 0.963924890028253
0.457178796684543 0.135617416007867
0.967848743731803 0.457178796684543
0.121358456655436 0.967848743731803
0.415859268449222 0.121358456655436
0.94738931544699 0.415859268449222
0.194386921653109 0.94738931544699
0.610742520739014 0.194386921653109
0.927170766991439 0.610742520739014
0.263348029728988 0.927170766991439
0.756583795370707 0.263348029728988
0.718242548217336 0.756583795370707
0.789243741575654 0.718242548217336
0.648718426041333 0.789243741575654
0.888743036047575 0.648718426041333
0.385627522505604 0.888743036047575
0.923983851927996 0.385627522505604
0.273927003886766 0.923983851927996
0.775674901670704 0.273927003886766
0.678613059496506 0.775674901670704
0.850579762411461 0.678613059496506
0.495665937731343 0.850579762411461
0.974926742026581 0.495665937731343
0.0953338998612693 0.974926742026581
0.336356854854192 0.0953338998612693
0.870561592082473 0.336356854854192
0.43946801524588 0.870561592082473
0.960709927404735 0.43946801524588
0.147210814883822 0.960709927404735
0.489605184373747 0.147210814883822
0.974578596451575 0.489605184373747
0.0966231075812061 0.974578596451575
0.340419622383971 0.0966231075812061
0.87568300201171 0.340419622383971
0.42456289979793 0.87568300201171
0.952806051261101 0.42456289979793
0.175370051771183 0.952806051261101
0.564000047180526 0.175370051771183
0.959025576447473 0.564000047180526
0.153252528651555 0.959025576447473
0.506088145342492 0.153252528651555
0.974855444496526 0.506088145342492
0.0955979966448264 0.974855444496526
0.337190176761056 0.0955979966448264
0.871622549681925 0.337190176761056
0.43639705421483 0.871622549681925
0.95922319462105 0.43639705421483
0.152544824335946 0.95922319462105
0.50417211352662 0.152544824335946
0.974932114528012 0.50417211352662
0.0953139977008105 0.974932114528012
0.336294034218092 0.0953139977008105
0.870481391392909 0.336294034218092
0.439699800663139 0.870481391392909
0.960819155243746 0.439699800663139
0.146818254025719 0.960819155243746
0.488524351811177 0.146818254025719
0.974486407044721 0.488524351811177
0.096964333166187 0.974486407044721
0.341492779913312 0.096964333166187
0.877014298603522 0.341492779913312
0.420654852729127 0.877014298603522
0.950446955657797 0.420654852729127
0.183680406540627 0.950446955657797
0.584773467695419 0.183680406540627
0.946972490782086 0.584773467695419
0.195840810687827 0.946972490782086
0.61420003147215 0.195840810687827
0.924137575965864 0.61420003147215
0.273418534949824 0.924137575965864
0.774777274813292 0.273418534949824
0.680540052060104 0.774777274813292
0.847880629448326 0.680540052060104
0.503018363853083 0.847880629448326
0.974964468970637 0.503018363853083
0.0951941375402206 0.974964468970637
0.335915633500957 0.0951941375402206
0.86999765061537 0.335915633500957
0.441096780302514 0.86999765061537
0.961468601766152 0.441096780302514
0.144482245377572 0.961468601766152
0.482067791978085 0.144482245377572
0.973745900070289 0.482067791978085
0.0997028064497072 0.973745900070289
0.350072411659462 0.0997028064497072
0.887334701192121 0.350072411659462
0.389890134084399 0.887334701192121
0.927715687969389 0.389890134084399
0.261531232033003 0.927715687969389
0.75321732214678 0.261531232033003
0.724935852282774 0.75321732214678
0.777675063195506 0.724935852282774
0.674296581189549 0.777675063195506
0.856520736963976 0.674296581189549
0.479282560047188 0.856520736963976
0.973326071959026 0.479282560047188
0.10125347545497 0.973326071959026
0.354904715736714 0.10125347545497
0.892894698089769 0.354904715736714
0.372971649230501 0.892894698089769
0.912068812593046 0.372971649230501
0.312777245383812 0.912068812593046
0.838295796600313 0.312777245383812
0.528668220609983 0.838295796600313
0.971794719195524 0.528668220609983
0.106897997462946 0.971794719195524
0.372336180845296 0.106897997462946
0.911437602187457 0.372336180845296
0.314804488074321 0.911437602187457
0.841240227214084 0.314804488074321
0.520864918590432 0.841240227214084
0.973302155171637 0.520864918590432
0.101341772648546 0.973302155171637
0.355179309283487 0.101341772648546
0.893205173407529 0.355179309283487
0.372019797261665 0.893205173407529
0.911122164057513 0.372019797261665
0.315816408260602 0.911122164057513
0.842697977682502 0.315816408260602
0.516976574760061 0.842697977682502
0.973876004046723 0.516976574760061
0.0992219778759708 0.973876004046723
0.348570210231169 0.0992219778759708
0.885569173205214 0.348570210231169
0.395212009428005 0.885569173205214
0.932175960424346 0.395212009428005
0.246573363002048 0.932175960424346
0.724522264673642 0.246573363002048
0.778400035396694 0.724522264673642
0.672724339135367 0.778400035396694
0.858648580413978 0.672724339135367
0.47334766349145 0.858648580413978
0.972229646538677 0.47334766349145
0.105296727626831 0.972229646538677
0.367416374433855 0.105296727626831
0.906444170703772 0.367416374433855
0.3307322307936 0.906444170703772
0.863258847000767 0.3307322307936
0.460367739295125 0.863258847000767
0.968874207254541 0.460367739295125
0.11761221330856 0.968874207254541
0.404740364297964 0.11761221330856
0.93960984704307 0.404740364297964
0.221298411292798 0.93960984704307
0.672069155363119 0.221298411292798
0.85952960251323 0.672069155363119
0.470880013175049 0.85952960251323
0.971692902832527 0.470880013175049
0.107272641127954 0.971692902832527
0.37348436421421 0.107272641127954
0.912575796216697 0.37348436421421
0.311146728267017 0.912575796216697
0.835904322847417 0.311146728267017
0.53495631498043 0.835904322847417
0.970234418567657 0.53495631498043
0.112630407217917 0.970234418567657
0.389784714492588 0.112630407217917
0.927625104278026 0.389784714492588
0.261833403745708 0.927625104278026
0.753779022471715 0.261833403745708
0.723825210237873 0.753779022471715
0.779618873521691 0.723825210237873
0.670071813724797 0.779618873521691
0.862194754887796 0.670071813724797
0.4633783421739 0.862194754887796
0.969769531294465 0.4633783421739
0.114334691122776 0.969769531294465
0.394922851161677 0.114334691122776
0.931939291888835 0.394922851161677
0.24737094767803 0.931939291888835
0.726096391499323 0.24737094767803
0.77563364482884 0.726096391499323
0.678701765969638 0.77563364482884
0.850456147473397 0.678701765969638
0.496003905922607 0.850456147473397
0.974937721805286 0.496003905922607
0.0952932255849784 0.974937721805286
0.336228464296096 0.0952932255849784
0.870397647963422 0.336228464296096
0.439941771294344 0.870397647963422
0.960932735742483 0.439941771294344
0.146409951171707 0.960932735742483
0.487398901741462 0.146409951171707
0.974380728058447 0.487398901741462
0.0973554069022781 0.974380728058447
0.342721593431763 0.0973554069022781
0.878527661026689 0.342721593431763
0.416195558366888 0.878527661026689
0.947609580693993 0.416195558366888
0.193618086756695 0.947609580693993
0.608907480625751 0.193618086756695
0.928742726588631 0.608907480625751
0.25810073014965 0.928742726588631
0.746790498658884 0.25810073014965
0.737468354109628 0.746790498658884
0.755074245106211 0.737468354109628
0.721254804985637 0.755074245106211
0.784080613955994 0.721254804985637
0.660262998620103 0.784080613955994
0.874831507965843 0.660262998620103
0.427055228480605 0.874831507965843
0.954248335201134 0.427055228480605
0.170267954871292 0.954248335201134
0.550979435819434 0.170267954871292
0.964864278781776 0.550979435819434
0.132214689019528 0.964864278781776
0.447462463605292 0.132214689019528
0.964235248351341 0.447462463605292
0.134493973333857 0.964235248351341
0.453980843435842 0.134493973333857
0.966740725193582 0.453980843435842
0.125397072238501 0.966740725193582
0.427723321398802 0.125397072238501
0.954626718748479 0.427723321398802
0.168926731739967 0.954626718748479
0.54752291507012 0.168926731739967
0.966192132918629 0.54752291507012
0.12739309129864 0.966192132918629
0.433539957193263 0.12739309129864
0.957773944569494 0.433539957193263
0.157727761125815 0.957773944569494
0.518113886534605 0.157727761125815
0.973720359746985 0.518113886534605
0.0997971809686305 0.973720359746985
0.350367044193445 0.0997971809686305
0.887678916292715 0.350367044193445
0.388849725662213 0.887678916292715
0.926817904407075 0.388849725662213
0.264523258262457 0.926817904407075
0.758747745992614 0.264523258262457
0.713893455380592 0.758747745992614
0.796573400006867 0.713893455380592
0.631972451792631 0.796573400006867
0.907074760674583 0.631972451792631
0.328731542964752 0.907074760674583
0.860601750936585 0.328731542964752
0.467868871163731 0.860601750936585
0.970973603182858 0.467868871163731
0.109917073909329 0.970973603182858
0.381557712012908 0.109917073909329
0.920288555223892 0.381557712012908
0.286094368356471 0.920288555223892
0.796553084929614 0.286094368356471
0.632019444493044 0.796553084929614
0.907026378475417 0.632019444493044
0.328885156178227 0.907026378475417
0.860806869873016 0.328885156178227
0.467291770344503 0.860806869873016
0.970827669679933 0.467291770344503
0.11045309130868 0.970827669679933
0.383187503123243 0.11045309130868
0.921783878236328 0.383187503123243
0.281183604233696 0.921783878236328
0.788265601280993 0.281183604233696
0.650921478160619 0.788265601280993
0.886168558976273 0.650921478160619
0.393407991826935 0.886168558976273
0.930688760795169 0.393407991826935
0.251578046166413 0.930688760795169
0.734317478128663 0.251578046166413
0.760871745829352 0.734317478128663
0.709589135688989 0.760871745829352
0.803682337384455 0.709589135688989
0.615330448046784 0.803682337384455
0.923125662237979 0.615330448046784
0.276762228427118 0.923125662237979
0.780643099638378 0.276762228427118
0.667833857438917 0.780643099638378
0.865144005558975 0.667833857438917
0.455012435296954 0.865144005558975
0.967106864186148 0.455012435296954
0.124063591977716 0.967106864186148
0.42382008678088 0.124063591977716
0.952366821205517 0.42382008678088
0.176920610382423 0.952366821205517
0.567916861216903 0.176920610382423
0.957010469853531 0.567916861216903
0.160451578732592 0.957010469853531
0.525356791497763 0.160451578732592
0.972492429187262 0.525356791497763
0.104328527006808 0.972492429187262
0.364431933291653 0.104328527006808
0.903323067226948 0.364431933291653
0.340588963426322 0.903323067226948
0.875893673532171 0.340588963426322
0.423945390174109 0.875893673532171
0.952441215664503 0.423945390174109
0.176658090835304 0.952441215664503
0.567255038133146 0.176658090835304
0.957359363398265 0.567255038133146
0.159207409576541 0.957359363398265
0.522055600218626 0.159207409576541
0.973102846946085 0.522055600218626
0.102077415224886 0.973102846946085
0.357464704450981 0.102077415224886
0.895766389138739 0.357464704450981
0.364138964389528 0.895766389138739
0.903012938111115 0.364138964389528
0.341564229688683 0.903012938111115
0.877102616074852 0.341564229688683
0.42039510610306 0.877102616074852
0.950285937383862 0.42039510610306
0.184246040917914 0.950285937383862
0.586167805563554 0.184246040917914
0.94604292620901 0.586167805563554
0.199078261117725 0.94604292620901
0.621839817565464 0.199078261117725
0.917104729536896 0.621839817565464
0.296492213932015 0.917104729536896
0.813479865937859 0.296492213932015
0.59174845724117 0.813479865937859
0.942170660316074 0.59174845724117
0.212491917907023 0.942170660316074
0.652622500651751 0.212491917907023
0.884154851949745 0.652622500651751
0.399457693921532 0.884154851949745
0.935575854284852 0.399457693921532
0.235067333139683 0.935575854284852
0.701261659918092 0.235067333139683
0.817025602563355 0.701261659918092
0.583029592545432 0.817025602563355
0.948113738370784 0.583029592545432
0.191856902185106 0.948113738370784
0.604686541949258 0.191856902185106
0.932258838945355 0.604686541949258
0.246293954999094 0.932258838945355
0.723969646646994 0.246293954999094
0.779366629785201 0.723969646646994
0.670621716033588 0.779366629785201
0.861464097069239 0.670621716033588
0.465440455466688 0.861464097069239
0.970341987738435 0.465440455466688
0.112235816824005 0.970341987738435
0.38859185915883 0.112235816824005
0.926594082001824 0.38859185915883
0.265268207883981 0.926594082001824
0.760113844503008 0.265268207883981
0.711129072801674 0.760113844503008
0.801155607009831 0.711129072801674
0.621290671429506 0.801155607009831
0.9176254347943 0.621290671429506
0.2947970852302 0.9176254347943
0.810777878702915 0.2947970852302
0.598326729424774 0.810777878702915
0.937294231694446 0.598326729424774
0.229217644214379 0.937294231694446
0.689039971601238 0.229217644214379
};
\addplot [semithick, steelblue31119180, mark=*, mark size=3, mark options={solid}, only marks]
table {%
0.368330771049688 0.897966504908837
0.918133537704314 0.368330771049688
0.295558417320631 0.918133537704314
0.809465280754736 0.295558417320631
0.59023008351814 0.809465280754736
0.931459699162453 0.59023008351814
0.231345126069112 0.931459699162453
0.672944916027848 0.231345126069112
0.844248764746705 0.672944916027848
0.480769223014381 0.844248764746705
0.937353482509415 0.480769223014381
0.195798547476823 0.937353482509415
0.57250615734559 0.195798547476823
0.926520907909534 0.57250615734559
0.229458379402281 0.926520907909534
0.64990392056038 0.229458379402281
0.86281083697337 0.64990392056038
0.424999926431383 0.86281083697337
0.914582062930172 0.424999926431383
0.266108118015229 0.914582062930172
0.726131326192984 0.266108118015229
0.739983641175623 0.726131326192984
0.706930227524104 0.739983641175623
0.755018997426433 0.706930227524104
0.66084159117678 0.755018997426433
0.80776177657274 0.66084159117678
0.523595577744072 0.80776177657274
0.895787247231855 0.523595577744072
0.278330035890999 0.895787247231855
0.70281829591552 0.278330035890999
0.745752673732866 0.70281829591552
0.659295169889397 0.745752673732866
};
\addlegendentry{classical}
\addplot [semithick, darkorange25512714, mark=square*, mark size=3, mark options={solid}, only marks]
table {%
0.491849795938633 0.85070360927882
0.967859190865356 0.491849795938633
0.113581380978474 0.967859190865356
0.357563583227189 0.113581380978474
0.884406841068515 0.357563583227189
0.387934250897835 0.884406841068515
0.893505596217641 0.387934250897835
0.407356878231103 0.893505596217641
0.97372037923968 0.407356878231103
0.148448703282967 0.97372037923968
0.42801585048125 0.148448703282967
0.851660063078141 0.42801585048125
0.396094135115538 0.851660063078141
0.905405285431639 0.396094135115538
0.345024295839106 0.905405285431639
0.924027509489122 0.345024295839106
0.216987350658235 0.924027509489122
0.611743173725331 0.216987350658235
0.727729657421753 0.611743173725331
0.482711696711145 0.727729657421753
0.79291333171595 0.482711696711145
0.429404082078727 0.79291333171595
0.826136736140733 0.429404082078727
0.35584976249855 0.826136736140733
0.677572670556818 0.35584976249855
0.5335182162737 0.677572670556818
0.655284104541455 0.5335182162737
0.593265757648044 0.655284104541455
0.503783242906937 0.593265757648044
0.722096453266802 0.503783242906937
0.509275739574639 0.722096453266802
0.566919279753287 0.509275739574639
};
\addlegendentry{quantum $\langle Z_1, Z_2, Z_3\rangle$}
\addplot [semithick, forestgreen4416044, mark=triangle*, mark size=3, mark options={solid}, only marks]
table {%
0.510900530212432 0.829258099060368
0.961528769765003 0.510900530212432
0.109186231062077 0.961528769765003
0.363434279689993 0.109186231062077
0.891293709062423 0.363434279689993
0.386944774330152 0.891293709062423
0.865115992785717 0.386944774330152
0.43977225704009 0.865115992785717
0.997209000396607 0.43977225704009
0.136506904232594 0.997209000396607
0.379168272930296 0.136506904232594
0.889786273211007 0.379168272930296
0.37715592163589 0.889786273211007
0.912020787025167 0.37715592163589
0.37430164020688 0.912020787025167
0.9692997964073 0.37430164020688
0.177888732309289 0.9692997964073
0.486792928648562 0.177888732309289
0.809929846316222 0.486792928648562
0.421386668911068 0.809929846316222
0.822802776726386 0.421386668911068
0.450184290459215 0.822802776726386
0.870530118641021 0.450184290459215
0.276433015689181 0.870530118641021
0.612068637559899 0.276433015689181
0.700037123474407 0.612068637559899
0.537091575878353 0.700037123474407
0.659659055059914 0.537091575878353
0.562145096746528 0.659659055059914
0.626955482439796 0.562145096746528
0.485148377646148 0.626955482439796
0.50642065413148 0.485148377646148
};
\addlegendentry{quantum $\langle Z_1Z_2, Z_2Z_3, Z_3Z_4\rangle$}
\end{axis}

\end{tikzpicture}

%% file: figures/convergence_ep4_4_num_qubits4_num_meas3_degree3_num_reservoirs20_timeplex10_methodquantum_stab_noiseTrue.pickle.tex
\begin{tikzpicture}

\definecolor{crimson2143940}{RGB}{214,39,40}
\definecolor{darkgray176}{RGB}{176,176,176}
\definecolor{darkorange25512714}{RGB}{255,127,14}
\definecolor{forestgreen4416044}{RGB}{44,160,44}
\definecolor{gray}{RGB}{128,128,128}
\definecolor{lightgray204}{RGB}{204,204,204}
\definecolor{steelblue31119180}{RGB}{31,119,180}

\begin{axis}[
width=1.1\textwidth,
height=.7\textwidth,
legend cell align={left},
legend style={
  fill opacity=0,
  draw opacity=0,
  text opacity=0,
  at={(0.03,0.03)},
  anchor=south west,
  draw=lightgray204
},
tick align=outside,
tick pos=left,
x grid style={darkgray176},
xmin=166, xmax=200,
xtick style={color=black},
y grid style={darkgray176},
ymin=0, ymax=1.1,
ytick style={color=black},
xlabel=$n$,
ylabel=$x_n$
]
\path [draw=gray, fill=gray, opacity=0.5]
(axis cs:0,2)
--(axis cs:0,-2)
--(axis cs:1,-2)
--(axis cs:2,-2)
--(axis cs:3,-2)
--(axis cs:4,-2)
--(axis cs:5,-2)
--(axis cs:6,-2)
--(axis cs:7,-2)
--(axis cs:8,-2)
--(axis cs:9,-2)
--(axis cs:10,-2)
--(axis cs:11,-2)
--(axis cs:12,-2)
--(axis cs:13,-2)
--(axis cs:14,-2)
--(axis cs:15,-2)
--(axis cs:16,-2)
--(axis cs:17,-2)
--(axis cs:18,-2)
--(axis cs:19,-2)
--(axis cs:20,-2)
--(axis cs:21,-2)
--(axis cs:22,-2)
--(axis cs:23,-2)
--(axis cs:24,-2)
--(axis cs:25,-2)
--(axis cs:26,-2)
--(axis cs:27,-2)
--(axis cs:28,-2)
--(axis cs:29,-2)
--(axis cs:30,-2)
--(axis cs:31,-2)
--(axis cs:32,-2)
--(axis cs:33,-2)
--(axis cs:34,-2)
--(axis cs:35,-2)
--(axis cs:36,-2)
--(axis cs:37,-2)
--(axis cs:38,-2)
--(axis cs:39,-2)
--(axis cs:40,-2)
--(axis cs:41,-2)
--(axis cs:42,-2)
--(axis cs:43,-2)
--(axis cs:44,-2)
--(axis cs:45,-2)
--(axis cs:46,-2)
--(axis cs:47,-2)
--(axis cs:48,-2)
--(axis cs:49,-2)
--(axis cs:50,-2)
--(axis cs:51,-2)
--(axis cs:52,-2)
--(axis cs:53,-2)
--(axis cs:54,-2)
--(axis cs:55,-2)
--(axis cs:56,-2)
--(axis cs:57,-2)
--(axis cs:58,-2)
--(axis cs:59,-2)
--(axis cs:60,-2)
--(axis cs:61,-2)
--(axis cs:62,-2)
--(axis cs:63,-2)
--(axis cs:64,-2)
--(axis cs:65,-2)
--(axis cs:66,-2)
--(axis cs:67,-2)
--(axis cs:68,-2)
--(axis cs:69,-2)
--(axis cs:70,-2)
--(axis cs:71,-2)
--(axis cs:72,-2)
--(axis cs:73,-2)
--(axis cs:74,-2)
--(axis cs:75,-2)
--(axis cs:76,-2)
--(axis cs:77,-2)
--(axis cs:78,-2)
--(axis cs:79,-2)
--(axis cs:80,-2)
--(axis cs:81,-2)
--(axis cs:82,-2)
--(axis cs:83,-2)
--(axis cs:84,-2)
--(axis cs:85,-2)
--(axis cs:86,-2)
--(axis cs:87,-2)
--(axis cs:88,-2)
--(axis cs:89,-2)
--(axis cs:90,-2)
--(axis cs:91,-2)
--(axis cs:92,-2)
--(axis cs:93,-2)
--(axis cs:94,-2)
--(axis cs:95,-2)
--(axis cs:96,-2)
--(axis cs:97,-2)
--(axis cs:98,-2)
--(axis cs:99,-2)
--(axis cs:100,-2)
--(axis cs:101,-2)
--(axis cs:102,-2)
--(axis cs:103,-2)
--(axis cs:104,-2)
--(axis cs:105,-2)
--(axis cs:106,-2)
--(axis cs:107,-2)
--(axis cs:108,-2)
--(axis cs:109,-2)
--(axis cs:110,-2)
--(axis cs:111,-2)
--(axis cs:112,-2)
--(axis cs:113,-2)
--(axis cs:114,-2)
--(axis cs:115,-2)
--(axis cs:116,-2)
--(axis cs:117,-2)
--(axis cs:118,-2)
--(axis cs:119,-2)
--(axis cs:120,-2)
--(axis cs:121,-2)
--(axis cs:122,-2)
--(axis cs:123,-2)
--(axis cs:124,-2)
--(axis cs:125,-2)
--(axis cs:126,-2)
--(axis cs:127,-2)
--(axis cs:128,-2)
--(axis cs:129,-2)
--(axis cs:130,-2)
--(axis cs:131,-2)
--(axis cs:132,-2)
--(axis cs:133,-2)
--(axis cs:134,-2)
--(axis cs:135,-2)
--(axis cs:136,-2)
--(axis cs:137,-2)
--(axis cs:138,-2)
--(axis cs:139,-2)
--(axis cs:140,-2)
--(axis cs:141,-2)
--(axis cs:142,-2)
--(axis cs:143,-2)
--(axis cs:144,-2)
--(axis cs:145,-2)
--(axis cs:146,-2)
--(axis cs:147,-2)
--(axis cs:148,-2)
--(axis cs:149,-2)
--(axis cs:150,-2)
--(axis cs:151,-2)
--(axis cs:152,-2)
--(axis cs:153,-2)
--(axis cs:154,-2)
--(axis cs:155,-2)
--(axis cs:156,-2)
--(axis cs:157,-2)
--(axis cs:158,-2)
--(axis cs:159,-2)
--(axis cs:160,-2)
--(axis cs:161,-2)
--(axis cs:162,-2)
--(axis cs:163,-2)
--(axis cs:164,-2)
--(axis cs:165,-2)
--(axis cs:166,-2)
--(axis cs:166,2)
--(axis cs:166,2)
--(axis cs:165,2)
--(axis cs:164,2)
--(axis cs:163,2)
--(axis cs:162,2)
--(axis cs:161,2)
--(axis cs:160,2)
--(axis cs:159,2)
--(axis cs:158,2)
--(axis cs:157,2)
--(axis cs:156,2)
--(axis cs:155,2)
--(axis cs:154,2)
--(axis cs:153,2)
--(axis cs:152,2)
--(axis cs:151,2)
--(axis cs:150,2)
--(axis cs:149,2)
--(axis cs:148,2)
--(axis cs:147,2)
--(axis cs:146,2)
--(axis cs:145,2)
--(axis cs:144,2)
--(axis cs:143,2)
--(axis cs:142,2)
--(axis cs:141,2)
--(axis cs:140,2)
--(axis cs:139,2)
--(axis cs:138,2)
--(axis cs:137,2)
--(axis cs:136,2)
--(axis cs:135,2)
--(axis cs:134,2)
--(axis cs:133,2)
--(axis cs:132,2)
--(axis cs:131,2)
--(axis cs:130,2)
--(axis cs:129,2)
--(axis cs:128,2)
--(axis cs:127,2)
--(axis cs:126,2)
--(axis cs:125,2)
--(axis cs:124,2)
--(axis cs:123,2)
--(axis cs:122,2)
--(axis cs:121,2)
--(axis cs:120,2)
--(axis cs:119,2)
--(axis cs:118,2)
--(axis cs:117,2)
--(axis cs:116,2)
--(axis cs:115,2)
--(axis cs:114,2)
--(axis cs:113,2)
--(axis cs:112,2)
--(axis cs:111,2)
--(axis cs:110,2)
--(axis cs:109,2)
--(axis cs:108,2)
--(axis cs:107,2)
--(axis cs:106,2)
--(axis cs:105,2)
--(axis cs:104,2)
--(axis cs:103,2)
--(axis cs:102,2)
--(axis cs:101,2)
--(axis cs:100,2)
--(axis cs:99,2)
--(axis cs:98,2)
--(axis cs:97,2)
--(axis cs:96,2)
--(axis cs:95,2)
--(axis cs:94,2)
--(axis cs:93,2)
--(axis cs:92,2)
--(axis cs:91,2)
--(axis cs:90,2)
--(axis cs:89,2)
--(axis cs:88,2)
--(axis cs:87,2)
--(axis cs:86,2)
--(axis cs:85,2)
--(axis cs:84,2)
--(axis cs:83,2)
--(axis cs:82,2)
--(axis cs:81,2)
--(axis cs:80,2)
--(axis cs:79,2)
--(axis cs:78,2)
--(axis cs:77,2)
--(axis cs:76,2)
--(axis cs:75,2)
--(axis cs:74,2)
--(axis cs:73,2)
--(axis cs:72,2)
--(axis cs:71,2)
--(axis cs:70,2)
--(axis cs:69,2)
--(axis cs:68,2)
--(axis cs:67,2)
--(axis cs:66,2)
--(axis cs:65,2)
--(axis cs:64,2)
--(axis cs:63,2)
--(axis cs:62,2)
--(axis cs:61,2)
--(axis cs:60,2)
--(axis cs:59,2)
--(axis cs:58,2)
--(axis cs:57,2)
--(axis cs:56,2)
--(axis cs:55,2)
--(axis cs:54,2)
--(axis cs:53,2)
--(axis cs:52,2)
--(axis cs:51,2)
--(axis cs:50,2)
--(axis cs:49,2)
--(axis cs:48,2)
--(axis cs:47,2)
--(axis cs:46,2)
--(axis cs:45,2)
--(axis cs:44,2)
--(axis cs:43,2)
--(axis cs:42,2)
--(axis cs:41,2)
--(axis cs:40,2)
--(axis cs:39,2)
--(axis cs:38,2)
--(axis cs:37,2)
--(axis cs:36,2)
--(axis cs:35,2)
--(axis cs:34,2)
--(axis cs:33,2)
--(axis cs:32,2)
--(axis cs:31,2)
--(axis cs:30,2)
--(axis cs:29,2)
--(axis cs:28,2)
--(axis cs:27,2)
--(axis cs:26,2)
--(axis cs:25,2)
--(axis cs:24,2)
--(axis cs:23,2)
--(axis cs:22,2)
--(axis cs:21,2)
--(axis cs:20,2)
--(axis cs:19,2)
--(axis cs:18,2)
--(axis cs:17,2)
--(axis cs:16,2)
--(axis cs:15,2)
--(axis cs:14,2)
--(axis cs:13,2)
--(axis cs:12,2)
--(axis cs:11,2)
--(axis cs:10,2)
--(axis cs:9,2)
--(axis cs:8,2)
--(axis cs:7,2)
--(axis cs:6,2)
--(axis cs:5,2)
--(axis cs:4,2)
--(axis cs:3,2)
--(axis cs:2,2)
--(axis cs:1,2)
--(axis cs:0,2)
--cycle;

\addplot [ultra thick, black]
table {%
0 0.5
1 0.975
2 0.0950625000000001
3 0.335499922265625
4 0.869464925259
5 0.442633109113109
6 0.962165255336889
7 0.141972779361614
8 0.475084386199614
9 0.972578927536905
10 0.104009713267468
11 0.363447601972601
12 0.90227842611257
13 0.343871064749135
14 0.879932646751981
15 0.412039617334933
16 0.944825587217519
17 0.2033077681307
18 0.631697506238813
19 0.907357490716863
20 0.327833511551757
21 0.859398930996064
22 0.471246392755656
23 0.971775597274708
24 0.10696836468276
25 0.372551921195438
26 0.911652250115202
27 0.314115457402856
28 0.840243053611457
29 0.52351519142969
30 0.972843439510898
31 0.103034418674866
32 0.360431476248473
33 0.899030445993496
34 0.354021342363904
35 0.891891902907578
36 0.376040872098363
37 0.915073124978476
38 0.30308577359035
39 0.823776671006208
40 0.566157802517338
41 0.9579302661477
42 0.157169498248997
43 0.516622263569707
44 0.973922431379895
45 0.0990503632363789
46 0.348033616238569
47 0.884934251005247
48 0.397119927371816
49 0.933721193558477
50 0.24135511240702
51 0.714101006275858
52 0.796226960535493
53 0.632773392622424
54 0.906247782224973
55 0.331354683805434
56 0.864079153569976
57 0.458040842749502
58 0.968133773579029
59 0.120318003135167
60 0.412782166901259
61 0.945332893399284
62 0.201546594820824
63 0.627609703254123
64 0.911491478178039
65 0.314631577208725
66 0.840990336544313
67 0.521529802495244
68 0.973192223657611
69 0.101747565932864
70 0.356440495162445
71 0.894623607426105
72 0.367661613001829
73 0.906697550174217
74 0.329928700460932
75 0.862195436985061
76 0.463376415166083
77 0.96976898083226
78 0.114336708126496
79 0.394928918675041
80 0.931944264689895
81 0.247354193585875
82 0.72606337635529
83 0.775691864496302
84 0.678576583817155
85 0.85063057447756
86 0.495526980941992
87 0.974921969191976
88 0.0953515803973814
89 0.336412660401102
90 0.870632811059524
91 0.439262145527849
92 0.960612560833067
93 0.147560668330936
94 0.490567418221388
95 0.97465300296386
96 0.0963476544318197
97 0.339552657277495
98 0.874600935831819
99 0.427729141608309
100 0.95462999980658
101 0.16891509677589
102 0.547492868742619
103 0.966203266932526
104 0.127352604215616
105 0.433422281818955
106 0.957712889023041
107 0.157945753766508
108 0.518695681271186
109 0.973636838857044
110 0.100105765022275
111 0.351329943243161
112 0.888799135473281
113 0.385457405795393
114 0.923831977040227
115 0.274429175428874
116 0.776559432098978
117 0.676708034016653
118 0.853219655784481
119 0.488419911593223
120 0.974477016054716
121 0.0969990888196709
122 0.341602035792516
123 0.877149331246205
124 0.420257689568106
125 0.950200539315235
126 0.184545950161058
127 0.586906095516819
128 0.945544589191884
129 0.200811074263158
130 0.625895348194095
131 0.913186409082042
132 0.309180266264562
133 0.832992533946832
134 0.542552292109105
135 0.967938279501419
136 0.121031689651281
137 0.414893777030605
138 0.946752030166346
139 0.196609231814768
140 0.616020762941902
141 0.922502812008878
142 0.278816357993905
143 0.784203406384786
144 0.659990852817209
145 0.875171415357807
146 0.426060995483839
147 0.953678792083521
148 0.172284659093187
149 0.556150355803892
150 0.962703836418076
151 0.14003012307958
152 0.4696445820686
153 0.971406339548619
154 0.108326645830151
155 0.376708736170027
156 0.915717130626511
157 0.300999142484275
158 0.820554768961224
159 0.574254096375244
160 0.953496683768834
161 0.172928955461613
162 0.557795674115817
163 0.961972674208644
164 0.142666868309607
165 0.47702082868571
166 0.972940634974263
167 0.102675907581549
168 0.359320905777032
169 0.897816630549561
170 0.357793520986957
171 0.896131537574181
172 0.363011239262551
173 0.90181290978331
174 0.345330903572732
175 0.881702135380453
176 0.406783571399409
177 0.941111720011909
178 0.216139756825728
179 0.660751113145668
180 0.874220410527471
181 0.428840428944138
182 0.955251630243868
183 0.16670921732517
184 0.541777291317597
185 0.968193165927642
186 0.120101121576741
187 0.412139684473726
188 0.944894203326932
189 0.203069676599769
190 0.631146293877194
191 0.907922533448879
192 0.326036906148129
193 0.856973683712192
194 0.478022177634096
195 0.973116203763806
196 0.102028125170887
197 0.357311708695503
198 0.895596200945952
199 0.364664218608807
};
\addplot [semithick, steelblue31119180, mark=*, mark size=3, mark options={solid}]
table {%
0 0.5
1 0.975
2 0.0950625000000001
3 0.335499922265625
4 0.869464925259
5 0.442633109113109
6 0.962165255336889
7 0.141972779361614
8 0.475084386199614
9 0.972578927536905
10 0.104009713267468
11 0.363447601972601
12 0.90227842611257
13 0.343871064749135
14 0.879932646751981
15 0.412039617334933
16 0.944825587217519
17 0.2033077681307
18 0.631697506238813
19 0.907357490716863
20 0.327833511551757
21 0.859398930996064
22 0.471246392755656
23 0.971775597274708
24 0.10696836468276
25 0.372551921195438
26 0.911652250115202
27 0.314115457402856
28 0.840243053611457
29 0.52351519142969
30 0.972843439510898
31 0.103034418674866
32 0.360431476248473
33 0.899030445993496
34 0.354021342363904
35 0.891891902907578
36 0.376040872098363
37 0.915073124978476
38 0.30308577359035
39 0.823776671006208
40 0.566157802517338
41 0.9579302661477
42 0.157169498248997
43 0.516622263569707
44 0.973922431379895
45 0.0990503632363789
46 0.348033616238569
47 0.884934251005247
48 0.397119927371816
49 0.933721193558477
50 0.24135511240702
51 0.714101006275858
52 0.796226960535493
53 0.632773392622424
54 0.906247782224973
55 0.331354683805434
56 0.864079153569976
57 0.458040842749502
58 0.968133773579029
59 0.120318003135167
60 0.412782166901259
61 0.945332893399284
62 0.201546594820824
63 0.627609703254123
64 0.911491478178039
65 0.314631577208725
66 0.840990336544313
67 0.521529802495244
68 0.973192223657611
69 0.101747565932864
70 0.356440495162445
71 0.894623607426105
72 0.367661613001829
73 0.906697550174217
74 0.329928700460932
75 0.862195436985061
76 0.463376415166083
77 0.96976898083226
78 0.114336708126496
79 0.394928918675041
80 0.931944264689895
81 0.247354193585875
82 0.72606337635529
83 0.775691864496302
84 0.678576583817155
85 0.85063057447756
86 0.495526980941992
87 0.974921969191976
88 0.0953515803973814
89 0.336412660401102
90 0.870632811059524
91 0.439262145527849
92 0.960612560833067
93 0.147560668330936
94 0.490567418221388
95 0.97465300296386
96 0.0963476544318197
97 0.339552657277495
98 0.874600935831819
99 0.427729141608309
100 0.95462999980658
101 0.16891509677589
102 0.547492868742619
103 0.966203266932526
104 0.127352604215616
105 0.433422281818955
106 0.957712889023041
107 0.157945753766508
108 0.518695681271186
109 0.973636838857044
110 0.100105765022275
111 0.351329943243161
112 0.888799135473281
113 0.385457405795393
114 0.923831977040227
115 0.274429175428874
116 0.776559432098978
117 0.676708034016653
118 0.853219655784481
119 0.488419911593223
120 0.974477016054716
121 0.0969990888196709
122 0.341602035792516
123 0.877149331246205
124 0.420257689568106
125 0.950200539315235
126 0.184545950161058
127 0.586906095516819
128 0.945544589191884
129 0.200811074263158
130 0.625895348194095
131 0.913186409082042
132 0.309180266264562
133 0.832992533946832
134 0.542552292109105
135 0.967938279501419
136 0.121031689651281
137 0.414893777030605
138 0.946752030166346
139 0.196609231814768
140 0.616020762941902
141 0.922502812008878
142 0.278816357993905
143 0.784203406384786
144 0.659990852817209
145 0.875171415357807
146 0.426060995483839
147 0.953678792083521
148 0.172284659093187
149 0.556150355803892
150 0.962703836418076
151 0.14003012307958
152 0.4696445820686
153 0.971406339548619
154 0.108326645830151
155 0.376708736170027
156 0.915717130626511
157 0.300999142484275
158 0.820554768961224
159 0.574254096375244
160 0.953496683768834
161 0.172928955461613
162 0.557795674115817
163 0.961972674208644
164 0.142666868309607
165 0.47702082868571
166 0.972940634974263
167 0.104295456887201
168 0.362725916435377
169 0.900801002907254
170 0.348806854897003
171 0.884944343487437
172 0.397395194974912
173 0.933718818797754
174 0.243169005699116
175 0.715837885320507
176 0.79348512799271
177 0.637602084401319
178 0.90194646168511
179 0.344842734314163
180 0.880922890047422
181 0.4113253839128
182 0.944378604486243
183 0.205667985868587
184 0.63492767909954
185 0.907167986582991
186 0.327478708054937
187 0.859061772293909
188 0.473966079316195
189 0.97228463319275
190 0.106838774316053
191 0.372371587498467
192 0.911227677570523
193 0.314493556914741
194 0.841996457355791
195 0.516578791647294
196 0.974698834499304
197 0.0987994363977904
198 0.348440511085709
199 0.883918150670901
};
\addlegendentry{classical}
\addplot [semithick, darkorange25512714, mark=square*, mark size=3, mark options={solid}]
table {%
0 0.5
1 0.975
2 0.0950625000000001
3 0.335499922265625
4 0.869464925259
5 0.442633109113109
6 0.962165255336889
7 0.141972779361614
8 0.475084386199614
9 0.972578927536905
10 0.104009713267468
11 0.363447601972601
12 0.90227842611257
13 0.343871064749135
14 0.879932646751981
15 0.412039617334933
16 0.944825587217519
17 0.2033077681307
18 0.631697506238813
19 0.907357490716863
20 0.327833511551757
21 0.859398930996064
22 0.471246392755656
23 0.971775597274708
24 0.10696836468276
25 0.372551921195438
26 0.911652250115202
27 0.314115457402856
28 0.840243053611457
29 0.52351519142969
30 0.972843439510898
31 0.103034418674866
32 0.360431476248473
33 0.899030445993496
34 0.354021342363904
35 0.891891902907578
36 0.376040872098363
37 0.915073124978476
38 0.30308577359035
39 0.823776671006208
40 0.566157802517338
41 0.9579302661477
42 0.157169498248997
43 0.516622263569707
44 0.973922431379895
45 0.0990503632363789
46 0.348033616238569
47 0.884934251005247
48 0.397119927371816
49 0.933721193558477
50 0.24135511240702
51 0.714101006275858
52 0.796226960535493
53 0.632773392622424
54 0.906247782224973
55 0.331354683805434
56 0.864079153569976
57 0.458040842749502
58 0.968133773579029
59 0.120318003135167
60 0.412782166901259
61 0.945332893399284
62 0.201546594820824
63 0.627609703254123
64 0.911491478178039
65 0.314631577208725
66 0.840990336544313
67 0.521529802495244
68 0.973192223657611
69 0.101747565932864
70 0.356440495162445
71 0.894623607426105
72 0.367661613001829
73 0.906697550174217
74 0.329928700460932
75 0.862195436985061
76 0.463376415166083
77 0.96976898083226
78 0.114336708126496
79 0.394928918675041
80 0.931944264689895
81 0.247354193585875
82 0.72606337635529
83 0.775691864496302
84 0.678576583817155
85 0.85063057447756
86 0.495526980941992
87 0.974921969191976
88 0.0953515803973814
89 0.336412660401102
90 0.870632811059524
91 0.439262145527849
92 0.960612560833067
93 0.147560668330936
94 0.490567418221388
95 0.97465300296386
96 0.0963476544318197
97 0.339552657277495
98 0.874600935831819
99 0.427729141608309
100 0.95462999980658
101 0.16891509677589
102 0.547492868742619
103 0.966203266932526
104 0.127352604215616
105 0.433422281818955
106 0.957712889023041
107 0.157945753766508
108 0.518695681271186
109 0.973636838857044
110 0.100105765022275
111 0.351329943243161
112 0.888799135473281
113 0.385457405795393
114 0.923831977040227
115 0.274429175428874
116 0.776559432098978
117 0.676708034016653
118 0.853219655784481
119 0.488419911593223
120 0.974477016054716
121 0.0969990888196709
122 0.341602035792516
123 0.877149331246205
124 0.420257689568106
125 0.950200539315235
126 0.184545950161058
127 0.586906095516819
128 0.945544589191884
129 0.200811074263158
130 0.625895348194095
131 0.913186409082042
132 0.309180266264562
133 0.832992533946832
134 0.542552292109105
135 0.967938279501419
136 0.121031689651281
137 0.414893777030605
138 0.946752030166346
139 0.196609231814768
140 0.616020762941902
141 0.922502812008878
142 0.278816357993905
143 0.784203406384786
144 0.659990852817209
145 0.875171415357807
146 0.426060995483839
147 0.953678792083521
148 0.172284659093187
149 0.556150355803892
150 0.962703836418076
151 0.14003012307958
152 0.4696445820686
153 0.971406339548619
154 0.108326645830151
155 0.376708736170027
156 0.915717130626511
157 0.300999142484275
158 0.820554768961224
159 0.574254096375244
160 0.953496683768834
161 0.172928955461613
162 0.557795674115817
163 0.961972674208644
164 0.142666868309607
165 0.47702082868571
166 0.972940634974263
167 0.112371991969281
168 0.380355945065228
169 0.914751270716367
170 0.298587077356125
171 0.819844950474381
172 0.577764171506969
173 0.950598279885858
174 0.19062165065039
175 0.59561937484085
176 0.936569353082301
177 0.221274335591213
178 0.686667725522728
179 0.836463137549119
180 0.52910384620363
181 0.97826495594534
182 0.126066433017273
183 0.39187625353578
184 0.947499273462323
185 0.228765009263566
186 0.70680796581077
187 0.830608055952972
188 0.579752563658229
189 0.946308805431367
190 0.180487485637981
191 0.5920806486143
192 0.930685752137337
193 0.219999457990021
194 0.677142890993029
195 0.841985534323281
196 0.519459410789605
197 0.9766316763145
198 0.128836980075621
199 0.395799713251612
};
\addlegendentry{quantum $\langle Z_1, Z_2, Z_3\rangle$}
\addplot [semithick, forestgreen4416044, mark=triangle*, mark size=3, mark options={solid}]
table {%
0 0.5
1 0.975
2 0.0950625000000001
3 0.335499922265625
4 0.869464925259
5 0.442633109113109
6 0.962165255336889
7 0.141972779361614
8 0.475084386199614
9 0.972578927536905
10 0.104009713267468
11 0.363447601972601
12 0.90227842611257
13 0.343871064749135
14 0.879932646751981
15 0.412039617334933
16 0.944825587217519
17 0.2033077681307
18 0.631697506238813
19 0.907357490716863
20 0.327833511551757
21 0.859398930996064
22 0.471246392755656
23 0.971775597274708
24 0.10696836468276
25 0.372551921195438
26 0.911652250115202
27 0.314115457402856
28 0.840243053611457
29 0.52351519142969
30 0.972843439510898
31 0.103034418674866
32 0.360431476248473
33 0.899030445993496
34 0.354021342363904
35 0.891891902907578
36 0.376040872098363
37 0.915073124978476
38 0.30308577359035
39 0.823776671006208
40 0.566157802517338
41 0.9579302661477
42 0.157169498248997
43 0.516622263569707
44 0.973922431379895
45 0.0990503632363789
46 0.348033616238569
47 0.884934251005247
48 0.397119927371816
49 0.933721193558477
50 0.24135511240702
51 0.714101006275858
52 0.796226960535493
53 0.632773392622424
54 0.906247782224973
55 0.331354683805434
56 0.864079153569976
57 0.458040842749502
58 0.968133773579029
59 0.120318003135167
60 0.412782166901259
61 0.945332893399284
62 0.201546594820824
63 0.627609703254123
64 0.911491478178039
65 0.314631577208725
66 0.840990336544313
67 0.521529802495244
68 0.973192223657611
69 0.101747565932864
70 0.356440495162445
71 0.894623607426105
72 0.367661613001829
73 0.906697550174217
74 0.329928700460932
75 0.862195436985061
76 0.463376415166083
77 0.96976898083226
78 0.114336708126496
79 0.394928918675041
80 0.931944264689895
81 0.247354193585875
82 0.72606337635529
83 0.775691864496302
84 0.678576583817155
85 0.85063057447756
86 0.495526980941992
87 0.974921969191976
88 0.0953515803973814
89 0.336412660401102
90 0.870632811059524
91 0.439262145527849
92 0.960612560833067
93 0.147560668330936
94 0.490567418221388
95 0.97465300296386
96 0.0963476544318197
97 0.339552657277495
98 0.874600935831819
99 0.427729141608309
100 0.95462999980658
101 0.16891509677589
102 0.547492868742619
103 0.966203266932526
104 0.127352604215616
105 0.433422281818955
106 0.957712889023041
107 0.157945753766508
108 0.518695681271186
109 0.973636838857044
110 0.100105765022275
111 0.351329943243161
112 0.888799135473281
113 0.385457405795393
114 0.923831977040227
115 0.274429175428874
116 0.776559432098978
117 0.676708034016653
118 0.853219655784481
119 0.488419911593223
120 0.974477016054716
121 0.0969990888196709
122 0.341602035792516
123 0.877149331246205
124 0.420257689568106
125 0.950200539315235
126 0.184545950161058
127 0.586906095516819
128 0.945544589191884
129 0.200811074263158
130 0.625895348194095
131 0.913186409082042
132 0.309180266264562
133 0.832992533946832
134 0.542552292109105
135 0.967938279501419
136 0.121031689651281
137 0.414893777030605
138 0.946752030166346
139 0.196609231814768
140 0.616020762941902
141 0.922502812008878
142 0.278816357993905
143 0.784203406384786
144 0.659990852817209
145 0.875171415357807
146 0.426060995483839
147 0.953678792083521
148 0.172284659093187
149 0.556150355803892
150 0.962703836418076
151 0.14003012307958
152 0.4696445820686
153 0.971406339548619
154 0.108326645830151
155 0.376708736170027
156 0.915717130626511
157 0.300999142484275
158 0.820554768961224
159 0.574254096375244
160 0.953496683768834
161 0.172928955461613
162 0.557795674115817
163 0.961972674208644
164 0.142666868309607
165 0.47702082868571
166 0.972940634974263
167 0.106679971114714
168 0.36923063193854
169 0.914505395723844
170 0.320817262551907
171 0.837890180153985
172 0.520784490010515
173 0.978787661639612
174 0.104026751793486
175 0.363253739604343
176 0.921051139726275
177 0.320214018704041
178 0.842693189164104
179 0.520527255608948
180 0.974240210640877
181 0.0983758632565404
182 0.360079558829734
183 0.90868160083447
184 0.347508641767356
185 0.879755846261265
186 0.416356070538921
187 0.93943730383201
188 0.198781149277007
189 0.618693848648849
190 0.909354610419203
191 0.29014636148072
192 0.805178137870637
193 0.610026106071778
194 0.91814603085141
195 0.267428745496928
196 0.780593873568116
197 0.695106099103756
198 0.823811298432036
199 0.546528501634934
};
\addlegendentry{quantum $\langle Z_1Z_2, Z_2Z_3, Z_3Z_4\rangle$}
\end{axis}

\end{tikzpicture}

%% file: figures/convergence2_ep4_4_num_qubits4_num_meas3_degree3_num_reservoirs20_timeplex10_methodquantum_stab_noiseTrue.pickle.tex
\begin{tikzpicture}

\definecolor{darkgray176}{RGB}{176,176,176}
\definecolor{darkorange25512714}{RGB}{255,127,14}
\definecolor{forestgreen4416044}{RGB}{44,160,44}
\definecolor{gray}{RGB}{128,128,128}
\definecolor{lightgray204}{RGB}{204,204,204}
\definecolor{steelblue31119180}{RGB}{31,119,180}

\begin{axis}[
width=.8\textwidth,
height=.8\textwidth,
legend cell align={left},
legend style={
  fill opacity=0,
  text opacity=1,
  at={(0.02,0.55)},
  anchor=west,
  draw=none
},
tick align=outside,
tick pos=left,
x grid style={darkgray176},
xmin=0.0508762419180195, xmax=1.02297391972159,
xtick style={color=black},
y grid style={darkgray176},
ymin=0.0508762419180195, ymax=1.02297391972159,
ytick style={color=black},
xlabel=$x_{n+1}$,
ylabel=$x_n$
]

\addplot [ultra thick, black] 
table {%
-10 10
10 -10
};
\addlegendentry{exact}

\addplot [only marks, black, opacity=0.25, mark=*, mark size=3, mark options={solid}, forget plot]
table {%
0.975 0.5
0.0950625000000001 0.975
0.335499922265625 0.0950625000000001
0.869464925259 0.335499922265625
0.442633109113109 0.869464925259
0.962165255336889 0.442633109113109
0.141972779361614 0.962165255336889
0.475084386199614 0.141972779361614
0.972578927536905 0.475084386199614
0.104009713267468 0.972578927536905
0.363447601972601 0.104009713267468
0.90227842611257 0.363447601972601
0.343871064749135 0.90227842611257
0.879932646751981 0.343871064749135
0.412039617334933 0.879932646751981
0.944825587217519 0.412039617334933
0.2033077681307 0.944825587217519
0.631697506238813 0.2033077681307
0.907357490716863 0.631697506238813
0.327833511551757 0.907357490716863
0.859398930996064 0.327833511551757
0.471246392755656 0.859398930996064
0.971775597274708 0.471246392755656
0.10696836468276 0.971775597274708
0.372551921195438 0.10696836468276
0.911652250115202 0.372551921195438
0.314115457402856 0.911652250115202
0.840243053611457 0.314115457402856
0.52351519142969 0.840243053611457
0.972843439510898 0.52351519142969
0.103034418674866 0.972843439510898
0.360431476248473 0.103034418674866
0.899030445993496 0.360431476248473
0.354021342363904 0.899030445993496
0.891891902907578 0.354021342363904
0.376040872098363 0.891891902907578
0.915073124978476 0.376040872098363
0.30308577359035 0.915073124978476
0.823776671006208 0.30308577359035
0.566157802517338 0.823776671006208
0.9579302661477 0.566157802517338
0.157169498248997 0.9579302661477
0.516622263569707 0.157169498248997
0.973922431379895 0.516622263569707
0.0990503632363789 0.973922431379895
0.348033616238569 0.0990503632363789
0.884934251005247 0.348033616238569
0.397119927371816 0.884934251005247
0.933721193558477 0.397119927371816
0.24135511240702 0.933721193558477
0.714101006275858 0.24135511240702
0.796226960535493 0.714101006275858
0.632773392622424 0.796226960535493
0.906247782224973 0.632773392622424
0.331354683805434 0.906247782224973
0.864079153569976 0.331354683805434
0.458040842749502 0.864079153569976
0.968133773579029 0.458040842749502
0.120318003135167 0.968133773579029
0.412782166901259 0.120318003135167
0.945332893399284 0.412782166901259
0.201546594820824 0.945332893399284
0.627609703254123 0.201546594820824
0.911491478178039 0.627609703254123
0.314631577208725 0.911491478178039
0.840990336544313 0.314631577208725
0.521529802495244 0.840990336544313
0.973192223657611 0.521529802495244
0.101747565932864 0.973192223657611
0.356440495162445 0.101747565932864
0.894623607426105 0.356440495162445
0.367661613001829 0.894623607426105
0.906697550174217 0.367661613001829
0.329928700460932 0.906697550174217
0.862195436985061 0.329928700460932
0.463376415166083 0.862195436985061
0.96976898083226 0.463376415166083
0.114336708126496 0.96976898083226
0.394928918675041 0.114336708126496
0.931944264689895 0.394928918675041
0.247354193585875 0.931944264689895
0.72606337635529 0.247354193585875
0.775691864496302 0.72606337635529
0.678576583817155 0.775691864496302
0.85063057447756 0.678576583817155
0.495526980941992 0.85063057447756
0.974921969191976 0.495526980941992
0.0953515803973814 0.974921969191976
0.336412660401102 0.0953515803973814
0.870632811059524 0.336412660401102
0.439262145527849 0.870632811059524
0.960612560833067 0.439262145527849
0.147560668330936 0.960612560833067
0.490567418221388 0.147560668330936
0.97465300296386 0.490567418221388
0.0963476544318197 0.97465300296386
0.339552657277495 0.0963476544318197
0.874600935831819 0.339552657277495
0.427729141608309 0.874600935831819
0.95462999980658 0.427729141608309
0.16891509677589 0.95462999980658
0.547492868742619 0.16891509677589
0.966203266932526 0.547492868742619
0.127352604215616 0.966203266932526
0.433422281818955 0.127352604215616
0.957712889023041 0.433422281818955
0.157945753766508 0.957712889023041
0.518695681271186 0.157945753766508
0.973636838857044 0.518695681271186
0.100105765022275 0.973636838857044
0.351329943243161 0.100105765022275
0.888799135473281 0.351329943243161
0.385457405795393 0.888799135473281
0.923831977040227 0.385457405795393
0.274429175428874 0.923831977040227
0.776559432098978 0.274429175428874
0.676708034016653 0.776559432098978
0.853219655784481 0.676708034016653
0.488419911593223 0.853219655784481
0.974477016054716 0.488419911593223
0.0969990888196709 0.974477016054716
0.341602035792516 0.0969990888196709
0.877149331246205 0.341602035792516
0.420257689568106 0.877149331246205
0.950200539315235 0.420257689568106
0.184545950161058 0.950200539315235
0.586906095516819 0.184545950161058
0.945544589191884 0.586906095516819
0.200811074263158 0.945544589191884
0.625895348194095 0.200811074263158
0.913186409082042 0.625895348194095
0.309180266264562 0.913186409082042
0.832992533946832 0.309180266264562
0.542552292109105 0.832992533946832
0.967938279501419 0.542552292109105
0.121031689651281 0.967938279501419
0.414893777030605 0.121031689651281
0.946752030166346 0.414893777030605
0.196609231814768 0.946752030166346
0.616020762941902 0.196609231814768
0.922502812008878 0.616020762941902
0.278816357993905 0.922502812008878
0.784203406384786 0.278816357993905
0.659990852817209 0.784203406384786
0.875171415357807 0.659990852817209
0.426060995483839 0.875171415357807
0.953678792083521 0.426060995483839
0.172284659093187 0.953678792083521
0.556150355803892 0.172284659093187
0.962703836418076 0.556150355803892
0.14003012307958 0.962703836418076
0.4696445820686 0.14003012307958
0.971406339548619 0.4696445820686
0.108326645830151 0.971406339548619
0.376708736170027 0.108326645830151
0.915717130626511 0.376708736170027
0.300999142484275 0.915717130626511
0.820554768961224 0.300999142484275
0.574254096375244 0.820554768961224
0.953496683768834 0.574254096375244
0.172928955461613 0.953496683768834
0.557795674115817 0.172928955461613
0.961972674208644 0.557795674115817
0.142666868309607 0.961972674208644
0.47702082868571 0.142666868309607
0.972940634974263 0.47702082868571
0.102675907581549 0.972940634974263
0.359320905777032 0.102675907581549
0.897816630549561 0.359320905777032
0.357793520986957 0.897816630549561
0.896131537574181 0.357793520986957
0.363011239262551 0.896131537574181
0.90181290978331 0.363011239262551
0.345330903572732 0.90181290978331
0.881702135380453 0.345330903572732
0.406783571399409 0.881702135380453
0.941111720011909 0.406783571399409
0.216139756825728 0.941111720011909
0.660751113145668 0.216139756825728
0.874220410527471 0.660751113145668
0.428840428944138 0.874220410527471
0.955251630243868 0.428840428944138
0.16670921732517 0.955251630243868
0.541777291317597 0.16670921732517
0.968193165927642 0.541777291317597
0.120101121576741 0.968193165927642
0.412139684473726 0.120101121576741
0.944894203326932 0.412139684473726
0.203069676599769 0.944894203326932
0.631146293877194 0.203069676599769
0.907922533448879 0.631146293877194
0.326036906148129 0.907922533448879
0.856973683712192 0.326036906148129
0.478022177634096 0.856973683712192
0.973116203763806 0.478022177634096
0.102028125170887 0.973116203763806
0.357311708695503 0.102028125170887
0.895596200945952 0.357311708695503
0.364664218608807 0.895596200945952
0.903568482473417 0.364664218608807
0.339816671821003 0.903568482473417
0.874931075356649 0.339816671821003
0.426764086054436 0.874931075356649
0.954082353543376 0.426764086054436
0.170855943181984 0.954082353543376
0.552490340459377 0.170855943181984
0.964254580217989 0.552490340459377
0.134423970511815 0.964254580217989
0.453781249988249 0.134423970511815
0.96666892587467 0.453781249988249
0.12565844312963 0.96666892587467
0.428486755319482 0.12565844312963
0.955054837757531 0.428486755319482
0.167407869070521 0.955054837757531
0.543591650330775 0.167407869070521
0.967589095283614 0.543591650330775
0.122305708090221 0.967589095283614
0.418653385249205 0.122305708090221
0.949192640247485 0.418653385249205
0.188081290595226 0.949192640247485
0.595556203020715 0.188081290595226
0.939389147050629 0.595556203020715
0.222054992071069 0.939389147050629
0.673711633012808 0.222054992071069
0.857314647368492 0.673711633012808
0.477072346826128 0.857314647368492
0.972949858607761 0.477072346826128
0.102641881847307 0.972949858607761
0.359215451158787 0.102641881847307
0.897700872149547 0.359215451158787
0.358152663536811 0.897700872149547
0.896529399239365 0.358152663536811
0.361781298601585 0.896529399239365
0.90049280327657 0.361781298601585
0.349461506642331 0.90049280327657
0.886618831868652 0.349461506642331
0.392050927493625 0.886618831868652
0.929553291205552 0.392050927493625
0.255387483056464 0.929553291205552
0.741642394562739 0.255387483056464
0.747274917284944 0.741642394562739
0.736534949598723 0.747274917284944
0.756799748711485 0.736534949598723
0.717810167340701 0.756799748711485
0.789979050911761 0.717810167340701
0.647057385126026 0.789979050911761
0.890659089371594 0.647057385126026
0.379803355976293 0.890659089371594
0.91865579038521 0.379803355976293
0.291436583790049 0.91865579038521
0.805355075533455 0.291436583790049
0.611357283599236 0.805355075533455
0.926638266018657 0.611357283599236
0.265121180877515 0.926638266018657
0.759844567277746 0.265121180877515
0.71167512333934 0.759844567277746
0.800255204421174 0.71167512333934
0.623402567650197 0.800255204421174
0.915610044559621 0.623402567650197
0.301346334358485 0.915610044559621
0.821093212396038 0.301346334358485
0.572906680917453 0.821093212396038
0.954270001922643 0.572906680917453
0.170191184877488 0.954270001922643
0.550781967323191 0.170191184877488
0.964942647999667 0.550781967323191
0.131930502877126 0.964942647999667
0.446646896622088 0.131930502877126
0.963898440803789 0.446646896622088
0.135713122817273 0.963898440803789
0.457450777338581 0.135713122817273
0.967939298238545 0.457450777338581
0.121027971339671 0.967939298238545
0.414882785822991 0.121027971339671
0.946744733417906 0.414882785822991
0.196634658337119 0.946744733417906
0.616080930963277 0.196634658337119
0.922448348120126 0.616080930963277
0.278995833365251 0.922448348120126
0.784512917487312 0.278995833365251
0.659304359153146 0.784512917487312
0.876026272503741 0.659304359153146
0.423556545309076 0.876026272503741
0.952209953116175 0.423556545309076
0.177474017380401 0.952209953116175
0.569310263087524 0.177474017380401
0.956264740979879 0.569310263087524
0.163107695940399 0.956264740979879
0.532363944315112 0.163107695940399
0.970915042922637 0.532363944315112
0.110132187161768 0.970915042922637
0.38221204519966 0.110132187161768
0.920891391045417 0.38221204519966
0.284116704081031 0.920891391045417
0.793238169918335 0.284116704081031
0.639644365241489 0.793238169918335
0.898947859899576 0.639644365241489
0.354278359818038 0.898947859899576
0.892184293972447 0.354278359818038
0.3751467702892 0.892184293972447
0.91420551702005 0.3751467702892
0.305891779713596 0.91420551702005
0.828055795387256 0.305891779713596
0.555279640940057 0.828055795387256
0.9630822290604 0.555279640940057
0.13866391160096 0.9630822290604
0.465801301759877 0.13866391160096
0.970438751250856 0.465801301759877
0.111880787153987 0.970438751250856
0.387517558817187 0.111880787153987
0.925656031659665 0.387517558817187
0.268386076575811 0.925656031659665
0.765784462856611 0.268386076575811
0.699498615285689 0.765784462856611
0.819781179746462 0.699498615285689
0.576185988611848 0.819781179746462
0.952363211043018 0.576185988611848
0.176933348649914 0.952363211043018
0.567948961263225 0.176933348649914
0.95699346078667 0.567948961263225
0.160512209513066 0.95699346078667
0.525517356430168 0.160512209513066
0.972460571631181 0.525517356430168
0.104445932190355 0.972460571631181
0.364794219813047 0.104445932190355
0.903705648315746 0.364794219813047
0.339384823120057 0.903705648315746
0.874390783327716 0.339384823120057
0.428343011301112 0.874390783327716
0.954974576285369 0.428343011301112
0.167692726242401 0.954974576285369
0.544330315650389 0.167692726242401
0.967335810145914 0.544330315650389
0.123229238165524 0.967335810145914
0.421370792804007 0.123229238165524
0.950888046325346 0.421370792804007
0.18212988175556 0.950888046325346
0.580938492916337 0.18212988175556
0.949450945421285 0.580938492916337
0.187176005873661 0.949450945421285
0.593350479925492 0.187176005873661
0.941014182800953 0.593350479925492
0.216475313216789 0.941014182800953
0.661493632739481 0.216475313216789
0.873287245679962 0.661493632739481
0.431560865629405 0.873287245679962
0.956732731057754 0.431560865629405
0.161441328284048 0.956732731057754
0.527974300643128 0.161441328284048
0.971948010163759 0.527974300643128
0.106333795239628 0.971948010163759
0.370604984995304 0.106333795239628
0.909702027358544 0.370604984995304
0.320362570235365 0.909702027358544
0.849148535927468 0.320362570235365
0.499571669452852 0.849148535927468
0.974999284478475 0.499571669452852
0.0950651510052528 0.974999284478475
0.335508295471445 0.0950651510052528
0.869475668651033 0.335508295471445
0.442602148072 0.869475668651033
0.962151397716801 0.442602148072
0.142022733794789 0.962151397716801
0.47522387983295 0.142022733794789
0.972605961090925 0.47522387983295
0.103910061611161 0.972605961090925
0.36313976675779 0.103910061611161
0.901950178571862 0.36313976675779
0.344900610389587 0.901950178571862
0.881182299435661 0.344900610389587
0.408330212928074 0.881182299435661
0.942226935538932 0.408330212928074
0.212297816186997 0.942226935538932
0.652187068373993 0.212297816186997
0.884672475256945 0.652187068373993
0.397905638440811 0.884672475256945
0.934349291217504 0.397905638440811
0.239228703553523 0.934349291217504
0.709793490703532 0.239228703553523
0.803348095907865 0.709793490703532
0.616121737565385 0.803348095907865
0.922411394052704 0.616121737565385
0.279117595280359 0.922411394052704
0.784722756812549 0.279117595280359
0.658838511836943 0.784722756812549
0.876604275913959 0.658838511836943
0.42185995551696 0.876604275913959
0.951187120447935 0.42185995551696
0.181077711133412 0.951187120447935
0.578325437289978 0.181077711133412
0.951073990906002 0.578325437289978
0.181475793439716 0.951073990906002
0.579315086357052 0.181475793439716
0.950465556597076 0.579315086357052
0.183615051046777 0.950465556597076
0.584612199895879 0.183615051046777
0.947079024952241 0.584612199895879
0.195469347246239 0.947079024952241
0.613318217980137 0.195469347246239
0.924920027747844 0.613318217980137
0.270827583073207 0.924920027747844
0.770172012947736 0.270827583073207
0.690327625337097 0.770172012947736
0.833724040630813 0.690327625337097
0.54065023234967 0.833724040630813
0.96855547857868 0.54065023234967
0.118777477626228 0.96855547857868
0.408210614896416 0.118777477626228
0.942141364250993 0.408210614896416
0.212592954671255 0.942141364250993
0.652849042152062 0.212592954671255
0.883884964221469 0.652849042152062
0.400266103554257 0.883884964221469
0.936207284609024 0.400266103554257
0.232920498930684 0.936207284609024
0.696807306423405 0.232920498930684
0.823940848139619 0.696807306423405
0.565743074935678 0.823940848139619
0.958143607582207 0.565743074935678
0.156407295843087 0.958143607582207
0.514581809235545 0.156407295843087
0.974170746273731 0.514581809235545
0.0981322031750315 0.974170746273731
0.345158868112679 0.0981322031750315
0.881494473115828 0.345158868112679
0.407401671230101 0.881494473115828
0.941559643085184 0.407401671230101
0.21459781823409 0.941559643085184
0.657327819108707 0.21459781823409
0.878467033604542 0.657327819108707
0.416374547450856 0.878467033604542
0.947726456375208 0.416374547450856
0.193209979020625 0.947726456375208
0.607931543807152 0.193209979020625
0.929568049220479 0.607931543807152
0.255338035246759 0.929568049220479
0.741548039712027 0.255338035246759
0.747452723593981 0.741548039712027
0.736191883385092 0.747452723593981
0.757432237469712 0.736191883385092
0.716541708134217 0.757432237469712
0.79212778568943 0.716541708134217
0.642179291629942 0.79212778568943
0.896161691223271 0.642179291629942
0.36291806618776 0.896161691223271
0.901713319346957 0.36291806618776
0.345642995331075 0.901713319346957
0.882078268927586 0.345642995331075
0.405663166011867 0.882078268927586
0.940292190837072 0.405663166011867
0.218956868082775 0.940292190837072
0.666957556208413 0.218956868082775
0.866288180257167 0.666957556208413
0.451748579115184 0.866288180257167
0.965920021492126 0.451748579115184
0.12838228093383 0.965920021492126
0.436411056416623 0.12838228093383
0.959230140390406 0.436411056416623
0.152519944812332 0.959230140390406
0.504104683862424 0.152519944812332
0.974934291124519 0.504104683862424
0.0953059345447966 0.974934291124519
0.336268582202824 0.0953059345447966
0.870448889021894 0.336268582202824
0.439793720427534 0.870448889021894
0.960863295210164 0.439793720427534
0.146659590199314 0.960863295210164
0.488087163727345 0.146659590199314
0.97444652889457 0.488087163727345
0.0971119157575874 0.97444652889457
0.341956647144367 0.0971119157575874
0.877586964610784 0.341956647144367
0.418969528208456 0.877586964610784
0.949392844300835 0.418969528208456
0.1873796788157 0.949392844300835
0.593847285652239 0.1873796788157
0.940651479205258 0.593847285652239
0.217722468109452 0.940651479205258
0.664245640460144 0.217722468109452
0.869791141400364 0.664245640460144
0.441692595793081 0.869791141400364
0.961740961797139 0.441692595793081
0.143501608374748 0.961740961797139
0.479344697397572 0.143501608374748
0.973336098050168 0.479344697397572
0.101216459302302 0.973336098050168
0.354789581907569 0.101216459302302
0.892764344461944 0.354789581907569
0.37337106190558 0.892764344461944
0.91246393694461 0.37337106190558
0.311506652808595 0.91246393694461
0.836434006451864 0.311506652808595
0.533567421280713 0.836434006451864
0.970605590091396 0.533567421280713
0.111268476441442 0.970605590091396
0.385662430108186 0.111268476441442
0.924014988433815 0.385662430108186
0.27382402937554 0.924014988433815
0.775493278217125 0.27382402937554
0.679003469263008 0.775493278217125
0.850035256168049 0.679003469263008
0.497153745813537 0.850035256168049
0.974968405464714 0.497153745813537
0.0951795538601992 0.974968405464714
0.335869584909973 0.0951795538601992
0.869938706685265 0.335869584909973
0.441266877854532 0.869938706685265
0.961546639415877 0.441266877854532
0.144201328611249 0.961546639415877
0.481288491208199 0.144201328611249
0.973634529811064 0.481288491208199
0.100114295465545 0.973634529811064
0.351356550905031 0.100114295465545
0.888829987660491 0.351356550905031
0.385363838714176 0.888829987660491
0.923748347050037 0.385363838714176
0.27470561965221 0.923748347050037
0.77704552451645 0.27470561965221
0.67565853164708 0.77704552451645
0.854661913012409 0.67565853164708
0.484438217087677 0.854661913012409
0.9740555405591 0.484438217087677
0.0985582434143563 0.9740555405591
0.346493612670788 0.0985582434143563
0.883099577291622 0.346493612670788
0.402615384128024 0.883099577291622
0.938013322706724 0.402615384128024
0.226762883612521 0.938013322706724
0.683831765090195 0.226762883612521
0.843202940360911 0.683831765090195
0.515625792737737 0.843202940360911
0.974047754945138 0.515625792737737
0.0985870315227519 0.974047754945138
0.346583752079307 0.0985870315227519
0.883207474008352 0.346583752079307
0.402292924270138 0.883207474008352
0.937767976674044 0.402292924270138
0.227600874534982 0.937767976674044
0.685614994138984 0.227600874534982
0.840633588408062 0.685614994138984
0.522478157948164 0.840633588408062
0.973029456419504 0.522478157948164
0.102348220101929 0.973029456419504
0.358304941581194 0.102348220101929
0.896697790636795 0.358304941581194
0.361260365325153 0.896697790636795
0.899930123704131 0.361260365325153
0.351217995000593 0.899930123704131
0.88866926845459 0.351217995000593
0.385851179059998 0.88866926845459
0.924183182044228 0.385851179059998
0.273267649476249 0.924183182044228
0.774510520781302 0.273267649476249
0.681111498523476 0.774510520781302
0.847074637900066 0.681111498523476
0.505202863333499 0.847074637900066
0.974894427831218 0.505202863333499
0.0954536014179529 0.974894427831218
0.336734624437752 0.0954536014179529
0.871043226855821 0.336734624437752
0.438075002837237 0.871043226855821
0.960044649432928 0.438075002837237
0.149599790059724 0.960044649432928
0.496156802207861 0.149599790059724
0.974942396339849 0.496156802207861
0.0952759086199513 0.974942396339849
0.336173798440716 0.0952759086199513
0.870327805162362 0.336173798440716
0.440143535222147 0.870327805162362
0.961027094134761 0.440143535222147
0.146070672047268 0.961027094134761
0.486462720178218 0.146070672047268
0.974285294014605 0.486462720178218
0.0977084935377726 0.974285294014605
0.343830020930572 0.0977084935377726
0.879882656786072 0.343830020930572
0.412187751585307 0.879882656786072
0.94492713521059 0.412187751585307
0.202955392977861 0.94492713521059
0.630881555612345 0.202955392977861
0.908193071761921 0.630881555612345
0.325175823045712 0.908193071761921
0.855802377893798 0.325175823045712
0.481278204751964 0.855802377893798
0.973633028092493 0.481278204751964
0.100119843329748 0.973633028092493
0.351373855175362 0.100119843329748
0.888850049390806 0.351373855175362
0.385302992446193 0.888850049390806
0.923693926186987 0.385302992446193
0.274885481957797 0.923693926186987
0.777361469689845 0.274885481957797
0.674975399012809 0.777361469689845
0.855596077987202 0.674975399012809
0.481850574348466 0.855596077987202
0.973715333559226 0.481850574348466
0.099815752728399 0.973715333559226
0.350425016119082 0.099815752728399
0.887746564368391 0.350425016119082
0.388645147099987 0.887746564368391
0.926640377268904 0.388645147099987
0.265114155087003 0.926640377268904
0.75983169545208 0.265114155087003
0.71170121115014 0.75983169545208
0.800212129070499 0.71170121115014
0.623503442479937 0.800212129070499
0.91551290881286 0.623503442479937
0.301661188178518 0.91551290881286
0.821580691328446 0.301661188178518
0.571684849962403 0.821580691328446
0.954959000914884 0.571684849962403
0.167747999197468 0.954959000914884
0.544473571054586 0.167747999197468
0.967286195762845 0.544473571054586
0.123410083873002 0.967286195762845
0.421902136778697 0.123410083873002
0.95121282266504 0.421902136778697
0.180987255784324 0.95121282266504
0.578100389209137 0.180987255784324
0.951211283900987 0.578100389209137
0.180992671393648 0.951211283900987
0.578113864752213 0.180992671393648
0.951203074120545 0.578113864752213
0.181021565026263 0.951203074120545
0.578185756284652 0.181021565026263
0.951159251305368 0.578185756284652
0.181175786850161 0.951159251305368
0.578569372326607 0.181175786850161
0.950924729555592 0.578569372326607
0.182000864273343 0.950924729555592
0.580618543740687 0.182000864273343
0.949652536580011 0.580618543740687
0.186469125753933 0.949652536580011
0.591623724488537 0.186469125753933
0.94225986313231 0.591623724488537
0.212184232502544 0.94225986313231
0.651932127521417 0.212184232502544
0.884974851644582 0.651932127521417
0.396998018044804 0.884974851644582
0.933623307681875 0.396998018044804
0.241686225436615 0.933623307681875
0.714768576096186 0.241686225436615
0.795110389014305 0.714768576096186
0.63534844735372 0.795110389014305
0.903555111415856 0.63534844735372
0.339858760995531 0.903555111415856
0.874983655923443 0.339858760995531
0.426610305382128 0.874983655923443
0.953994415623192 0.426610305382128
0.171167375273529 0.953994415623192
0.55328950917045 0.171167375273529
0.963924890028253 0.55328950917045
0.135617416007867 0.963924890028253
0.457178796684543 0.135617416007867
0.967848743731803 0.457178796684543
0.121358456655436 0.967848743731803
0.415859268449222 0.121358456655436
0.94738931544699 0.415859268449222
0.194386921653109 0.94738931544699
0.610742520739014 0.194386921653109
0.927170766991439 0.610742520739014
0.263348029728988 0.927170766991439
0.756583795370707 0.263348029728988
0.718242548217336 0.756583795370707
0.789243741575654 0.718242548217336
0.648718426041333 0.789243741575654
0.888743036047575 0.648718426041333
0.385627522505604 0.888743036047575
0.923983851927996 0.385627522505604
0.273927003886766 0.923983851927996
0.775674901670704 0.273927003886766
0.678613059496506 0.775674901670704
0.850579762411461 0.678613059496506
0.495665937731343 0.850579762411461
0.974926742026581 0.495665937731343
0.0953338998612693 0.974926742026581
0.336356854854192 0.0953338998612693
0.870561592082473 0.336356854854192
0.43946801524588 0.870561592082473
0.960709927404735 0.43946801524588
0.147210814883822 0.960709927404735
0.489605184373747 0.147210814883822
0.974578596451575 0.489605184373747
0.0966231075812061 0.974578596451575
0.340419622383971 0.0966231075812061
0.87568300201171 0.340419622383971
0.42456289979793 0.87568300201171
0.952806051261101 0.42456289979793
0.175370051771183 0.952806051261101
0.564000047180526 0.175370051771183
0.959025576447473 0.564000047180526
0.153252528651555 0.959025576447473
0.506088145342492 0.153252528651555
0.974855444496526 0.506088145342492
0.0955979966448264 0.974855444496526
0.337190176761056 0.0955979966448264
0.871622549681925 0.337190176761056
0.43639705421483 0.871622549681925
0.95922319462105 0.43639705421483
0.152544824335946 0.95922319462105
0.50417211352662 0.152544824335946
0.974932114528012 0.50417211352662
0.0953139977008105 0.974932114528012
0.336294034218092 0.0953139977008105
0.870481391392909 0.336294034218092
0.439699800663139 0.870481391392909
0.960819155243746 0.439699800663139
0.146818254025719 0.960819155243746
0.488524351811177 0.146818254025719
0.974486407044721 0.488524351811177
0.096964333166187 0.974486407044721
0.341492779913312 0.096964333166187
0.877014298603522 0.341492779913312
0.420654852729127 0.877014298603522
0.950446955657797 0.420654852729127
0.183680406540627 0.950446955657797
0.584773467695419 0.183680406540627
0.946972490782086 0.584773467695419
0.195840810687827 0.946972490782086
0.61420003147215 0.195840810687827
0.924137575965864 0.61420003147215
0.273418534949824 0.924137575965864
0.774777274813292 0.273418534949824
0.680540052060104 0.774777274813292
0.847880629448326 0.680540052060104
0.503018363853083 0.847880629448326
0.974964468970637 0.503018363853083
0.0951941375402206 0.974964468970637
0.335915633500957 0.0951941375402206
0.86999765061537 0.335915633500957
0.441096780302514 0.86999765061537
0.961468601766152 0.441096780302514
0.144482245377572 0.961468601766152
0.482067791978085 0.144482245377572
0.973745900070289 0.482067791978085
0.0997028064497072 0.973745900070289
0.350072411659462 0.0997028064497072
0.887334701192121 0.350072411659462
0.389890134084399 0.887334701192121
0.927715687969389 0.389890134084399
0.261531232033003 0.927715687969389
0.75321732214678 0.261531232033003
0.724935852282774 0.75321732214678
0.777675063195506 0.724935852282774
0.674296581189549 0.777675063195506
0.856520736963976 0.674296581189549
0.479282560047188 0.856520736963976
0.973326071959026 0.479282560047188
0.10125347545497 0.973326071959026
0.354904715736714 0.10125347545497
0.892894698089769 0.354904715736714
0.372971649230501 0.892894698089769
0.912068812593046 0.372971649230501
0.312777245383812 0.912068812593046
0.838295796600313 0.312777245383812
0.528668220609983 0.838295796600313
0.971794719195524 0.528668220609983
0.106897997462946 0.971794719195524
0.372336180845296 0.106897997462946
0.911437602187457 0.372336180845296
0.314804488074321 0.911437602187457
0.841240227214084 0.314804488074321
0.520864918590432 0.841240227214084
0.973302155171637 0.520864918590432
0.101341772648546 0.973302155171637
0.355179309283487 0.101341772648546
0.893205173407529 0.355179309283487
0.372019797261665 0.893205173407529
0.911122164057513 0.372019797261665
0.315816408260602 0.911122164057513
0.842697977682502 0.315816408260602
0.516976574760061 0.842697977682502
0.973876004046723 0.516976574760061
0.0992219778759708 0.973876004046723
0.348570210231169 0.0992219778759708
0.885569173205214 0.348570210231169
0.395212009428005 0.885569173205214
0.932175960424346 0.395212009428005
0.246573363002048 0.932175960424346
0.724522264673642 0.246573363002048
0.778400035396694 0.724522264673642
0.672724339135367 0.778400035396694
0.858648580413978 0.672724339135367
0.47334766349145 0.858648580413978
0.972229646538677 0.47334766349145
0.105296727626831 0.972229646538677
0.367416374433855 0.105296727626831
0.906444170703772 0.367416374433855
0.3307322307936 0.906444170703772
0.863258847000767 0.3307322307936
0.460367739295125 0.863258847000767
0.968874207254541 0.460367739295125
0.11761221330856 0.968874207254541
0.404740364297964 0.11761221330856
0.93960984704307 0.404740364297964
0.221298411292798 0.93960984704307
0.672069155363119 0.221298411292798
0.85952960251323 0.672069155363119
0.470880013175049 0.85952960251323
0.971692902832527 0.470880013175049
0.107272641127954 0.971692902832527
0.37348436421421 0.107272641127954
0.912575796216697 0.37348436421421
0.311146728267017 0.912575796216697
0.835904322847417 0.311146728267017
0.53495631498043 0.835904322847417
0.970234418567657 0.53495631498043
0.112630407217917 0.970234418567657
0.389784714492588 0.112630407217917
0.927625104278026 0.389784714492588
0.261833403745708 0.927625104278026
0.753779022471715 0.261833403745708
0.723825210237873 0.753779022471715
0.779618873521691 0.723825210237873
0.670071813724797 0.779618873521691
0.862194754887796 0.670071813724797
0.4633783421739 0.862194754887796
0.969769531294465 0.4633783421739
0.114334691122776 0.969769531294465
0.394922851161677 0.114334691122776
0.931939291888835 0.394922851161677
0.24737094767803 0.931939291888835
0.726096391499323 0.24737094767803
0.77563364482884 0.726096391499323
0.678701765969638 0.77563364482884
0.850456147473397 0.678701765969638
0.496003905922607 0.850456147473397
0.974937721805286 0.496003905922607
0.0952932255849784 0.974937721805286
0.336228464296096 0.0952932255849784
0.870397647963422 0.336228464296096
0.439941771294344 0.870397647963422
0.960932735742483 0.439941771294344
0.146409951171707 0.960932735742483
0.487398901741462 0.146409951171707
0.974380728058447 0.487398901741462
0.0973554069022781 0.974380728058447
0.342721593431763 0.0973554069022781
0.878527661026689 0.342721593431763
0.416195558366888 0.878527661026689
0.947609580693993 0.416195558366888
0.193618086756695 0.947609580693993
0.608907480625751 0.193618086756695
0.928742726588631 0.608907480625751
0.25810073014965 0.928742726588631
0.746790498658884 0.25810073014965
0.737468354109628 0.746790498658884
0.755074245106211 0.737468354109628
0.721254804985637 0.755074245106211
0.784080613955994 0.721254804985637
0.660262998620103 0.784080613955994
0.874831507965843 0.660262998620103
0.427055228480605 0.874831507965843
0.954248335201134 0.427055228480605
0.170267954871292 0.954248335201134
0.550979435819434 0.170267954871292
0.964864278781776 0.550979435819434
0.132214689019528 0.964864278781776
0.447462463605292 0.132214689019528
0.964235248351341 0.447462463605292
0.134493973333857 0.964235248351341
0.453980843435842 0.134493973333857
0.966740725193582 0.453980843435842
0.125397072238501 0.966740725193582
0.427723321398802 0.125397072238501
0.954626718748479 0.427723321398802
0.168926731739967 0.954626718748479
0.54752291507012 0.168926731739967
0.966192132918629 0.54752291507012
0.12739309129864 0.966192132918629
0.433539957193263 0.12739309129864
0.957773944569494 0.433539957193263
0.157727761125815 0.957773944569494
0.518113886534605 0.157727761125815
0.973720359746985 0.518113886534605
0.0997971809686305 0.973720359746985
0.350367044193445 0.0997971809686305
0.887678916292715 0.350367044193445
0.388849725662213 0.887678916292715
0.926817904407075 0.388849725662213
0.264523258262457 0.926817904407075
0.758747745992614 0.264523258262457
0.713893455380592 0.758747745992614
0.796573400006867 0.713893455380592
0.631972451792631 0.796573400006867
0.907074760674583 0.631972451792631
0.328731542964752 0.907074760674583
0.860601750936585 0.328731542964752
0.467868871163731 0.860601750936585
0.970973603182858 0.467868871163731
0.109917073909329 0.970973603182858
0.381557712012908 0.109917073909329
0.920288555223892 0.381557712012908
0.286094368356471 0.920288555223892
0.796553084929614 0.286094368356471
0.632019444493044 0.796553084929614
0.907026378475417 0.632019444493044
0.328885156178227 0.907026378475417
0.860806869873016 0.328885156178227
0.467291770344503 0.860806869873016
0.970827669679933 0.467291770344503
0.11045309130868 0.970827669679933
0.383187503123243 0.11045309130868
0.921783878236328 0.383187503123243
0.281183604233696 0.921783878236328
0.788265601280993 0.281183604233696
0.650921478160619 0.788265601280993
0.886168558976273 0.650921478160619
0.393407991826935 0.886168558976273
0.930688760795169 0.393407991826935
0.251578046166413 0.930688760795169
0.734317478128663 0.251578046166413
0.760871745829352 0.734317478128663
0.709589135688989 0.760871745829352
0.803682337384455 0.709589135688989
0.615330448046784 0.803682337384455
0.923125662237979 0.615330448046784
0.276762228427118 0.923125662237979
0.780643099638378 0.276762228427118
0.667833857438917 0.780643099638378
0.865144005558975 0.667833857438917
0.455012435296954 0.865144005558975
0.967106864186148 0.455012435296954
0.124063591977716 0.967106864186148
0.42382008678088 0.124063591977716
0.952366821205517 0.42382008678088
0.176920610382423 0.952366821205517
0.567916861216903 0.176920610382423
0.957010469853531 0.567916861216903
0.160451578732592 0.957010469853531
0.525356791497763 0.160451578732592
0.972492429187262 0.525356791497763
0.104328527006808 0.972492429187262
0.364431933291653 0.104328527006808
0.903323067226948 0.364431933291653
0.340588963426322 0.903323067226948
0.875893673532171 0.340588963426322
0.423945390174109 0.875893673532171
0.952441215664503 0.423945390174109
0.176658090835304 0.952441215664503
0.567255038133146 0.176658090835304
0.957359363398265 0.567255038133146
0.159207409576541 0.957359363398265
0.522055600218626 0.159207409576541
0.973102846946085 0.522055600218626
0.102077415224886 0.973102846946085
0.357464704450981 0.102077415224886
0.895766389138739 0.357464704450981
0.364138964389528 0.895766389138739
0.903012938111115 0.364138964389528
0.341564229688683 0.903012938111115
0.877102616074852 0.341564229688683
0.42039510610306 0.877102616074852
0.950285937383862 0.42039510610306
0.184246040917914 0.950285937383862
0.586167805563554 0.184246040917914
0.94604292620901 0.586167805563554
0.199078261117725 0.94604292620901
0.621839817565464 0.199078261117725
0.917104729536896 0.621839817565464
0.296492213932015 0.917104729536896
0.813479865937859 0.296492213932015
0.59174845724117 0.813479865937859
0.942170660316074 0.59174845724117
0.212491917907023 0.942170660316074
0.652622500651751 0.212491917907023
0.884154851949745 0.652622500651751
0.399457693921532 0.884154851949745
0.935575854284852 0.399457693921532
0.235067333139683 0.935575854284852
0.701261659918092 0.235067333139683
0.817025602563355 0.701261659918092
0.583029592545432 0.817025602563355
0.948113738370784 0.583029592545432
0.191856902185106 0.948113738370784
0.604686541949258 0.191856902185106
0.932258838945355 0.604686541949258
0.246293954999094 0.932258838945355
0.723969646646994 0.246293954999094
0.779366629785201 0.723969646646994
0.670621716033588 0.779366629785201
0.861464097069239 0.670621716033588
0.465440455466688 0.861464097069239
0.970341987738435 0.465440455466688
0.112235816824005 0.970341987738435
0.38859185915883 0.112235816824005
0.926594082001824 0.38859185915883
0.265268207883981 0.926594082001824
0.760113844503008 0.265268207883981
0.711129072801674 0.760113844503008
0.801155607009831 0.711129072801674
0.621290671429506 0.801155607009831
0.9176254347943 0.621290671429506
0.2947970852302 0.9176254347943
0.810777878702915 0.2947970852302
0.598326729424774 0.810777878702915
0.937294231694446 0.598326729424774
0.229217644214379 0.937294231694446
0.689039971601238 0.229217644214379
};
\addplot [semithick, steelblue31119180, mark=*, mark size=3, mark options={solid}, only marks]
table {%
0.362725916435377 0.104295456887201
0.900801002907254 0.362725916435377
0.348806854897003 0.900801002907254
0.884944343487437 0.348806854897003
0.397395194974912 0.884944343487437
0.933718818797754 0.397395194974912
0.243169005699116 0.933718818797754
0.715837885320507 0.243169005699116
0.79348512799271 0.715837885320507
0.637602084401319 0.79348512799271
0.90194646168511 0.637602084401319
0.344842734314163 0.90194646168511
0.880922890047422 0.344842734314163
0.4113253839128 0.880922890047422
0.944378604486243 0.4113253839128
0.205667985868587 0.944378604486243
0.63492767909954 0.205667985868587
0.907167986582991 0.63492767909954
0.327478708054937 0.907167986582991
0.859061772293909 0.327478708054937
0.473966079316195 0.859061772293909
0.97228463319275 0.473966079316195
0.106838774316053 0.97228463319275
0.372371587498467 0.106838774316053
0.911227677570523 0.372371587498467
0.314493556914741 0.911227677570523
0.841996457355791 0.314493556914741
0.516578791647294 0.841996457355791
0.974698834499304 0.516578791647294
0.0987994363977904 0.974698834499304
0.348440511085709 0.0987994363977904
0.883918150670901 0.348440511085709
};
\addlegendentry{classical}
\addplot [semithick, darkorange25512714, mark=square*, mark size=3, mark options={solid}, only marks]
table {%
0.380355945065228 0.112371991969281
0.914751270716367 0.380355945065228
0.298587077356125 0.914751270716367
0.819844950474381 0.298587077356125
0.577764171506969 0.819844950474381
0.950598279885858 0.577764171506969
0.19062165065039 0.950598279885858
0.59561937484085 0.19062165065039
0.936569353082301 0.59561937484085
0.221274335591213 0.936569353082301
0.686667725522728 0.221274335591213
0.836463137549119 0.686667725522728
0.52910384620363 0.836463137549119
0.97826495594534 0.52910384620363
0.126066433017273 0.97826495594534
0.39187625353578 0.126066433017273
0.947499273462323 0.39187625353578
0.228765009263566 0.947499273462323
0.70680796581077 0.228765009263566
0.830608055952972 0.70680796581077
0.579752563658229 0.830608055952972
0.946308805431367 0.579752563658229
0.180487485637981 0.946308805431367
0.5920806486143 0.180487485637981
0.930685752137337 0.5920806486143
0.219999457990021 0.930685752137337
0.677142890993029 0.219999457990021
0.841985534323281 0.677142890993029
0.519459410789605 0.841985534323281
0.9766316763145 0.519459410789605
0.128836980075621 0.9766316763145
0.395799713251612 0.128836980075621
};
\addlegendentry{$\langle Z_1, Z_2, Z_3\rangle$}
\addplot [semithick, forestgreen4416044, mark=triangle*, mark size=3, mark options={solid}, only marks]
table {%
0.36923063193854 0.106679971114714
0.914505395723844 0.36923063193854
0.320817262551907 0.914505395723844
0.837890180153985 0.320817262551907
0.520784490010515 0.837890180153985
0.978787661639612 0.520784490010515
0.104026751793486 0.978787661639612
0.363253739604343 0.104026751793486
0.921051139726275 0.363253739604343
0.320214018704041 0.921051139726275
0.842693189164104 0.320214018704041
0.520527255608948 0.842693189164104
0.974240210640877 0.520527255608948
0.0983758632565404 0.974240210640877
0.360079558829734 0.0983758632565404
0.90868160083447 0.360079558829734
0.347508641767356 0.90868160083447
0.879755846261265 0.347508641767356
0.416356070538921 0.879755846261265
0.93943730383201 0.416356070538921
0.198781149277007 0.93943730383201
0.618693848648849 0.198781149277007
0.909354610419203 0.618693848648849
0.29014636148072 0.909354610419203
0.805178137870637 0.29014636148072
0.610026106071778 0.805178137870637
0.91814603085141 0.610026106071778
0.267428745496928 0.91814603085141
0.780593873568116 0.267428745496928
0.695106099103756 0.780593873568116
0.823811298432036 0.695106099103756
0.546528501634934 0.823811298432036
};
\addlegendentry{$\langle Z_1Z_2, Z_2Z_3, Z_3Z_4\rangle$}
\end{axis}

\end{tikzpicture}

%% file: figures/convergence_ep4_4_casenamehenon_num_qubits4_num_meas2_degree2_num_reservoirs20_timeplex10_methodquantum_stab_noiseNone.pickle.tex
\begin{tikzpicture}

\definecolor{darkgray176}{RGB}{176,176,176}
\definecolor{darkorange25512714}{RGB}{255,127,14}
\definecolor{forestgreen4416044}{RGB}{44,160,44}
\definecolor{lightgray204}{RGB}{204,204,204}
\definecolor{steelblue31119180}{RGB}{31,119,180}

\begin{axis}[
width=1.1\textwidth,
height=.7\textwidth,
legend cell align={left},
legend style={
  fill opacity=0,
  draw opacity=1,
  text opacity=1,
  at={(.79,1.01)},
  anchor=south west,
  draw=lightgray204
},
tick align=outside,
tick pos=left,
x grid style={darkgray176},
xmin=-1.6, xmax=33.6,
xtick style={color=black},
y grid style={darkgray176},
ymin=-1.5, ymax=1.5,
ytick style={color=black}
]
\addplot [ultra thick, black]
table {%
0 0.525609726081405
1 0.516509765960711
2 0.784187644158518
3 0.294022564036972
4 1.11422731781953
5 -0.649896752874333
6 0.742956090190615
7 0.0322537214058436
8 1.22143040349485
9 -1.07897300639247
10 -0.263426726884594
11 0.579157001470277
12 0.451380016641763
13 0.888505613248158
14 0.0301948903096506
15 1.26527526001331
16 -1.23223160994956
17 -0.746170058778433
18 -0.149147142249241
19 0.745006164308909
20 0.178207998523646
21 1.17904052225975
22 -0.892728774825658
23 0.237961625115515
24 0.652905338513639
25 0.474588954051202
26 0.880543056123712
27 0.0568781830525983
28 1.2596337380468
29 -1.20428456072029
30 -0.652531703050915
31 0.0425979589027746
32 0.80170006854097
};
\addlegendentry{exact}
\addplot [semithick, steelblue31119180, mark=x, mark size=3, mark options={solid}]
table {%
0 0.559580036070813
1 0.56975510971624
2 1.05614090670456
3 0.978797654649568
4 3.55036813368601
5 6.07610535919211
6 11.9335332899552
7 17.3319386869184
8 21.0864931400224
9 23.1537004589756
10 24.0282181296589
11 24.2887089773872
12 24.3645240004243
13 24.488690739929
14 24.7418108570852
15 25.1152437814075
16 25.5620098493739
17 26.0288628502264
18 26.4725794125089
19 26.8654776086593
20 27.1946720593663
21 27.4584838807925
22 27.6622552528593
23 27.8147787533587
24 27.9257785809387
25 28.0044149618673
26 28.0585749124608
27 28.0946730083534
28 28.1177311507444
29 28.1315781095364
30 28.1390758506496
31 28.142327846898
32 28.1428538407235
};
\addlegendentry{classical}
\addplot [semithick, darkorange25512714, mark=square*, mark size=3, mark options={solid}]
table {%
0 0.411366183480097
1 0.571950815910014
2 0.610200432074848
3 0.599968220312488
4 0.628297194601489
5 0.6169199886071
6 0.651075515099126
7 0.605189918296178
8 0.665926295606856
9 0.545663455517172
10 0.761664888510997
11 0.395644861358106
12 1.04203852875008
13 -0.399816006775749
14 1.05146526217818
15 -0.643384573894101
16 0.715219407611963
17 0.156133189831395
18 1.193351792301
19 -0.89273349331469
20 0.207044468249943
21 0.624115896017661
22 0.436952142693056
23 0.840688545482025
24 0.078671129283193
25 1.31548635342658
26 -1.14127310268159
27 -0.551764837823674
28 0.224298952114665
29 0.741396471778211
30 0.273348798238149
31 1.13885287868573
32 -0.733650932828304
};
\addlegendentry{$\langle Z_1, Z_2\rangle$}
\addplot [semithick, forestgreen4416044, mark=triangle*, mark size=3, mark options={solid}]
table {%
0 0.32995896183196
1 0.543557756583711
2 0.522357718244538
3 0.735037302516332
4 0.399083076153494
5 0.882261260924414
6 -0.169490449854514
7 1.1998633145822
8 -1.13764802369951
9 -0.319425653495932
10 0.394025313945383
11 0.57128079625161
12 0.570805377362703
13 0.708948196136798
14 0.401781843042258
15 0.877271928796473
16 -0.102124771531921
17 1.26299132563859
18 -1.21101004427286
19 -0.588708556868005
20 0.0755536812652631
21 0.73830144439868
22 0.172994657095135
23 1.14840325238402
24 -0.821171331563737
25 0.352531431189646
26 0.539772741986702
27 0.729036343276474
28 0.471410660028916
29 0.948480575777066
30 -0.161491211684226
31 1.26450325798132
32 -1.18595475008473
};
\addlegendentry{$\langle Z_1Z_2, Z_2Z_3\rangle$}
\end{axis}

\end{tikzpicture}